\documentclass[12pt]{article}

\usepackage{amsmath}
\usepackage[hidelinks]{hyperref}
\usepackage{graphicx,psfrag,epsf}
\usepackage{enumerate}
\usepackage{natbib}
\usepackage{url} 
\usepackage{amsmath} 
\usepackage{amssymb}
\usepackage[section]{placeins}
\usepackage{booktabs}
\usepackage[toc,page]{appendix}
\usepackage{multirow}
\usepackage{hyperref}
\usepackage{natbib}
\usepackage{float}
\usepackage{comment}
\usepackage{graphicx}
\usepackage{dsfont}
\usepackage{rotating}
\usepackage{lscape}
\usepackage{amsthm}
\usepackage{caption}
\usepackage{subcaption}
\usepackage{dsfont}
\usepackage[showframe=false]{geometry}
\usepackage{changepage}

\usepackage{tikz}
\usetikzlibrary{bayesnet}
\usepackage[frame,graph,arrow]{xypic}
\usepackage{pifont}
\newcommand{\cmark}{\ding{51}}%
\newcommand{\xmark}{\ding{55}}

\newtheorem{theorem}{Theorem}

\newtheorem{proposition}{Proposition}
\usepackage{setspace}
\usepackage{authblk}
\newcommand{\Tr}{\text{Tr}}

\usepackage[english]{babel}

\usepackage{geometry}
\usepackage{setspace} 
\usepackage{footmisc}

\title{Donor's Deferral and Return Behavior: Partial Identification from a Regression Discontinuity Design with Manipulation}
\author[1]{Evan T. R. Rosenman}
\author[2]{Karthik Rajkumar}
\author[3]{Romain Gauriot}
\author[4]{Robert Slonim}

\affil[1]{Mathematical Sciences Department, Claremont McKenna College}
\affil[2]{Data Science Applied Research, LinkedIn Corporation}
\affil[3]{Department of Economics, Deakin University}
\affil[4]{Department of Economics, University of Technology Sydney}

	\date{\today}

\begin{document}

\begin{titlepage}
\maketitle
\let\thefootnote\relax\footnotetext{An earlier version of the paper was circulated as "Optimized Partial Identification Bounds for Regression Discontinuity Designs with Manipulation". We thank Dr Eiman al Zaabi, Dr. Naima Oumeziane, Dr Jaishen Rajah, and the Abu Dhabi Blood Bank for sharing their data as well as for all their time helping us understand the institutional context and patiently answering our questions. We also thank Rebecca Diamond, Guido Imbens, Umair Khalil, Andrea La Nauze, Art Owen, Christoph Rothe, Stefan Wager and John Wooders for valuable feedback on our manuscript. Any remaining errors are our own. Gauriot is grateful for financial support from Tamkeen under the NYU Abu Dhabi Research Institute Award CG005 and from the Australian Research Council's Discovery Projects funding scheme (project number DP240100895). Views expressed in this work are personal and do not represent the opinion of LinkedIn or other institutions. This study uses preexisting data and was not pre-registered.}
\end{titlepage}

\newpage
\begin{abstract}
\doublespacing
Volunteer labor can temporarily yield lower benefits to charities than its costs. In such instances, organizations may wish to defer volunteer donations to a later date. Exploiting a discontinuity in blood donations’ eligibility criteria, we show that deferring donors reduces their future volunteerism. In our setting, medical staff manipulates donors’ reported hemoglobin levels over a threshold to facilitate donation. Such manipulation invalidates standard regression discontinuity design. To circumvent this issue, we propose a procedure for obtaining partial identification bounds where manipulation is present. Our procedure is applicable in various regression discontinuity settings where the running variable is manipulated.

\end{abstract}

\noindent%

\newpage
%
\section{Introduction}

Charitable organizations rely on volunteer labor, with Americans volunteering more than four billion hours per year, worth \$123 billion \citep{americops2023}. However, accepting volunteers may not always be optimal. The costs associated with managing volunteers—from training to insurance—may temporarily outweigh the benefits, particularly when volunteer demand is met or exceeded. In such instances, a potentially Pareto optimal solution is to temporarily reject donors. However, whether this is optimal in the long run is unclear and depends on how rejections affect future volunteerism. If temporary rejection reduces one's subsequent willingness to volunteer, then not rejecting volunteers, even if temporarily more costly, could still be long-term optimal.

Empirical evidence of the effect of turning away volunteers on future behavior is scant. Exploiting a discontinuity in blood donors' eligibility criteria, we show that, in the context of blood donation, turning away donors reduces their future volunteerism. Our estimation is complicated by an econometric issue often faced by researchers attempting to implement a regression discontinuity design: the running variable is manipulated. In our setting, this manipulation occurs as nurses bump up medical results of some donors above the eligibility threshold to allow the donation to occur. When the running variable is manipulated around the threshold, identifying the causal effect by comparing observations just below and just above the threshold is invalid.

To clarify the mechanisms driving this manipulation and their implications for identification, we develop two conceptual frameworks. These frameworks model how agents adjust the running variable to exceed the threshold, either to pursue an organizational objective or to correct perceived measurement error using observable donor characteristics.

To address the resulting identification challenge in regression discontinuity design, we develop a new procedure for obtaining partial identification bounds when manipulation is present. While we apply our method in the context of blood donation, the approach is broadly applicable to other settings where the running variable may be subject to manipulation—such as teachers inflating grades \citep{diamond2016long}, officers under-recording speeds \citep{goncalves2022should}, researchers inflating the value of reported statistical tests \citep{ brodeur2016star, brodeur2024preregistration}, or manipulation near union certification thresholds \citep{frandsen2017party}.

This paper focuses on causally identifying the effect of an organization’s decision to reject donors on their likelihood to volunteer again, ceteris paribus. A significant challenge in identifying this causal effect is that rejections occur for many reasons, including reasons affecting future donations. For instance, individuals who are rejected might, in part, be rejected because they are less prepared to be successful volunteers (e.g., not bringing the right equipment, not training properly, not understanding how hard the volunteer work might be). This under-preparedness could result from receiving less utility from volunteering than a person who is not rejected. In this case, rejections would be correlated with reduced future volunteering, but this would be due to those being rejected being less motivated volunteers rather than rejections, causally decreasing future donations.

Given the challenge to identification, empirical evidence on the effect of turning away volunteers on future behavior is limited. One context that has received some attention to identify temporary rejections on future volunteering is blood donations (e.g., \cite{custer2007consequences,bruhin2020sting}). Understanding how a temporary rejection (henceforth a deferral) affects future donation in the context of blood donation is crucial. The costs of collecting a unit of blood are non-trivial; the process of collecting a unit of blood typically requires more than an hour of a donor’s time (including about 12 uncomfortable minutes ‘needle in’ time), the marginal costs of staff time, equipment, needles, bags, and storage. If the collection is not used, there are also additional disposal costs.  

Despite these costs, in cases where there is already excess supply, the prevailing view across blood banks is that the risk of deferrals reducing future donations is too high, and thus donors will not be deferred unless there is a medical concern for the donor or if the donor is unable to provide a safe blood donation. Therefore, blood banks usually accept donations even when they know they do not need it. Such a policy led to wastage in periods of excess supply, since a whole blood donation can only be stored for a limited time (six weeks or less).  For example, after the 9/11 attacks, the American Red Cross allowed donors across the USA to donate in large numbers despite only 260 additional units of whole blood needed to treat victims of the 9/11 attacks. Blood banks accepted those donations, even though they did not need them, fearing that deferring donors would adversely deter future donations. This led to 200,000 units of blood being wasted \citep{korcok2002blood}.  

The evidence presented to date in the context of blood donations suggests that there is a negative correlation between receiving a temporary deferral and future donations. Early work in this direction was merely correlational and did not attempt to identify whether the relationship was causal (e.g., see \cite{custer2007consequences}). 

To identify a causal effect of deferrals on return behavior, we exploit a discontinuity in the blood donor’s eligibility criteria. Specifically, when a donor attempts to give blood, his hemoglobin level (henceforth h-level) is measured. If the h-level is below an eligibility threshold (13.5 g/dL for male donors and 12.5 g/dL for female donors), the donor is ineligible to donate blood, and the volunteer receives a temporary deferral. In contrast, if the donor’s h-level is at or above the threshold, the donor is eligible to donate. Thus, the h-level threshold provides a potentially good natural experiment to use with a regression discontinuity design to identify the causal effect of the temporary deferral.

 There is, however, a critical assumption in the RDD analysis that is a threat to using current RDD analyses. In particular, the RDD analyses requires the running variable, the h-level, to not be manipulated \citep{mccrary2008manipulation,frandsen2017party}. For instance, if the h-level is manipulated by a blood center staff member to increase the observed h-level to be at or just above the threshold, then the estimates from the RDD will be biased and unreliable. In section \ref{sec:discontinuitycovariates}, we  show that there is ample evidence of manipulation in our data.\footnote{We are not the first to exploit the h-level threshold to run a regression discontinuity analysis. \cite{bruhin2020sting} exploit the same discontinuity in h-level using a different dataset of blood donors to also try to establish a causal relationship between deferral and return behavior. We discuss this paper in Section \ref{sec:bruhinmanipulation}.}


To circumvent this issue, we introduce a new method to use in RDD designs in which the running variable has been manipulated. We propose a ``partial identification" method, where we bound the causal effect estimand rather than seeking to point-identify it \citep{manski1990nonparametric, manski2002inference}.  We propose a two-step procedure: first, we estimate the number of manipulators in the data using a continuity and log-concavity assumption on the un-manipulated density of the running variable.\footnote{By the un-manipulated density, we mean the counterfactual density of the running variable \textit{had there been} no manipulation.} Other than rounding effects, this step builds from the density model used in \cite{diamond2016long}. In the second step, we introduce a novel partial bounds estimator, which estimates best- and worst-case bounds consistent with the un-manipulated density. We also provide inferential tools that account for sampling uncertainty.

 We believe the proposed method to be useful outside of the blood donation context we are studying here. Indeed, as we discuss in Section \ref{sec:litReview} manipulation of the running variable is a key roadblock to inference in regression design.

 In a related paper, \cite{gerard2020bounds} 
 propose a method for bounding the causal effect when the running variable is manipulated. Their approach differs from ours in its assumptions about the density of the running variable if there were no manipulation. We impose a log-concavity assumption under which we can approximately recover this ``un-manipulated" distribution. 
As a result, while the method of \cite{gerard2020bounds} works in our setting, it provides wider bounds.\footnote{Section \ref{sec:comparaisongerard2020} shows the results of \cite{gerard2020bounds}'s method in our settings.}

 
By providing evidence that deferring donors reduces their future voluntarism, this paper contributes to the extensive literature on the determinants of charitable giving  (see \cite{list2011market,andreoni2013charitable} for reviews). For instance, prior studies have shown that risk \citep{exley2016excusing}, social pressure \citep{dellavigna2012testing,andreoni2017avoiding}, economic conditions \citep{list2011charitable,exley2023nonprofits} and gender \citep{andreoni2001fair,dellavigna2013importance,lilley2016gender} significantly influence donation behavior. Additionally, this literature explores donation-boosting strategies such as donation matching or seed money (e.g., \cite{list2002effects,eckel2003rebate,karlan2007does,gneezy2014avoiding,karlan2020can}).

This paper also contributes to our knowledge of the blood donations market (see \cite{slonim2014market} for an introduction to this market). This market is ideal for studying charitable giving. For instance, \cite{lacetera2012will,lacetera2013economic} and \cite{goette2020blood} explore the effect of incentives on blood donations. \cite{craig2017waiting} argue that the mechanism that causes the delay in return is due to donors adjusting their expectations of the cost to make a donation, with longer current wait times causing them to expect longer future wait times. A similar  mechanism could also provide an explanation for the current results. Specifically, a deferral, ceteris paribus, could cause donors to update their expectations of the probability of them being able to make a successful future donation, thus reducing their willingness to come back again.


The remainder of the paper is organized as follows. Section \ref{sec:application_correlation} describes our data and shows that, in our setting, the running variable is manipulated. Section \ref{sec:manipulation} presents two conceptual frameworks illustrating how manipulation can bias RD estimates and shows discontinuities in donors’ covariates. Section \ref{sec:litReview} reviews relevant literature on regression discontinuity designs, running variable manipulation, and partial identification of causal effects. Section \ref{sec:methodology} introduces our proposed method. Section \ref{sec:application} applies our method to the empirical setting and provides bounds for the causal effect of being deferred on future behavior. Section \ref{sec:discusson} concludes.

\section{The Effect of Deferrals on Donors' Return Behavior \label{sec:application_correlation}}

\subsection{Setting and Data}  

We study the effects of turning away donors (i.e., temporarily deferring)  using data from the Abu Dhabi Blood Bank. The Abu Dhabi Blood Bank has a monopoly over the collection of blood in the Emirates of Abu Dhabi (one of the seven emirates of the UAE). They collect donations in their Abu Dhabi and Al Ain centers as well as temporary mobile centers set up at people's workplaces, university campuses, and malls. Our data correspond to the universe of attempted donations from the $1^{st}$ of January 2016 to the $23^{rd}$ of July 2021, corresponding to 331,335 attempted donations from 172,161 unique donors.

The Abu Dhabi Blood Bank collects different types of blood donations: whole blood, plasma, and red-cell aspheris.\footnote{They also collect (i) Samples for medical tests which are not meant to be used for donations and (ii) Autologous donations, where a person donates blood for their own future use, typically before a scheduled surgery or medical procedure. We do not use these observations in our analyses since they are not voluntary donations in the sense of being directly intended to help a patient (N=4,327 and N=5, respectively).}  Whole blood donations, the most common type, involves giving all blood components—red cells, white cells, platelets, and plasma. In contrast, other donation types selectively extract one component, such as plasma, platelet, or red-cell donation. Blood is drawn and processed through a machine that separates the desired component(s), returning the rest to the donor.

Table \ref{tab:sum_stat} shows the sample sizes associated with each donation type (columns 1 to 4) and the number of days one needs to wait before attempting another donation (last column). For men, most donations are made for whole blood, some are for plasma, and only a few are for red-cell aspheris. For women, virtually all donations are made for whole blood.

\begin{table}[ht]
    \centering
    \begin{tabular}{l c cccc}
         \toprule
         Type of  &  \multicolumn{4}{c}{Sample Size} &  N Days\\
         Donations  & \multicolumn{2}{c}{N} & \multicolumn{2}{c}{\%} & Deferred    \\
         \cmidrule(lr){2-3} \cmidrule(lr){4-5} \cmidrule(lr){6-6}  
           & Male & Female & Male & Female & All \\
         \cmidrule(lr){2-3} \cmidrule(lr){4-5} \cmidrule(lr){6-6} 
         & \multicolumn{5}{c}{Successful Donations} \\
         Whole Blood        & 250,737 & 22,768 & 84.48\% & 65.95\% & 56 \\
         Plasma/Platelet    & 26,539 &  1 & 8.94\% & 0\%  & 14 \\
         Red Cell Apheresis         & 154 & 0 & 0.05\% & 0\%   & 112  \\
         \cmidrule(lr){2-3} \cmidrule(lr){4-5} \cmidrule(lr){6-6} 
         & \multicolumn{5}{c}{Unsuccessful Donations}\\
         Change of mind   & 3,491 & 1,043 & 1.18\% & 3.02\% & 0   \\
         Failed Phlebotomy & 1,477 & 387  & 0.50\% & 1.12\% & 1  \\
         Deferrals: \\
         \ \ Low h-level     & 8,675 & 8,796 & 2.92\%& 25.48\%  & 28  \\
         \ \ Other temporary  & 3,729 & 987 & 1.26\% & 2.86\%  & 136$^{\delta}$ \\
         \ \ Permanent       & 2,010 & 541 & 0.68\% &1.57\% & -   \\ 
         \midrule
         Total         & 296,812  & 34,523 &   100\% &   100\% & -  \\
         \bottomrule
    \end{tabular}
    \caption{Sample size and number of days deferred for the different donations' outcome. $^{\delta}$ The reported number of days deferred for temporary deferrals is the average number of days deferred for such deferral in our dataset.}
    \label{tab:sum_stat}
\end{table}

Not all attempted donations are successful.\footnote{In our data, for unsuccessful donations, we do not know what type of donations was attempted. We can safely assume that it was a whole blood donation for women. For men, we do not know whether the attempted donation was for whole blood, plasma, or red cell aspheris.} If the attempted donation is unsuccessful, it can be for several reasons. First, donors can start the process of donating\footnote{When one attempts to donate, one first goes through a medical screening, which involves a questionnaire as well as health checks such as measuring blood pressure and hemoglobin level. The process is detailed here: \url{www.seha.ae/bloodBank\#donate-carousel}.} (i.e., registering, health checks) and change their mind in the middle of it. In that case, they can come back on the same day. Second, it could be a failed phlebotomy, an unsuccessful attempt at drawing blood, often due to difficulties locating or accessing a vein, or the inability to collect sufficient blood for the donation. In case of a failed phlebotomy, the donor can return the next day.

Third, the donor can be deferred during the medical screening. The donor can be deferred for a variety of reasons. For instance, for having recently travelled to a country with yellow fever or malaria, having too high or low blood pressure, or having a too-high or low hemoglobin level. The reason for deferring is to ensure the safety of the collected blood and donor.

The main reason for deferral is low hemoglobin level. For women, 30\% of the attempted donations end up in a deferral, and 85\% of the deferrals are due to low hemoglobin. For men, 5\% of the attempted donation ends up in a deferral, and 60\% of the deferrals are due to low hemoglobin. The much higher deferral rate for women is due to women having a lower h-level, on average. When one is deferred for low hemoglobin, one is deferred for 28 days.

Besides being deferred for low h-level, one can receive a temporary deferral. Reasons for a temporary deferral include, for instance, having taken antibiotics recently (7 days), having too high/low blood pressure (14 days), and recently getting a tattoo (365 days). They can also receive a permanent deferral due to their medical history.

\FloatBarrier
\subsection{Return behavior: Correlational Evidences\label{sec:correlation}}

Before investigating the causal effect of deferrals on future behavior, we explore how return behavior depends on donors' and attempted donations' characteristics.

We define two measures of return behavior: (i) The probability that the donor returns in the next 12 months (i.e., the extensive margin) and (ii) the number of days the donor takes to return if he does return (i.e., the intensive margin).mn

Table \ref{tab:ols_return_short} shows how return behavior depends on the donors' characteristics and the attempted donations' outcome. It reports OLS regression where we regress our measures of return behavior on dummy variables for gender, whether it is a repeat  donation\footnote{Our dataset covers the period between January 2016 and July 2021. We do not know the donors' donation history before January 2016. We define a first time donation as the first time the donor appears in our dataset (i.e., the first donations made in the emirate of Abu Dhabi since January 2016).} the outcome of donation, controls for the height/weight of the donor as well as the donors' blood type. The baseline is a first-time male donor.\footnote{Table \ref{tab:ols_return_full_all} in Appendix shows the height, weight, and blood type coefficients and Tables \ref{tab:ols_return_full_F} and \ref{tab:ols_return_full_M} in Appendix report the results for each gender separately.} We only consider donations for which donors are eligible to come back one year before our dataset ends to allow time for donors to return. Hence, the sample sizes in Table \ref{tab:ols_return_short} are smaller than in Table \ref{tab:sum_stat}. Table \ref{tab:sum_stat_oneyear} in Appendix reproduces Table \ref{tab:sum_stat} with only donations for which donors are eligible to come back one year before our dataset ends.

In line with the literature, we find a large negative correlation between a deferral and one's return propensity. For instance, a donor who has been deferred for low hemoglobin is 12.4 percentage point\footnote{This difference becomes 4.1 percentage point when controlling for blood type (third row).} less likely to return in the next 12 months, and if he returns, it takes him on average 44 more days to do so. Unsurprisingly, repeat donors are more likely to return than first-time donors. We also find that women are less likely to return than men, but this difference disappears when we control for donors' height and weight.

Those correlations should not be interpreted as causal as they suffer from omitted variable bias.

\begin{table}[H]
\centering
\begin{tabular}{l ccc ccc}
\toprule
& \multicolumn{3}{c}{P(Return Next 12 Months)} & \multicolumn{3}{c}{Nbr. Days to Return}  \\
\cmidrule(lr){2-4} \cmidrule(lr){5-7} 
Constant & $0.256^{***} $  & $-0.391^{***} $  & $-0.304^{***} $  & $356.8^{***} $  & $611.3^{***} $  & $555.1^{***} $ \\
 & \footnotesize{($0.001$) }  & \footnotesize{($0.039$) }  & \footnotesize{($0.043$) }  & \footnotesize{($ 1.7$) }  & \footnotesize{($27.2$) }  & \footnotesize{($28.2$) } \\
Female & $-0.085^{***} $  & $-0.001 $  & $0.002 $  & $35.7^{***} $  & $ 1.3 $  & $ 2.1 $ \\
 & \footnotesize{($0.003$) }  & \footnotesize{($0.005$) }  & \footnotesize{($0.005$) }  & \footnotesize{($ 3.7$) }  & \footnotesize{($ 4.5$) }  & \footnotesize{($ 4.5$) } \\
Repeat Donation & $0.385^{***} $  & $0.362^{***} $  & $0.334^{***} $  & $-163.5^{***} $  & $-161.5^{***} $  & $-155.2^{***} $ \\
 & \footnotesize{($0.002$) }  & \footnotesize{($0.003$) }  & \footnotesize{($0.003$) }  & \footnotesize{($ 1.9$) }  & \footnotesize{($ 2.4$) }  & \footnotesize{($ 2.4$) } \\
\\
\textit{Successful Donations} \\
Plasma/Platelet & $0.332^{***} $  & $0.352^{***} $  & $0.385^{***} $  & $-175.2^{***} $  & $-177.5^{***} $  & $-192.2^{***} $ \\
 & \footnotesize{($0.003$) }  & \footnotesize{($0.004$) }  & \footnotesize{($0.004$) }  & \footnotesize{($ 1.4$) }  & \footnotesize{($ 1.9$) }  & \footnotesize{($ 2.1$) } \\
Apheresis & $0.058 $  & $0.042 $  & $0.037 $  & $-7.9 $  & $ 3.3 $  & $ 7.0 $ \\
 & \footnotesize{($0.069$) }  & \footnotesize{($0.068$) }  & \footnotesize{($0.069$) }  & \footnotesize{($44.7$) }  & \footnotesize{($43.8$) }  & \footnotesize{($45.1$) } \\
\\
\textit{Unsuccessful Donations}  \\
Failed Phlebotomy & $-0.135^{***} $  & $-0.125^{***} $  & $-0.108^{***} $  & $63.0^{***} $  & $65.1^{***} $  & $56.3^{***} $ \\
 & \footnotesize{($0.010$) }  & \footnotesize{($0.010$) }  & \footnotesize{($0.012$) }  & \footnotesize{($14.5$) }  & \footnotesize{($14.5$) }  & \footnotesize{($14.7$) } \\
Change of mind & $0.013 $  & $0.048^{***} $  & $0.172^{***} $  & $-40.4^{***} $  & $-44.4^{***} $  & $-50.0^{***} $ \\
 & \footnotesize{($0.008$) }  & \footnotesize{($0.008$) }  & \footnotesize{($0.011$) }  & \footnotesize{($ 7.1$) }  & \footnotesize{($ 7.3$) }  & \footnotesize{($ 7.4$) } \\
\ \ Low h-level  & $-0.124^{***} $  & $-0.124^{***} $  & $-0.044^{***} $  & $44.0^{***} $  & $44.8^{***} $  & $40.2^{***} $ \\
 & \footnotesize{($0.004$) }  & \footnotesize{($0.004$) }  & \footnotesize{($0.008$) }  & \footnotesize{($ 5.2$) }  & \footnotesize{($ 5.8$) }  & \footnotesize{($ 6.1$) } \\
\ \ Other temporary  & $-0.095^{***} $  & $-0.102^{***} $  & $0.025 $  & $-5.8 $  & $ 5.3 $  & $-6.3 $ \\
 & \footnotesize{($0.007$) }  & \footnotesize{($0.007$) }  & \footnotesize{($0.014$) }  & \footnotesize{($ 9.1$) }  & \footnotesize{($10.4$) } \\
\textbf{Control:}  \\
\textit{Height \& Weight \& Age} & \xmark & \cmark & \cmark & \xmark & \cmark & \cmark \\
\textit{Blood Types} & \xmark & \xmark & \cmark & \xmark & \xmark & \cmark \\
\cmidrule(lr){2-4} \cmidrule(lr){5-7} 
Mean  & $0.442 $  & $0.445 $  & $0.469 $  & $228.6 $  & $214.4 $  & $213.1 $ \\
Std. Dev.  & $0.497 $  & $0.497 $  & $0.499 $  & $291.8 $  & $293.2 $  & $292.3 $ \\
N  & $242,699 $  & $176,343 $  & $166,613 $  & $134,117 $  & $96,367 $  & $95,764 $ \\
\bottomrule
\end{tabular}
    \caption{\footnotesize{Donor's return behavior depending on the attempted donations' outcome and the donors' characteristics. OLS regression. On the left, the dependent variable is a dummy variable equal to 1 if the donors return in the next 12 months. On the right, the dependent variable is the number of days the donor takes to return if he returns. In parentheses, the standard errors clustered by donors. The rows labeled Mean and Std. Dev. report the mean and standard deviation of the outcome variable. Only donations made a year before the end of the dataset. $^{***}$ significance at the 0.1\% level, $^{**}$ significance at the 1\%, $^{*}$ significance at the 5\% level.\label{tab:ols_return_short}} }
\end{table}

\FloatBarrier
\section{Manipulation of the Running variable\label{sec:manipulation}}

Having shown a negative correlation between deferrals on return behavior, we now investigate whether it is causal by exploiting a discontinuity in the blood donor's eligibility criteria. When volunteers attempt to donate, their hemoglobin levels are measured. If the h-level is below an eligibility threshold (see Table \ref{tab:threshold}), the donor is ineligible to donate blood and receives a temporary deferral. Thus, the h-level threshold provides a natural experiment to identify the causal effect of the temporary deferral using a regression discontinuity design.

Table \ref{tab:threshold} summarizes the h-level eligibility threshold depending on the type of donations and donors' gender. Men need an h-level above 13.5g/dL to donate whole blood and 13g/dL to donate plasma or red-cell apheresis. Women need an h-level above 12.5g/dL to donate whole blood and 13g/dL to donate plasma or red-cell apheresis. However, as shown in Table \ref{tab:sum_stat}, the latter threshold is irrelevant in practice as the Abu Dhabi Blood Bank does not collect plasma and red-cell apheresis from women.

 \begin{table}[ht]
    \centering
    \begin{tabular}{l cc  }
         \toprule
         & \multicolumn{2}{c}{Threshold (g/dL)} \\
         \cmidrule(lr){2-3} 
         & Male & Female  \\
         \cmidrule(lr){2-2}  \cmidrule(lr){3-3} 
         Whole Blood       & 13.5 & 12.5 \\ 
         Plasma/Platelet   & 13   & 13   \\
         Apheresis         & 13   & 13   \\
         \bottomrule
    \end{tabular}
    \caption{H-level eligibility threshold by type of attempted donations and gender.}
    \label{tab:threshold}
\end{table}

\begin{figure}[ht]
	\centering
	\includegraphics[width=0.49\textwidth]{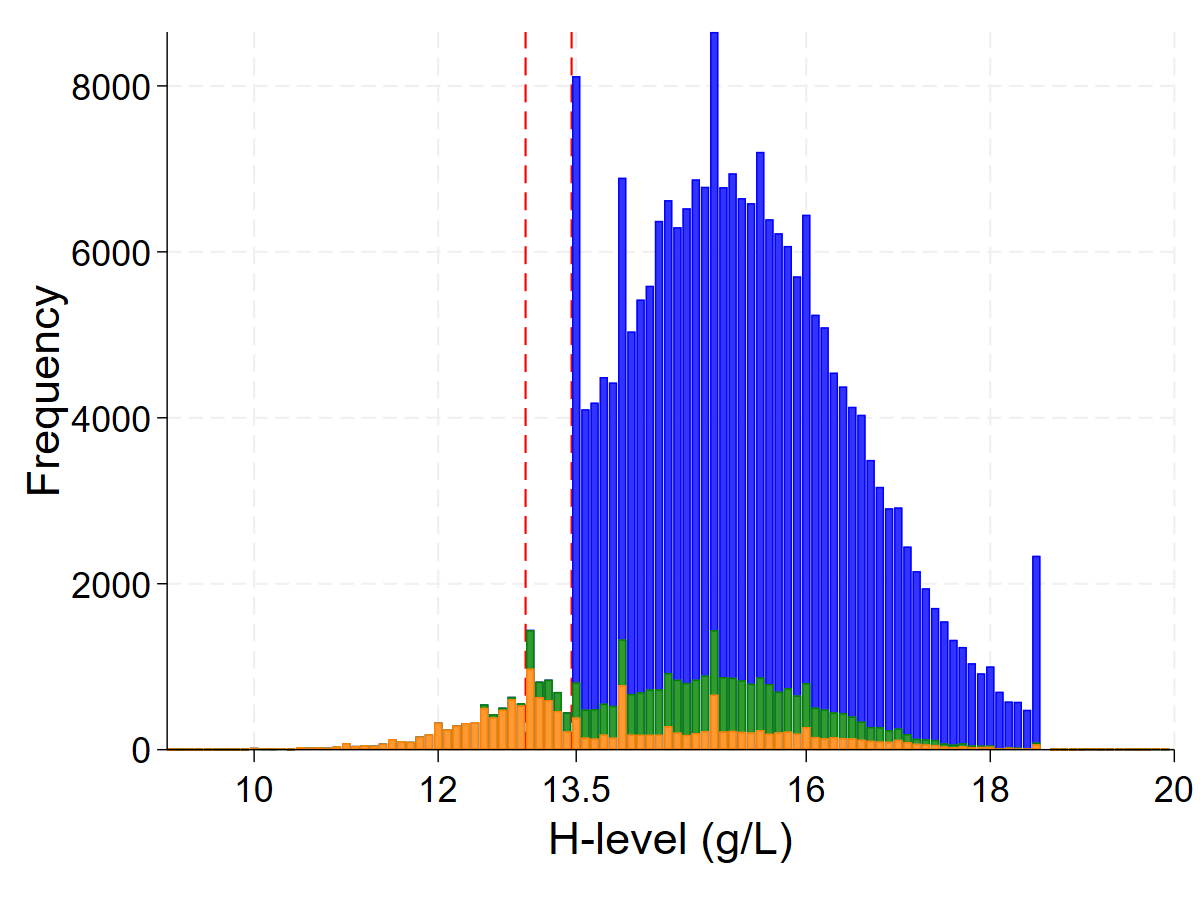}
	\includegraphics[width=0.49\textwidth]{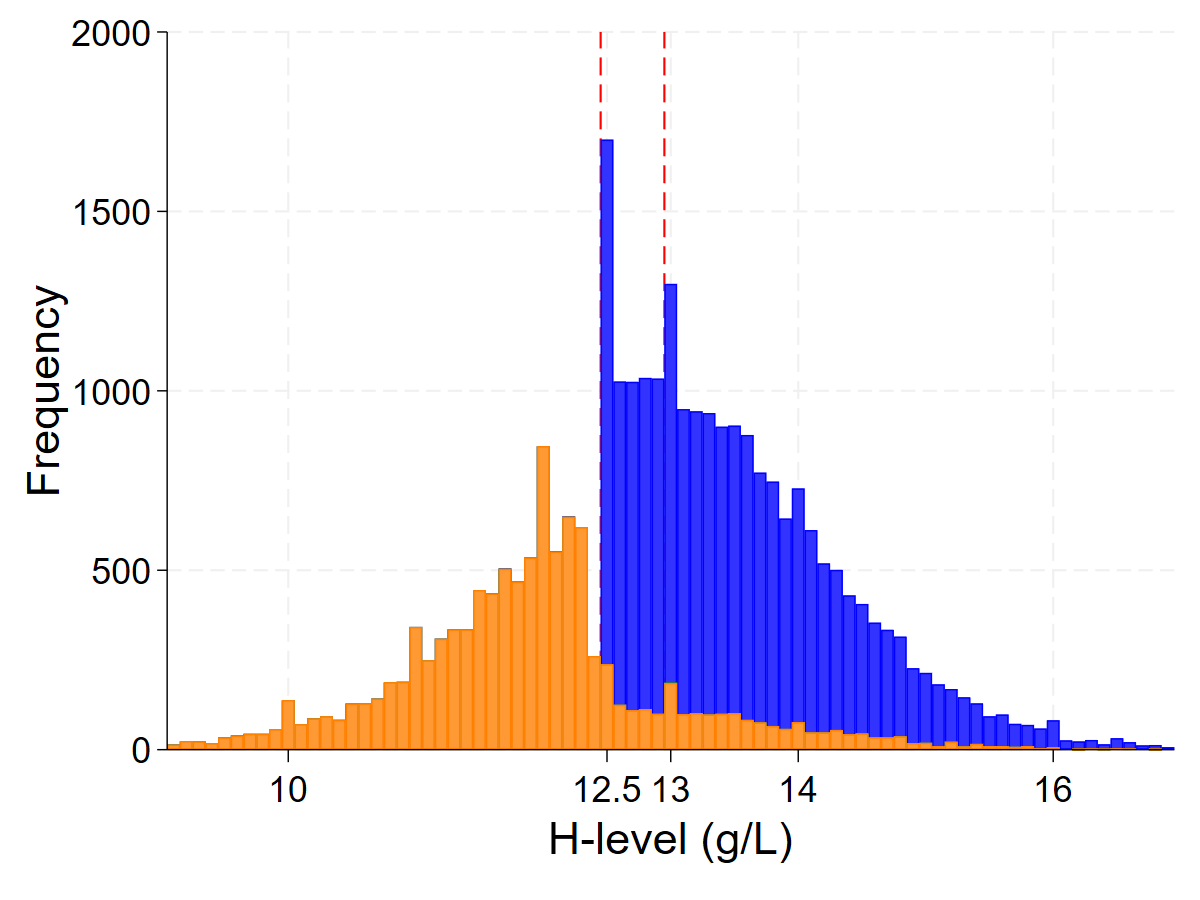}
\caption{Histogram of the h-level for men (left) and women (right). In orange, the unsuccessful donations; in green, the successful plasma/platelet and red cell aspheris donations; and in blue, the successful whole blood donations. The red dash lines show the eligibility cut-offs (see Table \ref{tab:threshold}).\label{fig:histhlevel}}
\end{figure}

Figure \ref{fig:histhlevel} shows histograms of the h-level for men and women. In orange, the unsuccessful donations; in green, the successful plasma/platelet and red cell aspheris donations; and in blue, the successful whole blood donations. These figures show how the running variable (i.e., the h-level) has been manipulated. First, although the h-levels should be smoothly distributed, the reported data shows extra mass at multiples of 0.5 (i.e., 12, 12.5, 13, etc.); when reporting the data, blood center staff are rounding up the h-level numbers they enter in the database. Second, and even more importantly for the RDD, the frequency of reported h-levels are lower just below the cut-off for whole blood donations, and spike at the cut-off (13.5g/dL for males and 12.5 g/dL for females). Thus, when the h-level is observed to be just below the threshold, staff for some (but not all) donors report an h-level just above the threshold to allow the donation to occur. 

For men, nurses not only have incentives to bump up the h-level above the threshold of 13.5 g/dL, to allow whole blood donation, but also have incentives to bump up the h-level above the threshold of 13g/dL to allow plasma and red cell apheresis donations. We do observe manipulations at both thresholds. In contrast, for women, nurses only have incentives to bump up the h-level above the threshold allowing whole blood donation of 12.5 as, as it is evident from Table \ref{tab:sum_stat}, they do not collect plasma or red cell apheresis donations from female donors. We do observe a spike at 13g/dL, but this spike is not larger than the one we observe at 12g/dL (where there is no incentive for manipulation) and is likely due to rounding.

\FloatBarrier

These manipulations are correlated with donors’ characteristics and, thus, invalidates the standard regression discontinuity design. Figures \ref{fig:discontinuitycovariates_M} and \ref{fig:discontinuitycovariates_F} in Section  \ref{sec:discontinuitycovariates} show how donor's characteristics change depending on the reported h-level for men and women. 

\subsection{Conceptual Frameworks \label{sec:framework1threshold}}

In our application, the running variable is subject to manipulation. This section presents two conceptual frameworks that model an agent’s decision to adjust the running variable to exceed the eligibility threshold. In both cases, such manipulation results in systematic differences between observations just above and just below the threshold. Consequently, a naïve regression discontinuity (RD) design ignoring manipulation would yield biased estimates.

While we develop these conceptual frameworks in the context of blood donations, the insights from these frameworks can be applied more broadly. For example, they can be adapted to settings where teachers manipulate grades to make students eligible for a course or scholarship, individuals adjust credit scores to qualify for financial products such as credit cards, or police officers reduce recorded speeding to keep drivers below a fine or license-loss threshold.

In the first conceptual framework the running variable is measured without error, but an agent manipulates it to satisfy a policy goal. In our case, nurses face a tradeoff between increasing the number of donations and decreasing the probability of an adverse event.

In the second conceptual framework, the running variable is measured with error. The agent does not act strategically but interprets noisy measurements using auxiliary information. When the observed value falls just below the threshold, yet they believe the true value exceeds it, they override the measurement to avoid a perceived false negative. This behaviour creates asymmetries around the threshold that can still bias RD estimates.

\subsubsection{Framework 1: Strategic Manipulation to Achieve Policy Goals}

In this first framework, nurses balance two objectives, (i) maximizing the blood bank's benefits from donations and (ii) minimizing the risk of adverse events for donors such as dizziness, fainting, or nausea.

Let $V_w$ denote the utility the blood bank derives from a successful whole blood donation, $V_r$ the utility from a rejection and $V_a$ the utility from an adverse event. We assume $V_w>V_r>V_a$ and, without loss of generality, set $V_a$ to 0. Let $p(s|h)$ be the probability that an accepted donation is successful given an observed h-level $h$.

The nurse observes $h$ and $\textbf{z}$ where $\textbf{z}$ is a set of co-variates such as gender, age, donor's experience, height, weight, and blood type and reports $h^*$. She has two possible strategies: report $h$ truthfully or report $ h^*=c$, where $c$ is the threshold, to allow the donation.

Figure \ref{fig:conceptualframework1} below describes the problem faced by the nurse.

\begin{figure}[ht]
$$
 \hspace*{-1.4cm}\begin{array}{ccccc}
\xymatrix@R-2pc{
& & h\geq c \ar[r] & \text{Report }h^*=h \ar[r] & \text{Accept Donation} \\
\text{Observe }h \ar[ur] \ar @{->} [dr]  \\
& & h<c \ar @{->} [r] \ar[ddr] & \text{report }h^*=c \ar[uur] &  \\
& \\
& & & \text{report }h^*=h \ar[r] & \text{Reject Donation} \\
}
\end{array}
$$
\caption{Strategic Reporting of Hemoglobin Levels}
\label{fig:conceptualframework1}
\end{figure}

Nurses are incentivized to round the h-level up to allow donations to occur. However, they have no incentives to misreport the h-level below the threshold for the donations not to take place. If they believe the donations to be unsafe, they can defer the donor for reasons other than being below the h-level. This is evidenced by the positive probability of being deferred above the threshold (see Figure \ref{fig:effectPlot} and \ref{fig:Men_WB}). Further, as seen in Figure \ref{fig:histhlevel}, there is considerable bunching just above the threshold consistent with nurses rounding the h-level up but not with rounding the h-level down

Given an observed h-level $h$ the expected utility the nurse gets is:

\begin{equation}
EU(h) =
\begin{cases} 
V_w*p(s|h) & \text{if attempt the donation} \\ 
V_r & \text{if reject donation} \\ 
\end{cases}
\end{equation}

Therefore the nurse decides to manipulate if:

$$p(s|h)>\frac{V_r}{V_w}$$

From the inequality above, we predict that 

\begin{proposition}[]
Nurses are more likely to manipulate when:
    \begin{enumerate}
        \item[a.] $V_w$, the utility of a successful donation increases.
        \item[b.] $V_r$, the utility of rejection decreases (i.e, the cost of a rejection increases).
        \item[c.] $p(s|h)$, the probability that the donation is successful increases.
    \end{enumerate}
\end{proposition}

\subsubsection*{Implications of Misreporting on RDD estimates\label{sec:biasinRDDframework1}}

We assume that the utility of a successful donation, $V_w$, is higher for high-demand blood types, such as O negative, which is the universal whole blood donor, and O positive, which is the most common and, therefore, needed blood type.\footnote{See here for an explanation of which blood types are in demand: \url{https://www.redcrossblood.org/donate-blood/blood-types.html}} We also assume that the utility of a successful donation is higher during periods of blood shortages.

Adverse events have been shown to be more likely for women, first-time donors, donors with lower weight, as well as younger donors \citep{philip2014single,almutairi2017incidence}. Furthermore, adverse events are more likely to occur at a lower h-level (hence the rule on minimum h-level needed to donate). Therefore, we assume $p(s|h)$ to be higher for men and to increase with experience, weight, height, and age, and the closer $h$ gets to the threshold.

We also assume the cost of a rejection to be higher for first-time donors than experienced donors. Indeed, among blood banks, there is a widely held belief that first-time donors are less likely to return if rejected than repeat donors. Therefore, we assume that $V_r$, the utility received from rejecting a donor, increases with experience. 

Given nurses' incentives, we expect nurses to manipulate the h-level more often for men, older and heavier donors (proposition 1.c), and donors of more in-demand blood types (proposition 1.a). This is indeed what we observe in our data (see Section \ref{sec:discontinuitycovariates}).\footnote{We also anticipate greater manipulation during periods of heightened shortages; however, our data does not include information on the level of shortages.}

When it comes to donors' experience, nurses face a trade-off. On the one hand, they may be more likely to manipulate first-time donors' h-level to avoid rejecting them as inexperienced donors may be less likely to return if rejected (proposition 1.b). On the other hand, they may be less likely to manipulate first-time donors' h-level to avoid adverse events (proposition 1.c). As we will see in Section \ref{sec:discontinuitycovariates}, facing this trade-off, nurses prioritize donors' safety and are less likely to misreport first-time donors' h-level.

Incidentally, Tables \ref{tab:ols_return_short}, \ref{tab:ols_return_full_all}, \ref{tab:ols_return_full_F} and \ref{tab:ols_return_full_M} show that the donors who are more likely to be bumped above the threshold are also the ones who are more likely to come back. For instance, O-negative donors are 10\% more likely to return, repeat donors are 40\% more likely to return and heavier and older donors are also more likely to return.

This selection of who gets bumped above the threshold introduces bias into our RDD estimates. Specifically, donors more likely to return will disproportionately fall just above the threshold, avoiding deferral. This selection process causes the estimated effect of a deferral on return behaviour to appear more negative than it truly is, leading to a downward bias.

\subsubsection{Framework 2: Non-Strategic Manipulation Due to Measurement Error}

In this second conceptual framework, nurses do not have strategic incentives to maximize donations. Instead, the h-level is measured with error, and nurses bump donors above the threshold if they believe the true value is above it. 

Let $h_x$ represent the true h-level, $\widetilde{h}_x$ the measured h-level, and $h_x^*$ the reported h-level. The true h-level is distributed as $h_x \sim N(x, \sigma^2)$, where $x$ is the predicted h-level, representing the portion of the h-level that can be explained by observable donor characteristics. Specifically, $x=f(z_1,...,z_n)$ where $z_1,...,z_n$ are observable characteristics such as age, gender, and prior donation experience.\footnote{Unobservable factors affecting the h-level are included in $\sigma^2$.}  A higher $x$ indicates a higher average h-level. The measured h-level is modeled as $\widetilde{h}_x = h_x + \epsilon$, where $\epsilon \sim N(0, \theta^2)$.

After measuring the h-level at $k$ and observing $x$ and $(z_1,...,z_n)$, nurses form a belief $P(h_x > c \mid \widetilde{h}_x = k)$ about whether the true h-level exceeds a threshold $c$.

\begin{itemize}
	\item If $\widetilde{h}_x\geq c$, the nurse reports truthfully $h_x^{*}=\widetilde{h}_x$.
	\item If $\widetilde{h}_x<c$, the nurse guesses whether $h_x>c$. If her belief $p(h_x>c|\widetilde{h}_x)$ exceeds a threshold $\alpha$, she reports $h_x^{*}=c$.
\end{itemize}

\begin{proposition}[]
$p(h_x>c|\widetilde{h}_x=k)$ is increasing with $k$ and $x$. Therefore, nurses are more likely to manipulate as $k$ and $x$ increase.
\end{proposition}

See Appendix \ref{sec:proofmeasurementerror} for the proof.

This mechanism shifts observations from below the threshold above it creating a bias for donors' characteristics correlated with h-level, such as age, weight, height and previous experience. 

Consider a donor characteristic $z_i$ that increases with the h-level, meaning $\frac{\partial f}{\partial z_i} > 0$.\footnote{The same logic applies inversely if $\frac{\partial f}{\partial z_i} < 0$, where the shift would artificially inflate the average just below the threshold.} Suppose nurses use $z_i$ to update their belief about a donor’s true h-level. In this case, donors with higher $z_i$ are more likely to be bump above the threshold because nurses expect them to have higher h-levels.

For example, if older donors tend to have higher h-levels ($\frac{\partial f}{\partial age} > 0$), then among those measured below the threshold, older donors are more likely to have been measured with an unusually low h-level due to random variation. This is a case of reversion to the mean: donors with lower-than-expected measured values tend to have true values closer to their expected average. Because nurses anticipate that older donors typically have higher h-levels, they are more likely to bump them above the threshold. This selective misreporting artificially inflates the average $z_i$ just above the threshold compared to just below it.

\begin{figure}[ht]
    \centering	
    \includegraphics[width=0.49\textwidth]{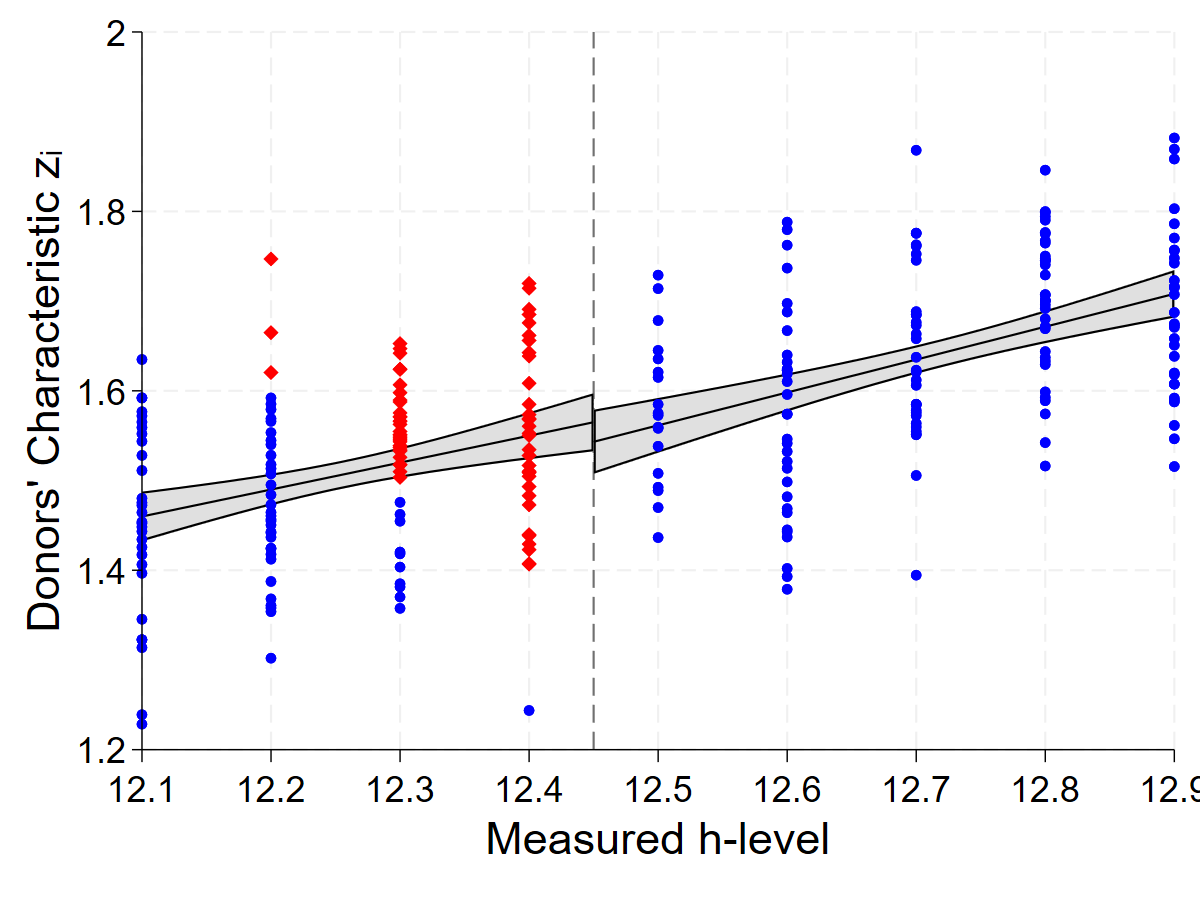}
    \includegraphics[width=0.49\textwidth]{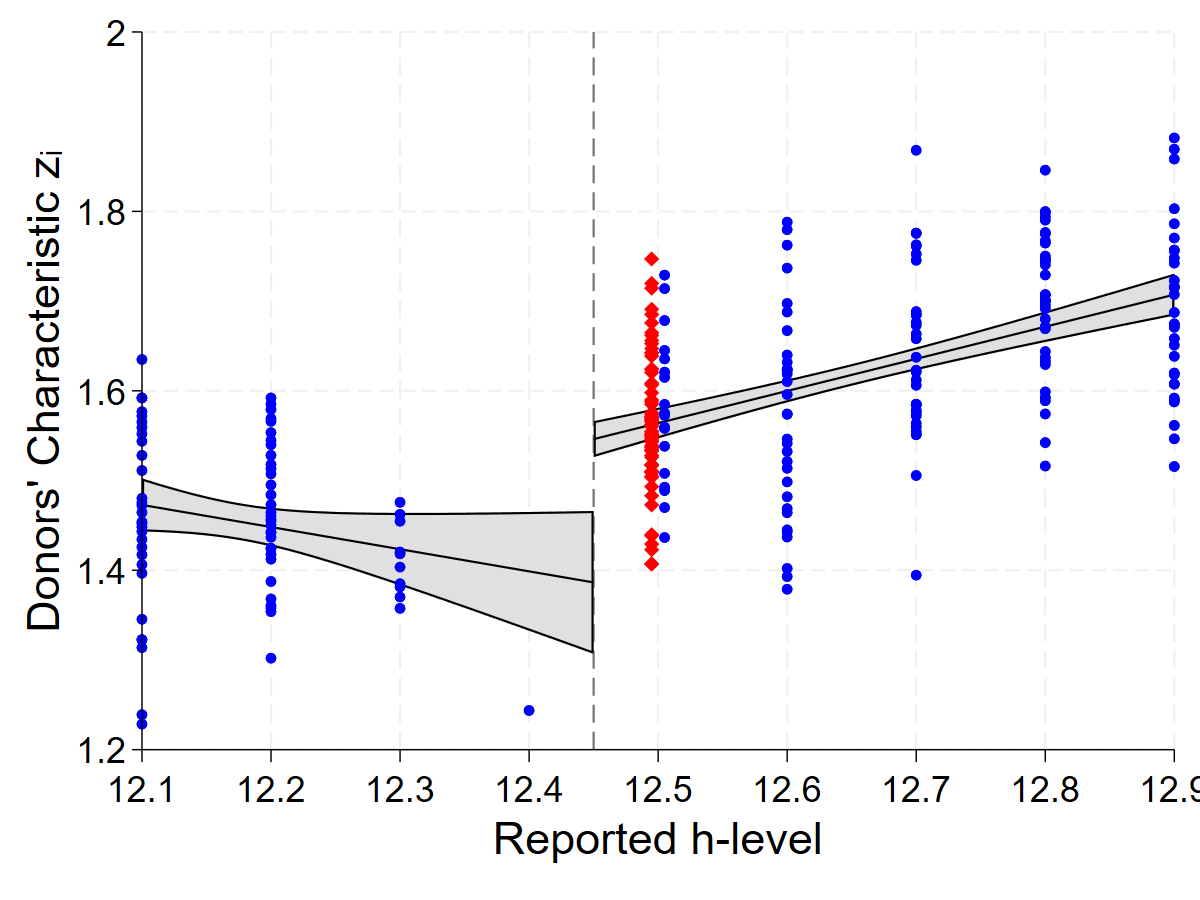}
\caption{Potential bias from misreporting. On the left, the x-axis is the measured h-level $\widetilde{h}$, on the right it is the reported h-level $h^*$. Red diamonds represents misreported observations. \label{fig:theory_h2}}
\end{figure}

Figure \ref{fig:theory_h2} illustrates this effect. In the left panel, red diamonds represent donors with high $z_i$ values who were measured below the threshold. Due to misreporting, they appear above the threshold in the right panel, creating a visible discontinuity.

\begin{figure}[ht]
	\centering
		\includegraphics[width=0.49\textwidth]{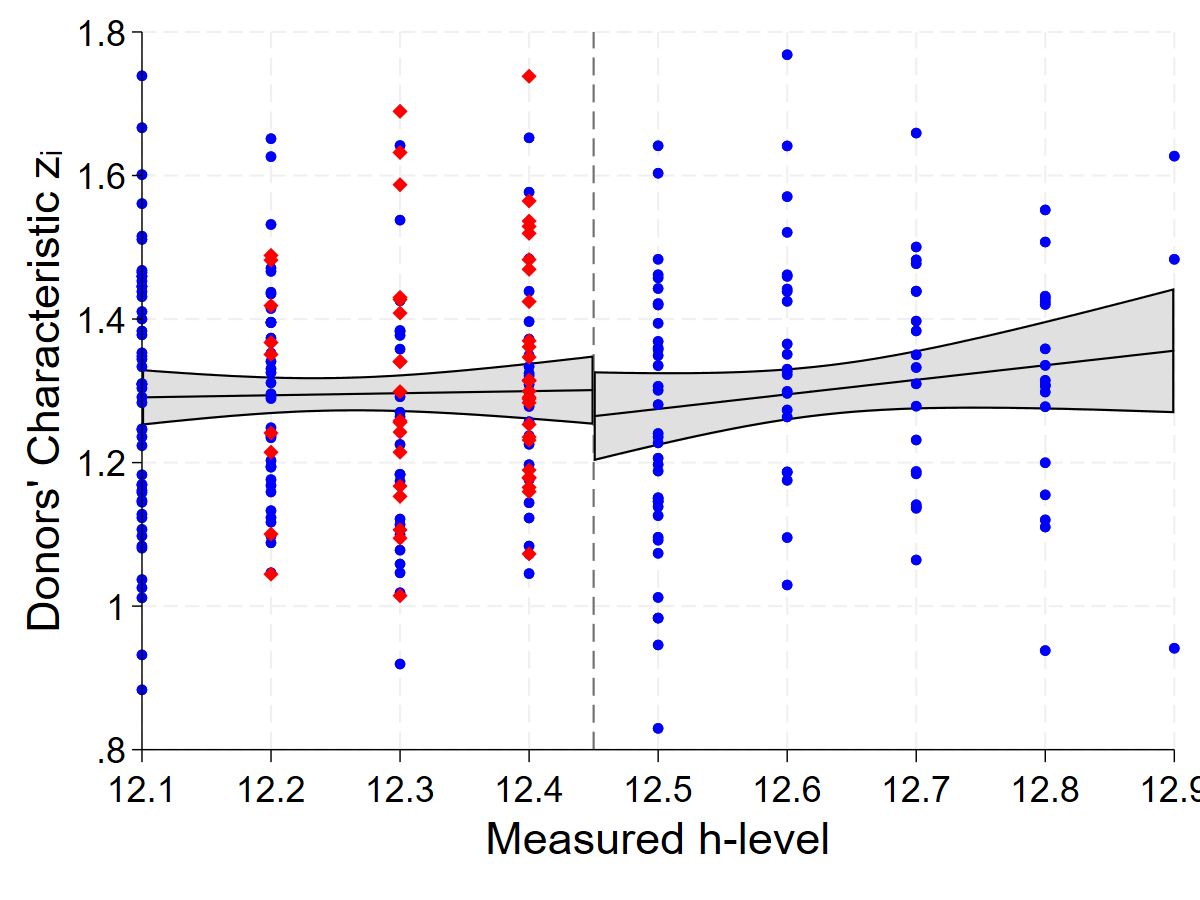}
		\includegraphics[width=0.49\textwidth]{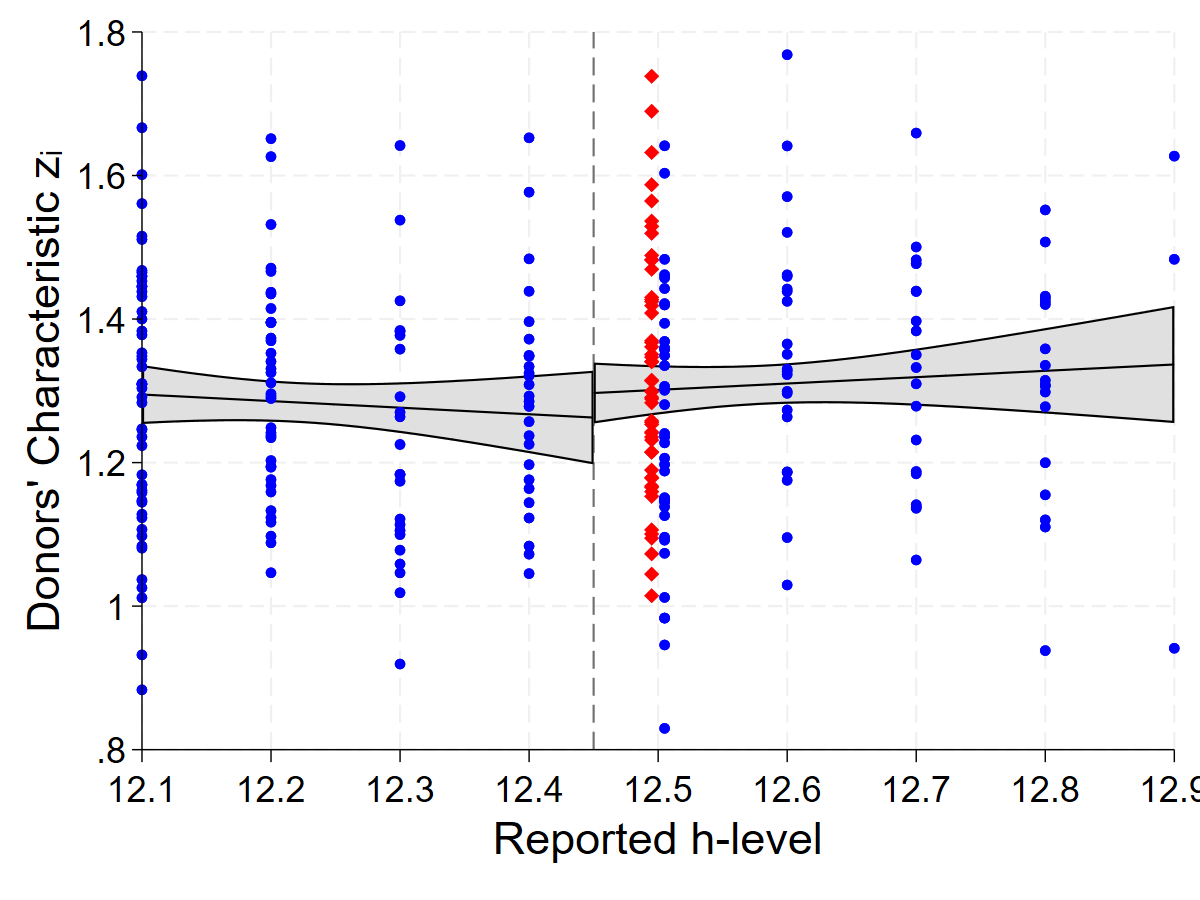}
\caption{Potential bias from misreporting. On the left, the x-axis is the measured h-level $\widetilde{h}$, on the right it is the reported h-level $h^*$. Red diamonds represents misreported observations.\label{fig:h_manipulated_notincreasing}}
\end{figure}

Note that this problem does not exist for the donor's characteristic not varying with the h-level (i.e., $\frac{\partial f}{\partial z_i}=0$) as illustrated in Figure \ref{fig:h_manipulated_notincreasing}. In this figure, the nurses do not manipulate donors with the highest $z_i$. Indeed, in this setting, a higher $z_i$ does not indicate a higher h-level. They randomly manipulate donors below the threshold with a higher probability to manipulate as  $\widetilde{h}$ gets closer to the threshold.

Appendix \ref{sec:theory2cutoff} shows the analogous figures when there is two cut-offs as is it the case for male donors.

\FloatBarrier
\subsubsection*{Implications of Misreporting on RDD estimates} 

We've seen that a bias is created for variables varying with the h-level. Hemoglobin level has been shown to depend on many factors such as gender, age, weight,  height, ethnicity, health, diet, menstrual cycle and smoking habits.\footnote{https://myacare.com/blog/factors-that-affect-hemoglobin-levels} Therefore, due to the misreporting, we expect donors below and above the threshold to differ in those characteristics. 

We now investigate the relationship between hemoglobin levels and donor characteristics in our data. However, in our data, near the threshold, we observe the reported hemoglobin levels and not the measured ones. To address this issue, we focus on the relationship between hemoglobin levels and donor characteristics away from the threshold, where hemoglobin levels are not subject to manipulation. We focus on the interval $(10,12)\cup(12.8,15)$ for women and $(11,12.7)\cup(13.7,17)$ for men.

We estimate $y=\alpha+\beta*\widetilde{h}+\epsilon$ where $y$ represents a donor characteristic. Table \ref{tab:ols_hlevelvariation} reports the results, and shows that, in our data, several donors' characteristics vary with h-level. For example, as  h-level increases, women are more experienced donors (i.e., they are less likely to be first-time donors and have more previous attempted donations) while men are less experienced.  As h-level rises, women tend to be shorter and men taller. Age also varies with h-level, with women tending to be older and men younger as h-levels rise. Finally, in our data, blood type varies with h-level with both men and women being less likely to be O negative, the universal donors' blood type, as the h-level increases.\footnote{Note that, our sample is not a random representation of the population; it consists of a selected group of donors. Individuals who are well-suited to being donors—such as young, healthy individuals with high hemoglobin levels and in-demand blood types—are over-represented in our sample, as they are actively encouraged to donate. Further, it only includes donor who passed the pre-screening and were not deferred before the hemoglobin test.}

All these characteristics correlate with return behaviour and introduce bias into our RDD estimates. However, the direction of the bias is not straightforward. For instance, according to Table \ref{tab:ols_hlevelvariation}, among women, h-level increases with experience and age but is lower for O-negative donors. As a result, more experienced and older donors will be disproportionately placed above the threshold. Since these donors are more likely to return, this selection process causes the estimated effect of a deferral on return behaviour to appear more negative than it truly is, leading to a downward bias. Conversely, O-negative donors will be underrepresented above the threshold. Since they are also more likely to return due to prioritization by blood banks, their underrepresentation causes the estimated effect of a deferral to appear less negative than it truly is, leading to an upward bias.\footnote{More experienced donors are generally more likely to return, while O-negative donors are often prioritized for follow-up by blood banks, increasing their likelihood of returning.}

\begin{table}[ht]
\centering
\begin{tabular}{l cc}
\toprule
& \multicolumn{1}{c}{Female} & \multicolumn{1}{c}{Male}  \\
\cmidrule(lr){2-2} \cmidrule(lr){3-3}
& $(10,12)\cup(12.8,15)$ & $(11,12.7)\cup(13.7,17)$ \\
\cmidrule(lr){2-2} \cmidrule(lr){3-3}
First-time Donor & $-0.017^{***} $  & $0.023^{***} $ \\
 & \footnotesize{($0.003$) }  & \footnotesize{($0.001$) } \\
\# Donations & $0.051^{***} $  & $-0.298^{***} $ \\
 & \footnotesize{($0.009$) }  & \footnotesize{($0.018$) } \\
Height (in cm) & $-0.126^{**} $  & $0.100^{***} $ \\
 & \footnotesize{($0.040$) }  & \footnotesize{($0.018$) } \\
Weight (in kg) & $-0.115 $  & $0.058 $ \\
 & \footnotesize{($0.077$) }  & \footnotesize{($0.039$) } \\
Age (in years) & $0.864^{***} $  & $-0.508^{***} $ \\
 & \footnotesize{($0.055$) }  & \footnotesize{($0.020$) } \\
O- & $-0.011^{***} $  & $-0.007^{***} $ \\
 & \footnotesize{($0.002$) }  & \footnotesize{($0.001$) } \\
O+ & $-0.010^{*} $  & $-0.008^{***} $ \\
 & \footnotesize{($0.004$) }  & \footnotesize{($0.001$) } \\
AB & $0.000 $  & $0.001^{*} $ \\
 & \footnotesize{($0.002$) }  & \footnotesize{($0.001$) } \\
A and B & $0.021^{***} $  & $0.014^{***} $ \\
 & \footnotesize{($0.004$) }  & \footnotesize{($0.001$) } \\
\midrule
N & $20,894$  & $188,126$ \\
\bottomrule
\end{tabular}
             \caption{Variation in donors' characteristics with h-level. Estimates of $\beta$ from the regression $y=\alpha+\beta*\widetilde{h}+\epsilon$ where $y$ is a donor characteristic. Standard errors clustered by donors in parenthesis.' $^{***}$ significance at the 0.1\% level, $^{**}$ significance at the 1\%, $^{*}$ significance at the 5\% level.\label{tab:ols_hlevelvariation}}
\end{table}

\FloatBarrier

\subsection{Discontinuity in covariates\label{sec:discontinuitycovariates}}

This section shows that, in our data, donors' covariates are not distributed smoothly around the threshold. Those discontinuities are in line with our first conceptual framework and not with the second one. If ignored, those discontinuities will bias RDD estimates of a deferral on return behaviour making it appear more negative than it truly is. 

\begin{figure}[ht]
    \centering
    \includegraphics[width = 0.32\textwidth]{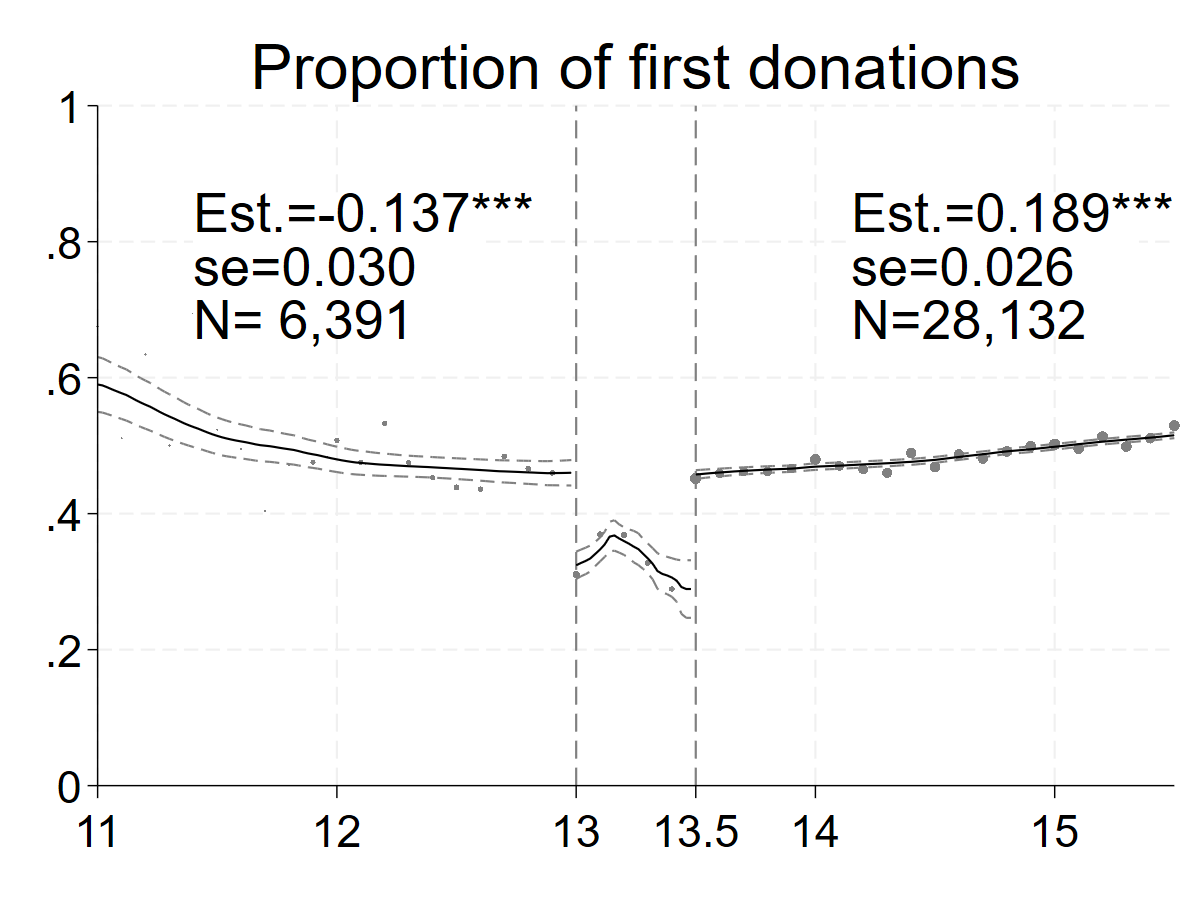}
    \includegraphics[width = 0.32\textwidth]{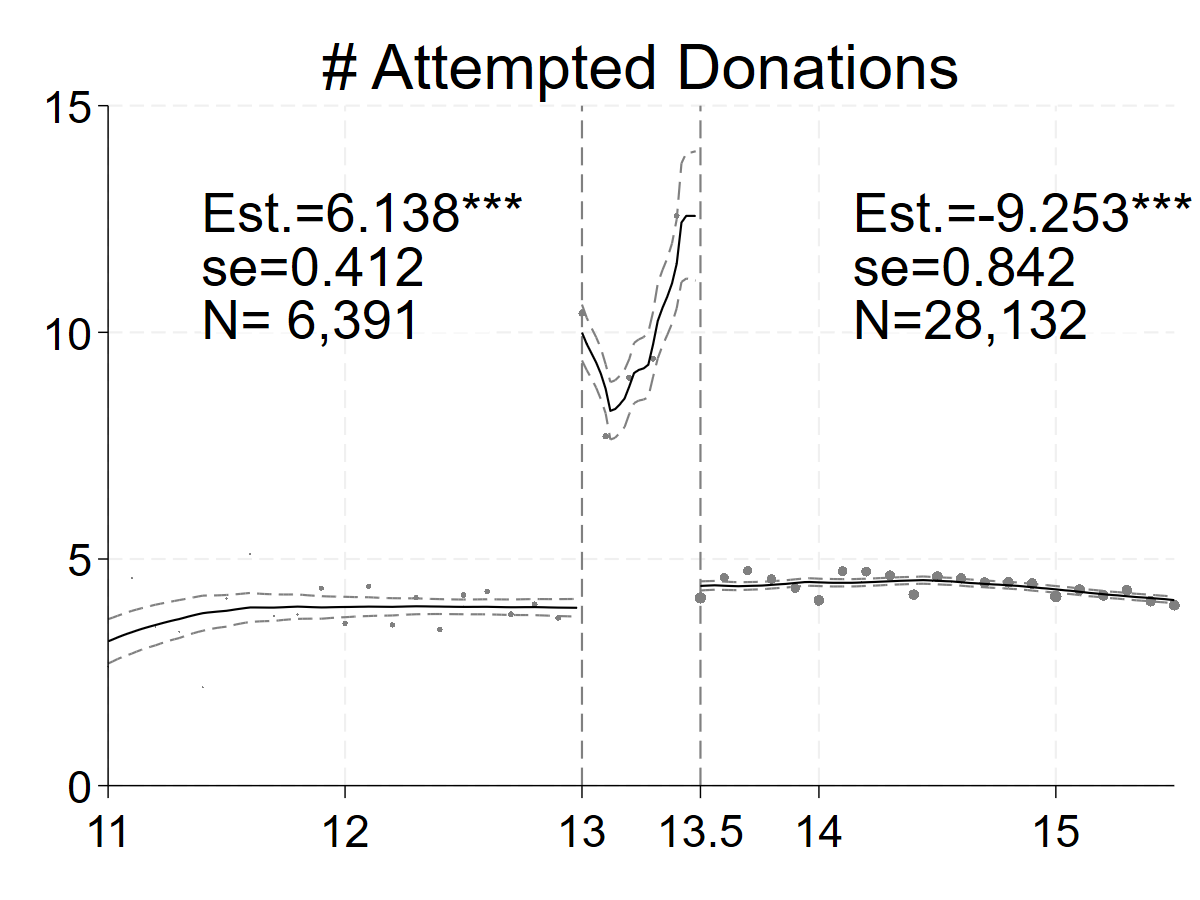}
    \includegraphics[width = 0.32\textwidth]{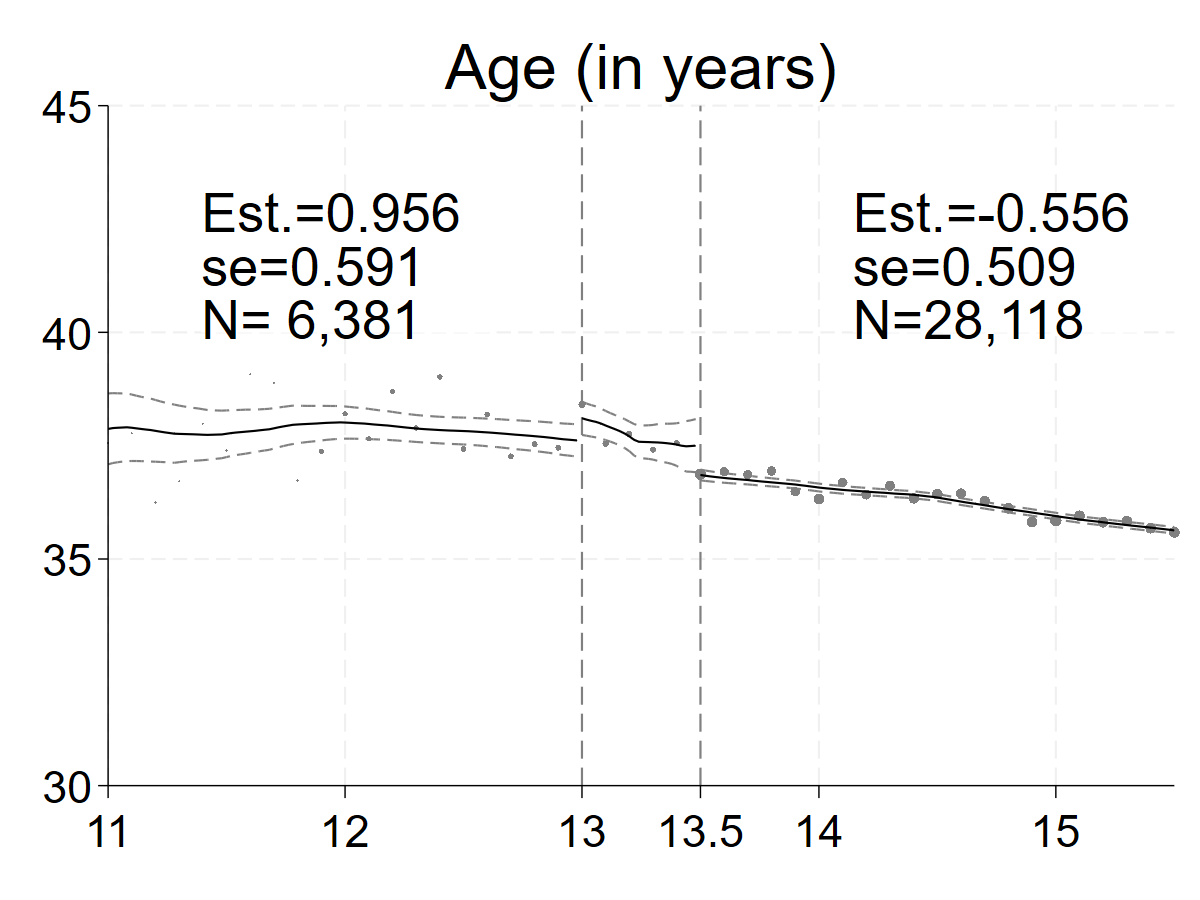}
    \includegraphics[width = 0.32\textwidth]{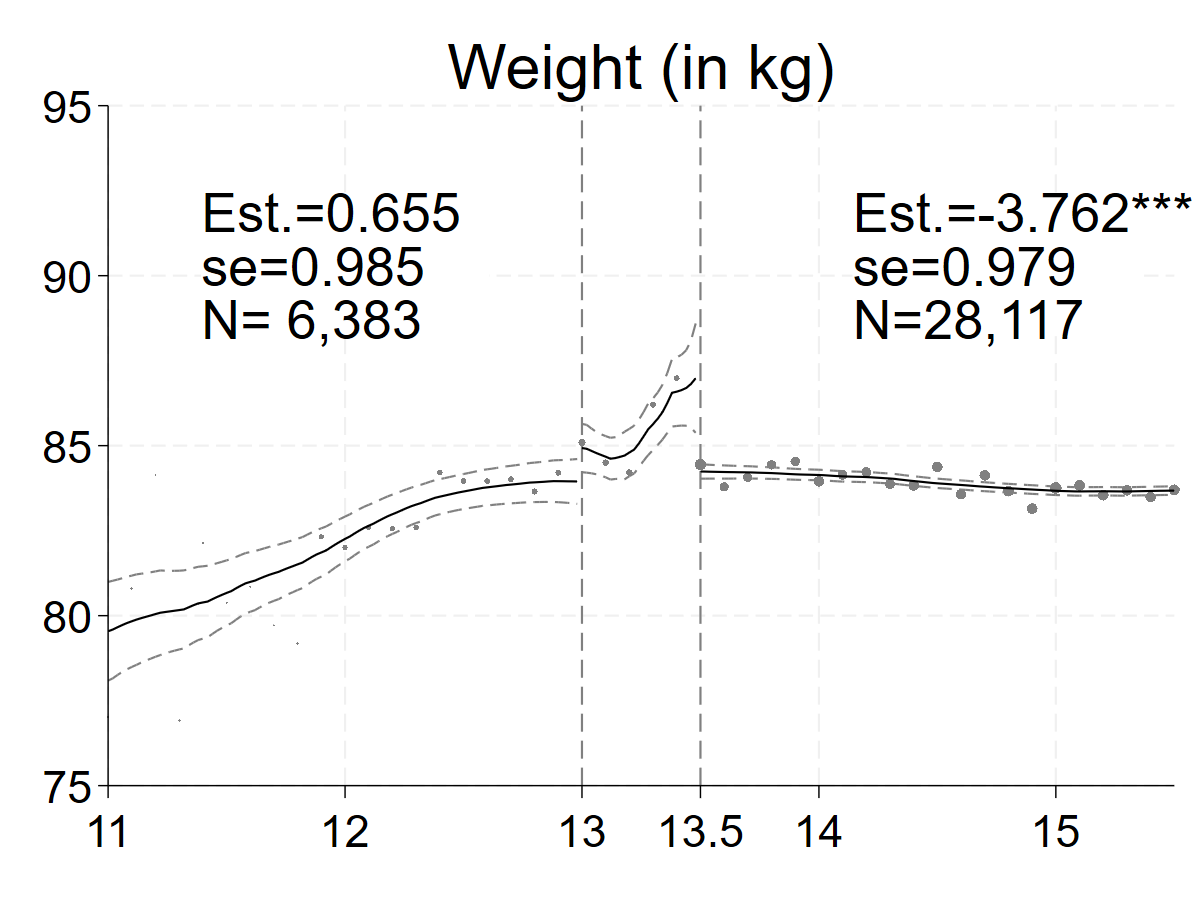}
    \includegraphics[width = 0.32\textwidth]{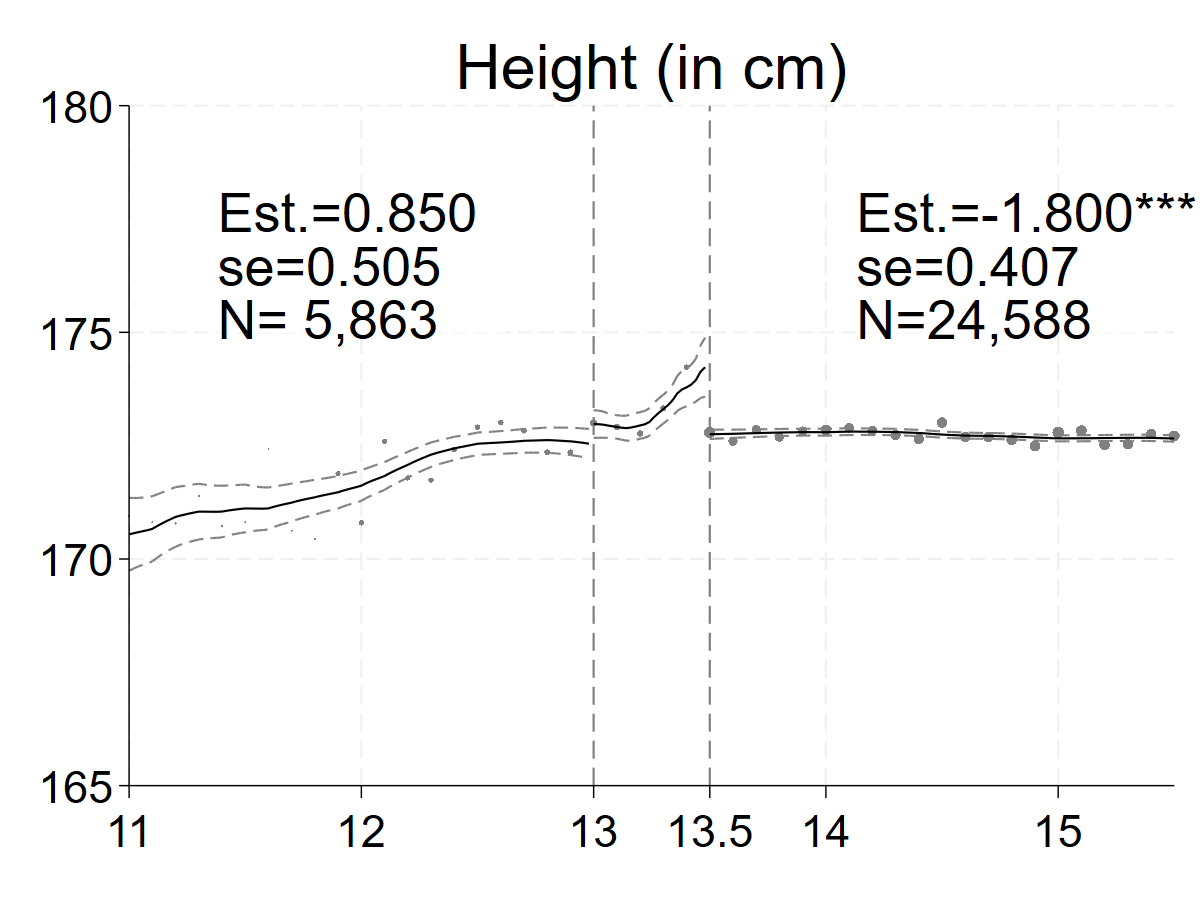}
    \includegraphics[width = 0.32\textwidth]{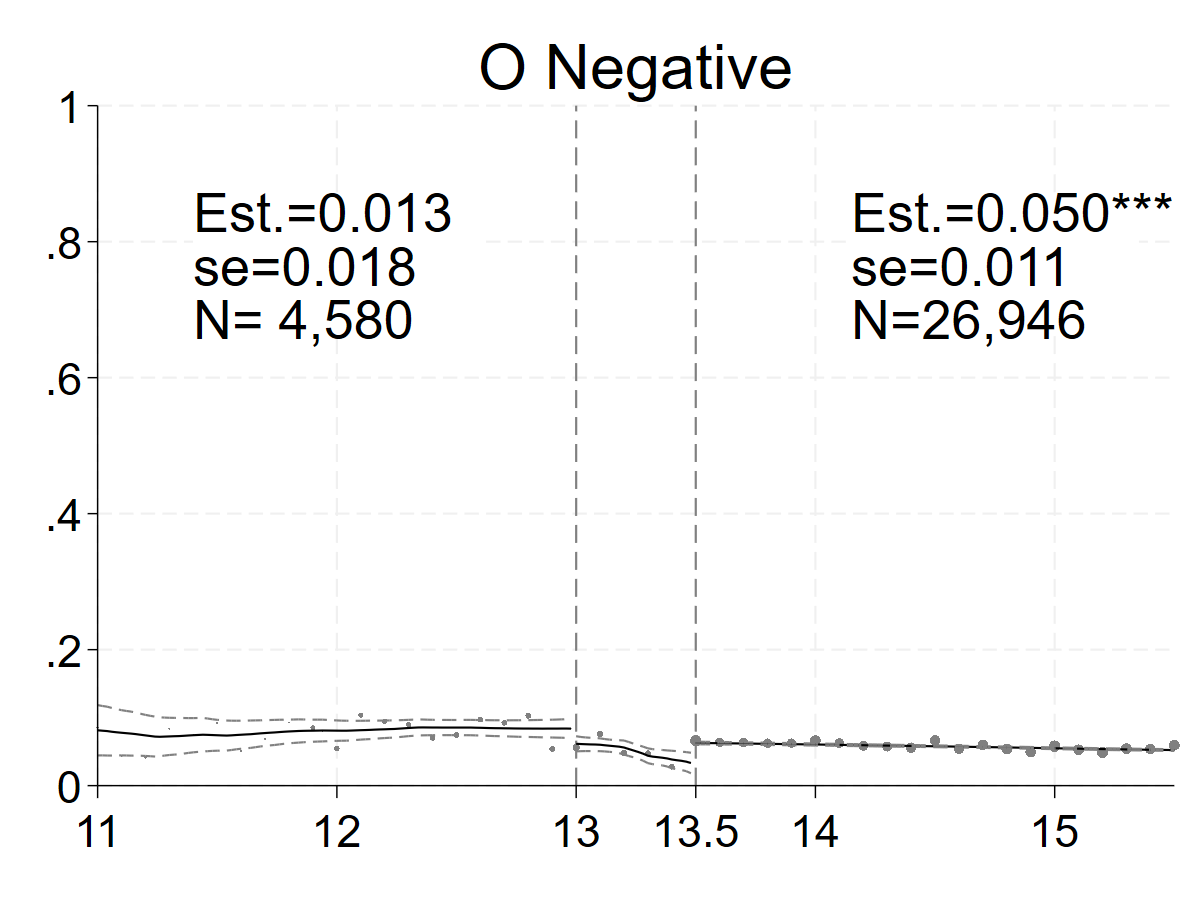}
     \includegraphics[width = 0.32\textwidth]{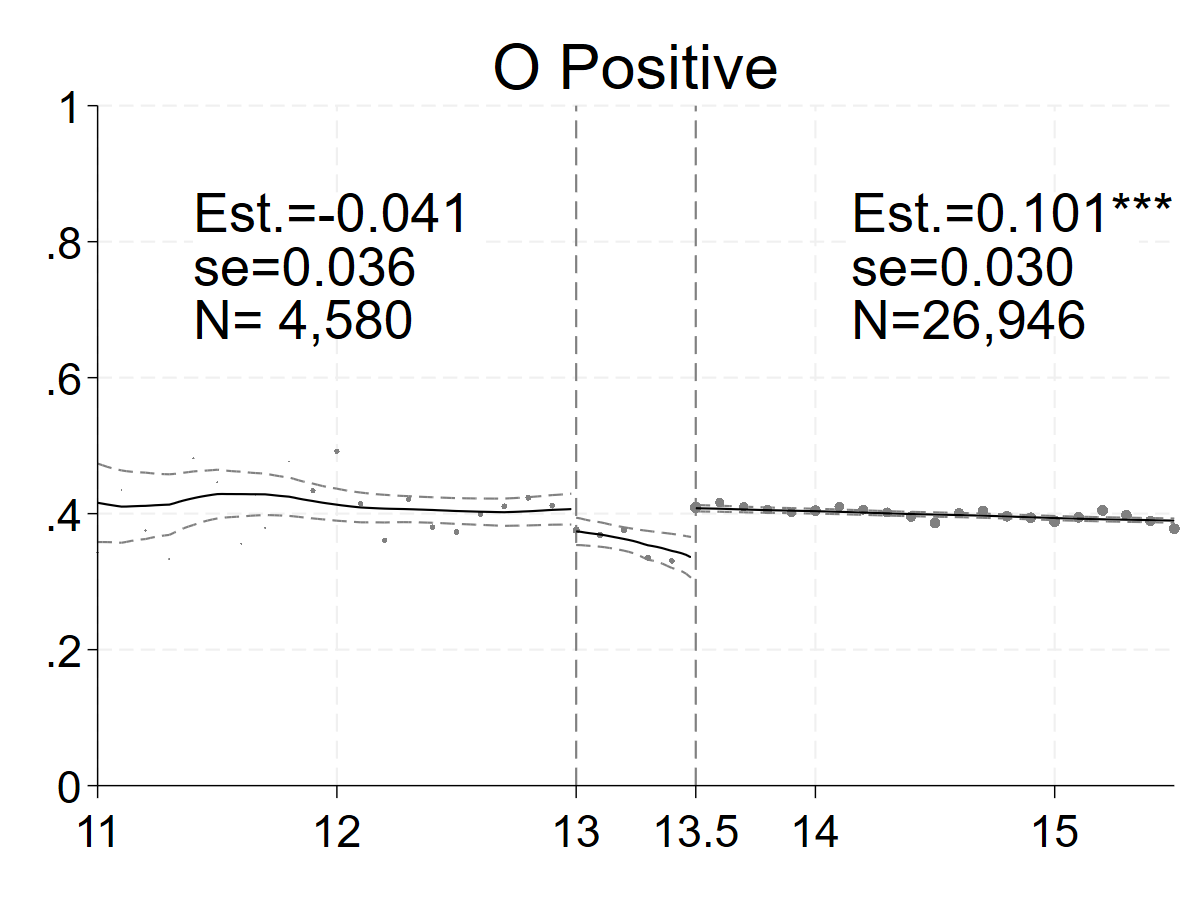}
    \includegraphics[width = 0.32\textwidth]{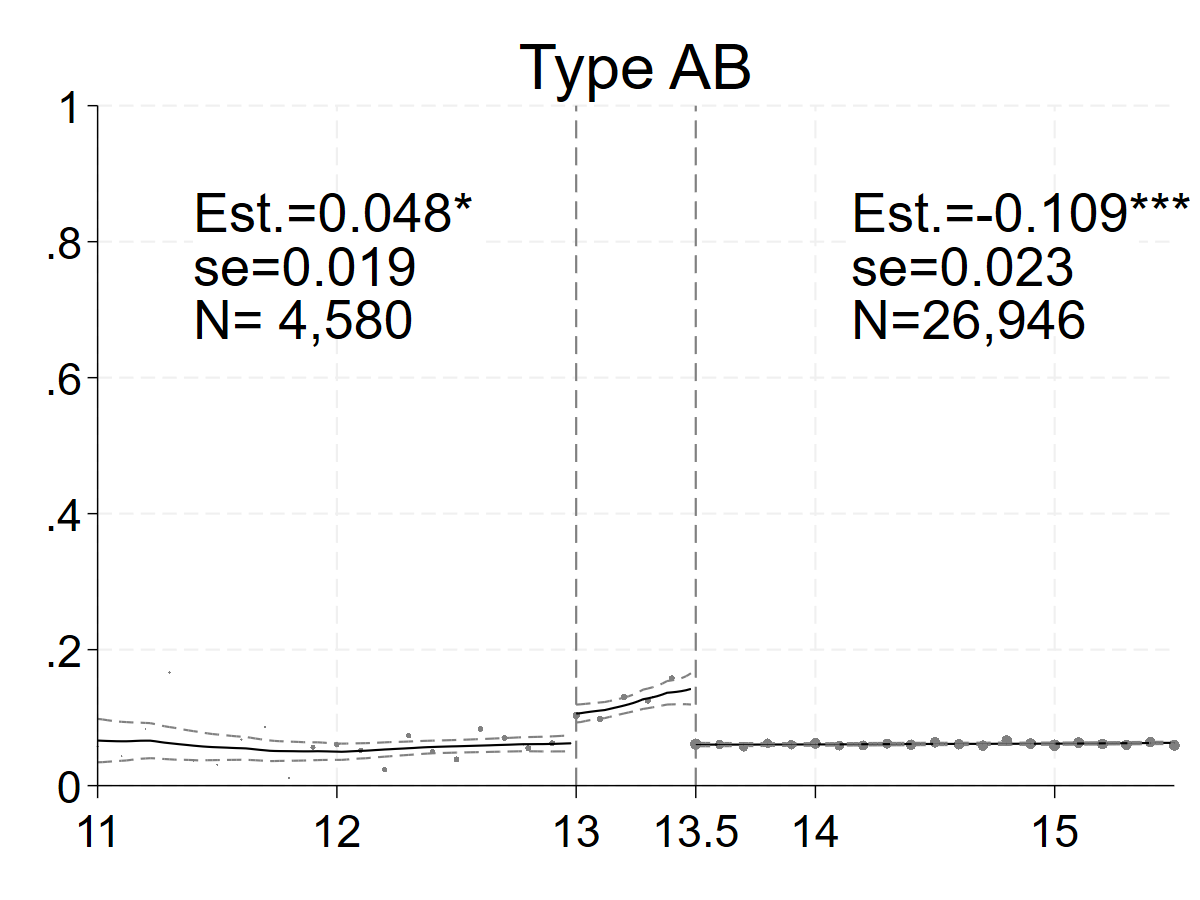}
    \includegraphics[width = 0.32\textwidth]{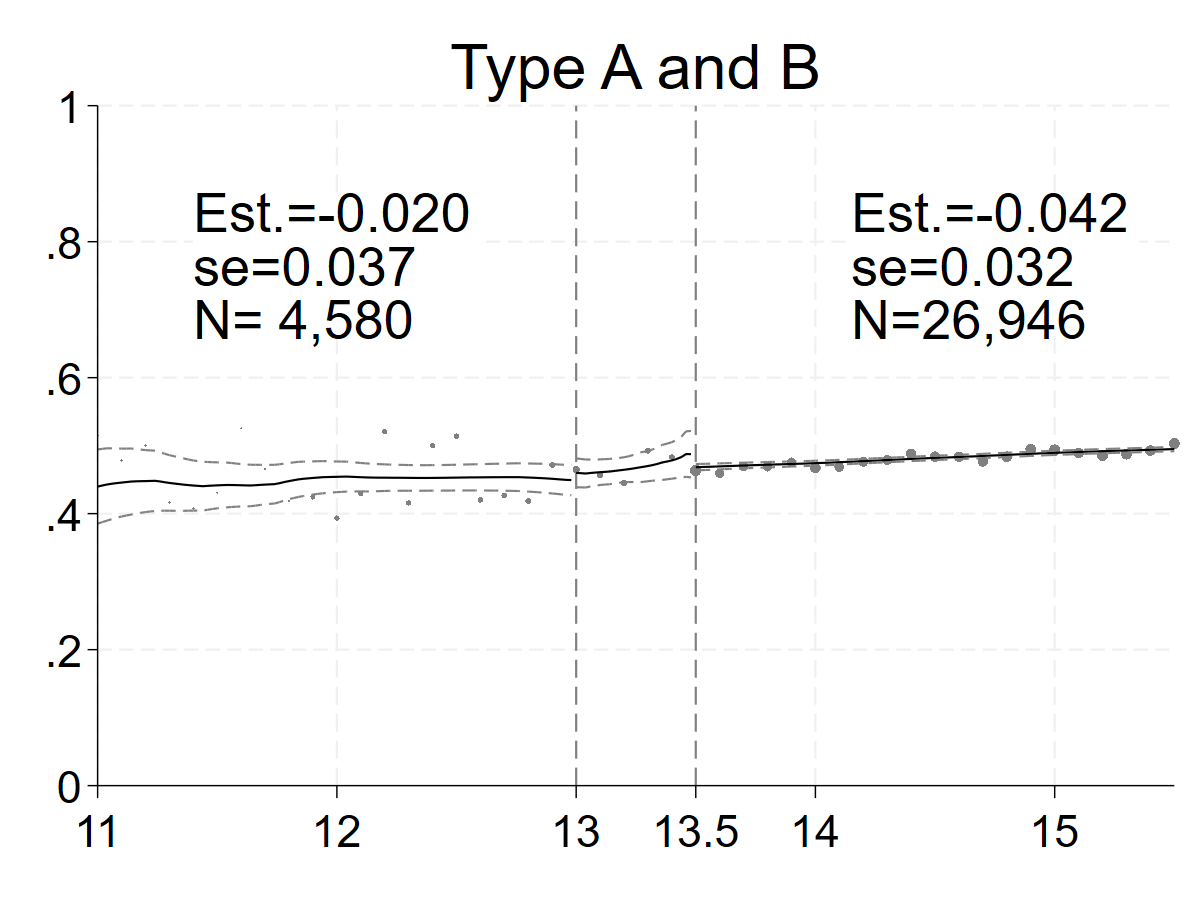}
    \caption{Difference in probability to be a first donor (first row, left), number of attempted donation previously (first row, middle), age (first row, right), weight (second row, left), height (second row, middle) and probability to have O-Negative blood type (i.e., Universal Whole Blood donor) (second row, right), O-Positive blood type (i.e., the most common blood type) (third row, left), AB blood type (i.e., Universal Plasma donor) (third row, middle) and A or B blood type (third row, right) around the eligibility threshold for men. Numbers on the figures shows estimate, standard error and sample size of RD estimates with a bandwidth of 0.5 for the 13g/dL threshold (left) and 13.5g/dL thresholds (right). The x-axis represents the measured h-level. \label{fig:discontinuitycovariates_M}}
\end{figure}

Figure \ref{fig:discontinuitycovariates_M} shows the distribution of covariates around the eligibility thresholds for men. The discontinuity in the probability of being a first-time donor and the number of attempted donations (top left and middle) are striking. Donors are markedly more experienced in the 13 to 13.4g/dL range, which is the range in which male donors can donate plasma but not whole blood. Furthermore, in that range they are more likely to have the AB blood type (the universal plasma donor), less likely to be O-Negative (the universal whole blood donor) and O-Positive (the most common blood type in high demand for whole blood donations). Furthermore, in that range, men are older, heavier and taller. These findings are in line with our theoretical framework in the case with two thresholds (Appendix \ref{sec:framework1_2threshold}), nurses allow plasma donations for more experienced donors and donors with less in-demand blood types.


Those discontinuities cannot be explained by our second theoretical framework. As Figure \ref{fig:theory_h2_2cutoff} in Appendix shows, if the covariates are strictly monotonic with the h-level, the jump in discontinuity should have the same sign at both thresholds which is not the case in Figure \ref{fig:discontinuitycovariates_M}.

\begin{figure}[ht]
    \centering
    \includegraphics[width = 0.32\textwidth]{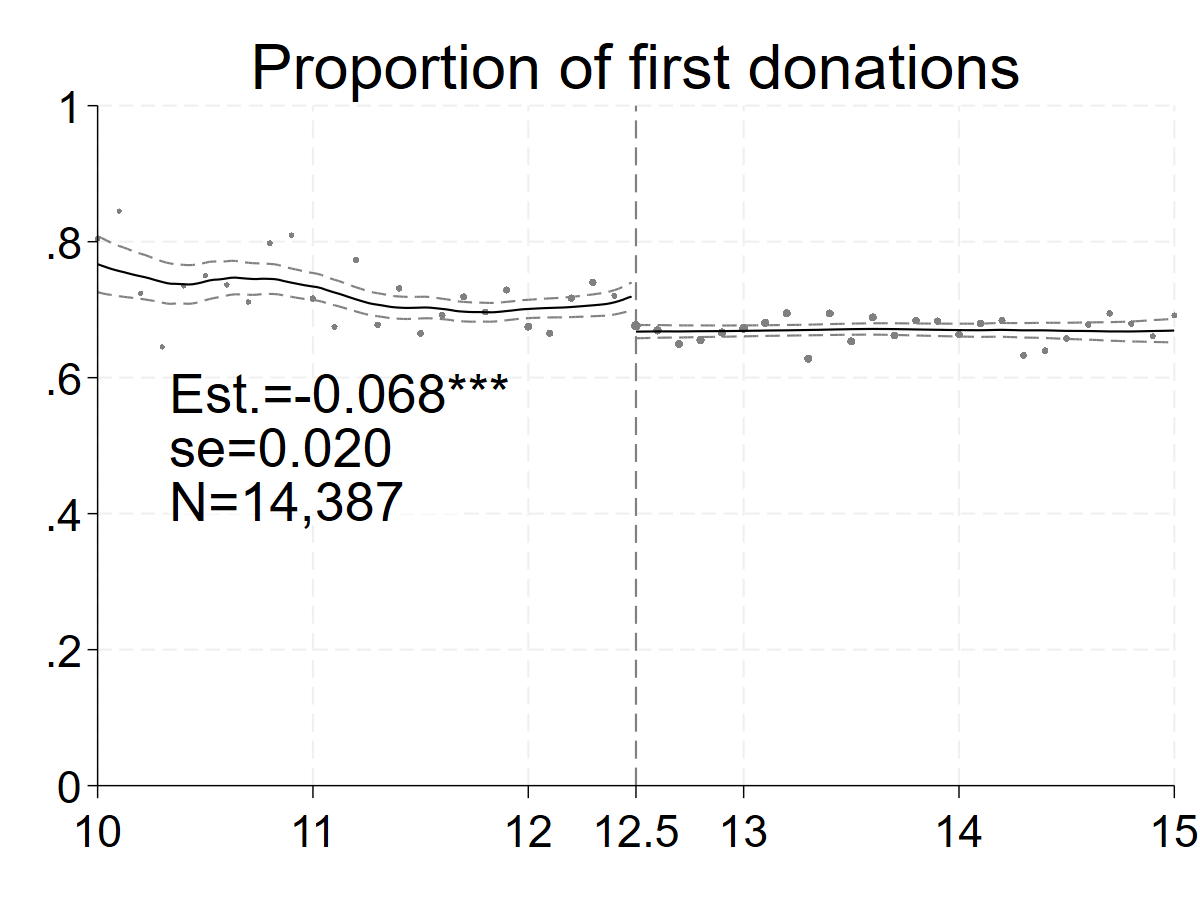}
    \includegraphics[width = 0.32\textwidth]{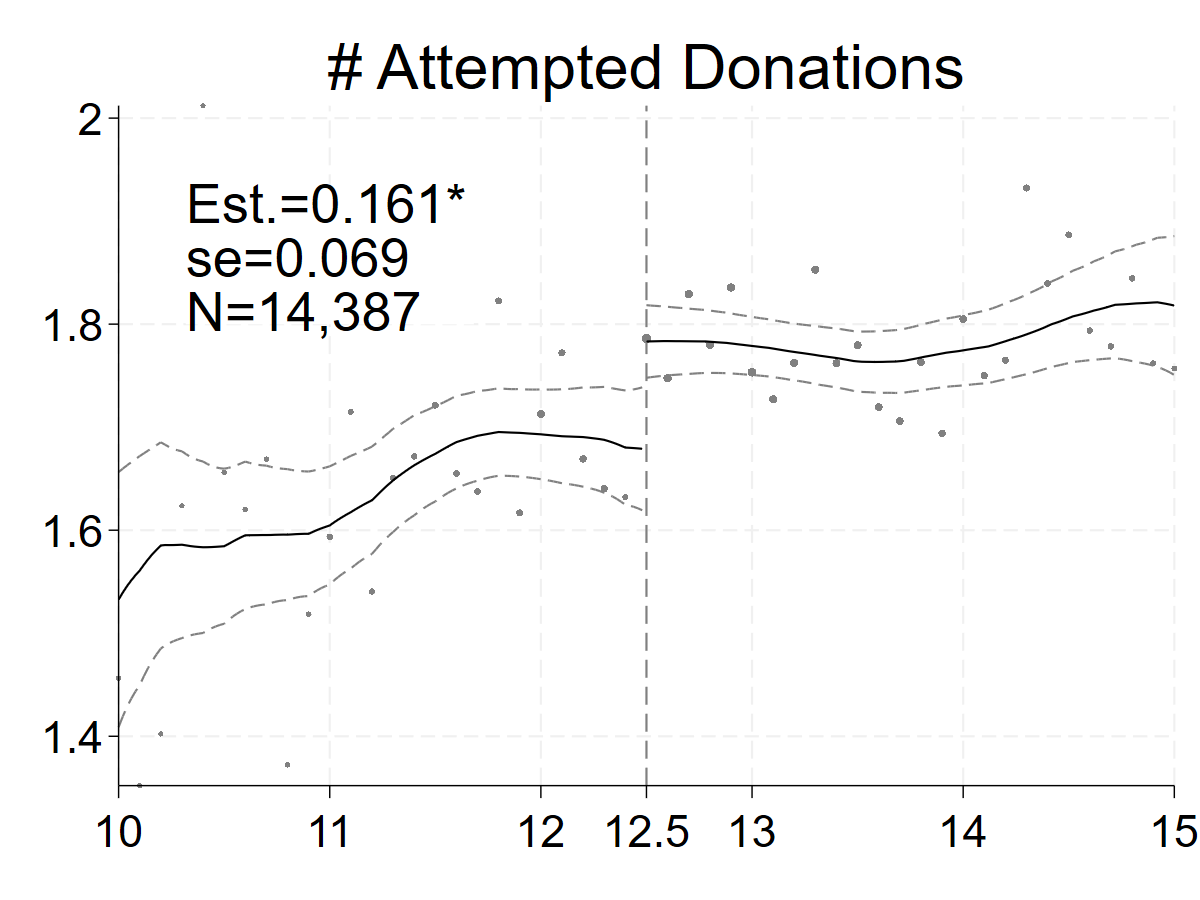}
    \includegraphics[width = 0.32\textwidth]{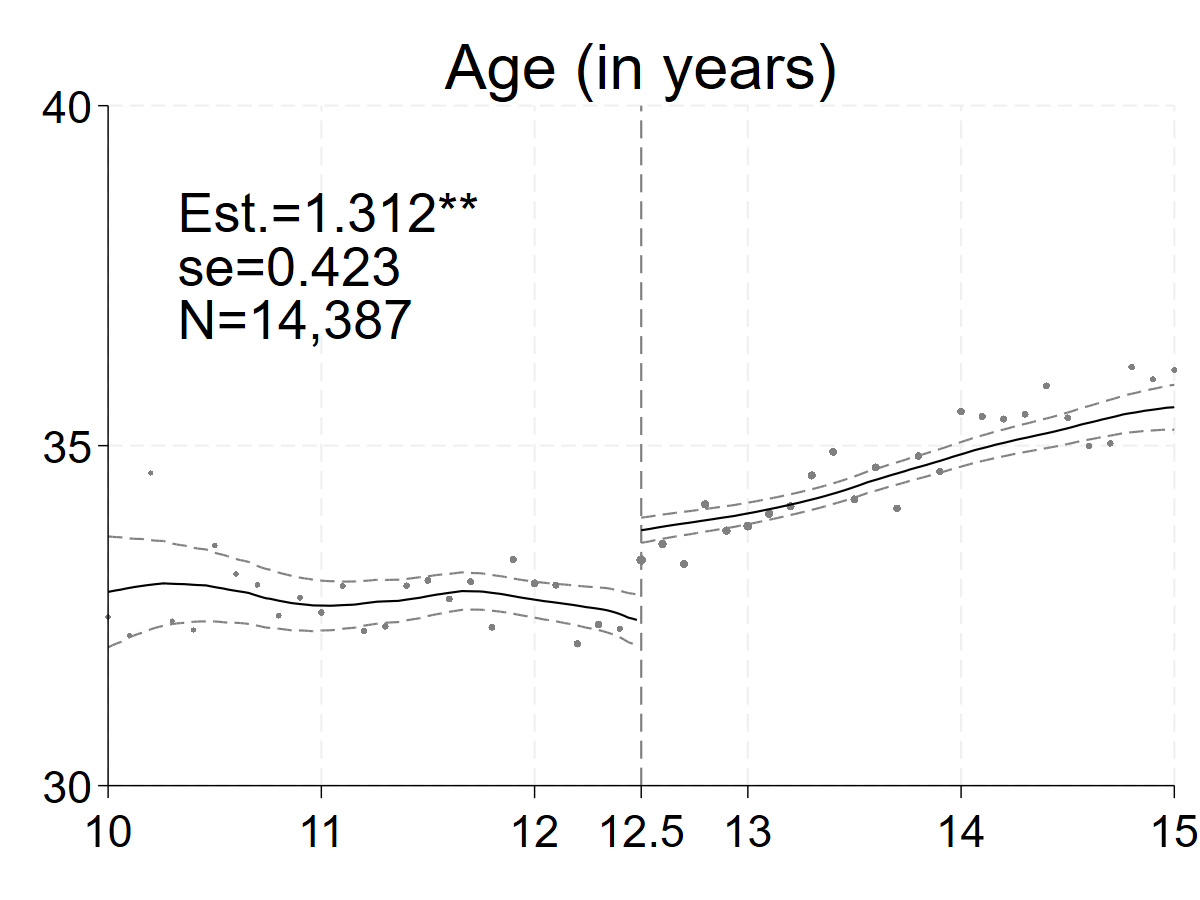}
    \includegraphics[width = 0.32\textwidth]{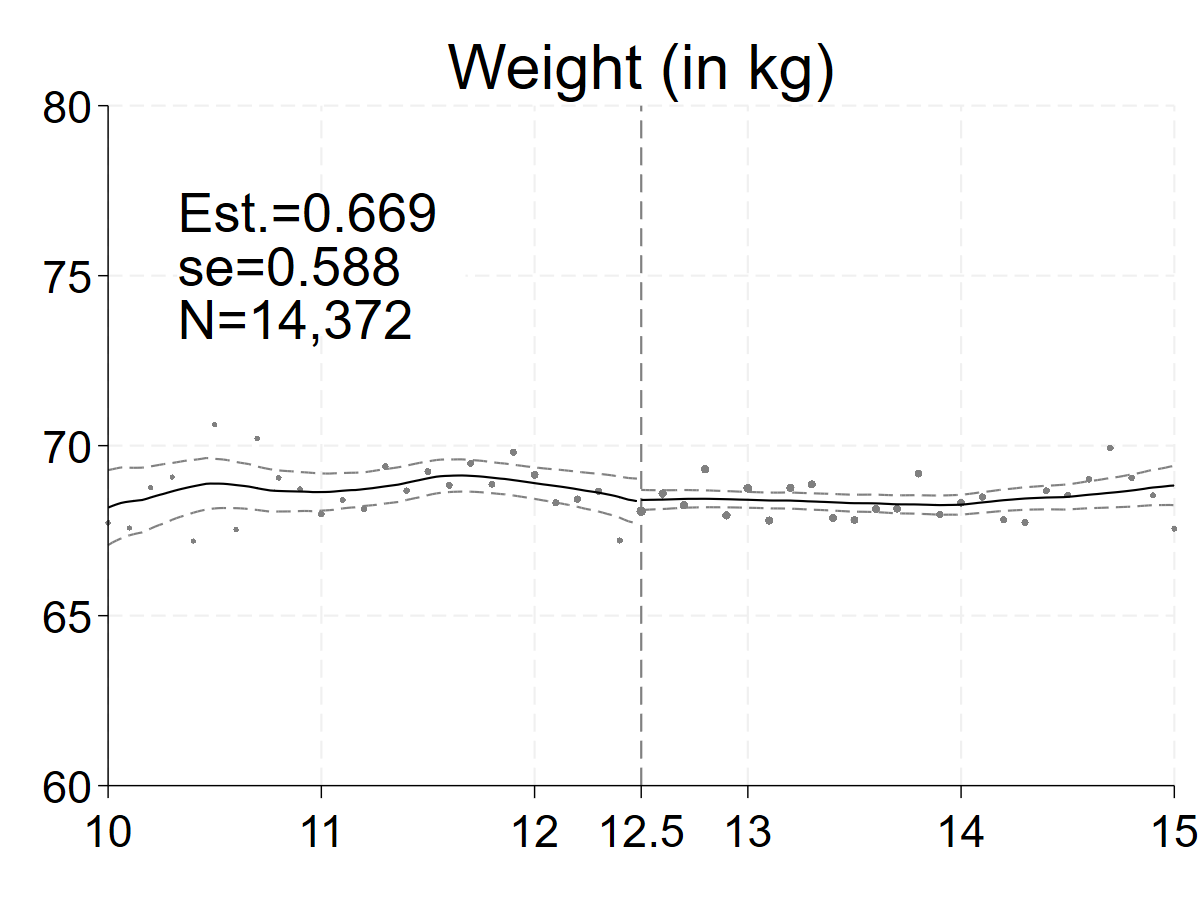}
    \includegraphics[width = 0.32\textwidth]{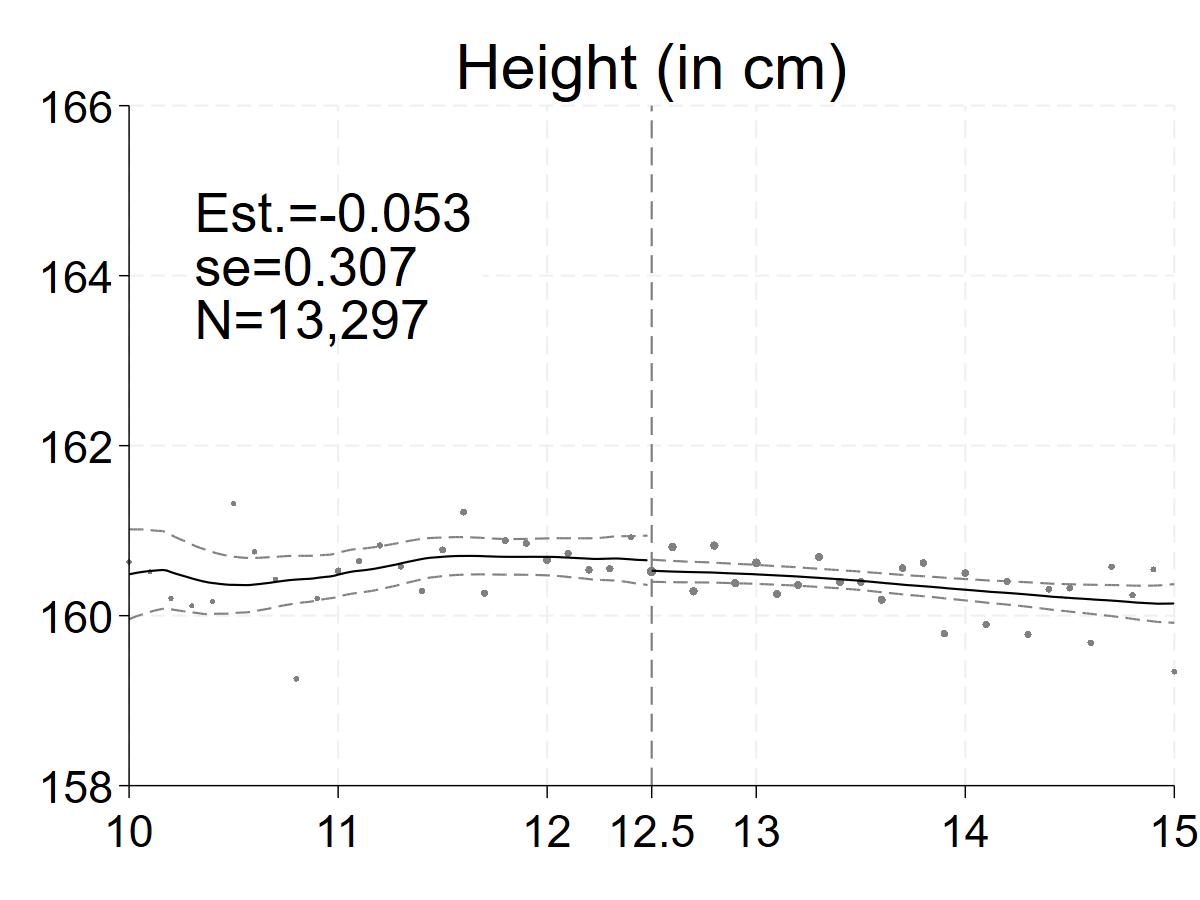}
    \includegraphics[width = 0.32\textwidth]{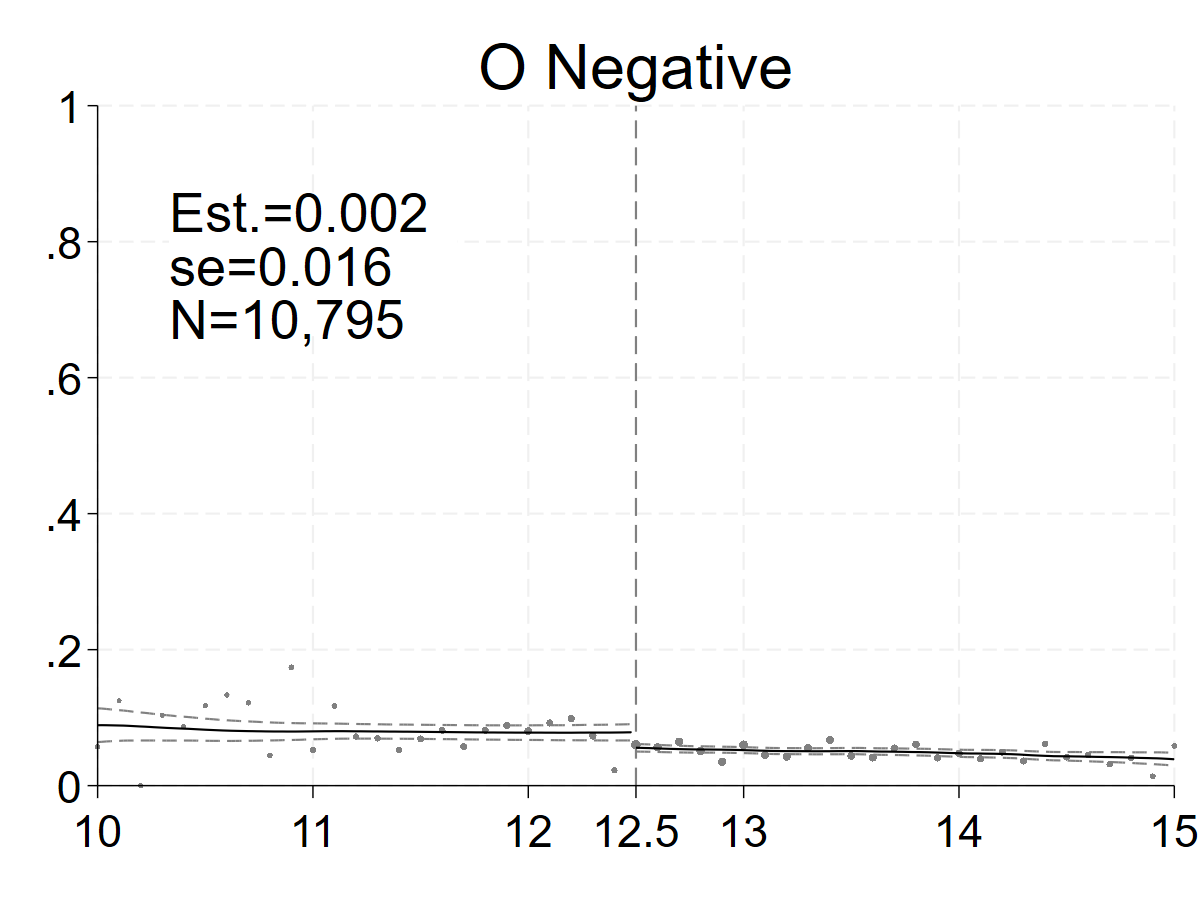}
    \includegraphics[width = 0.32\textwidth]{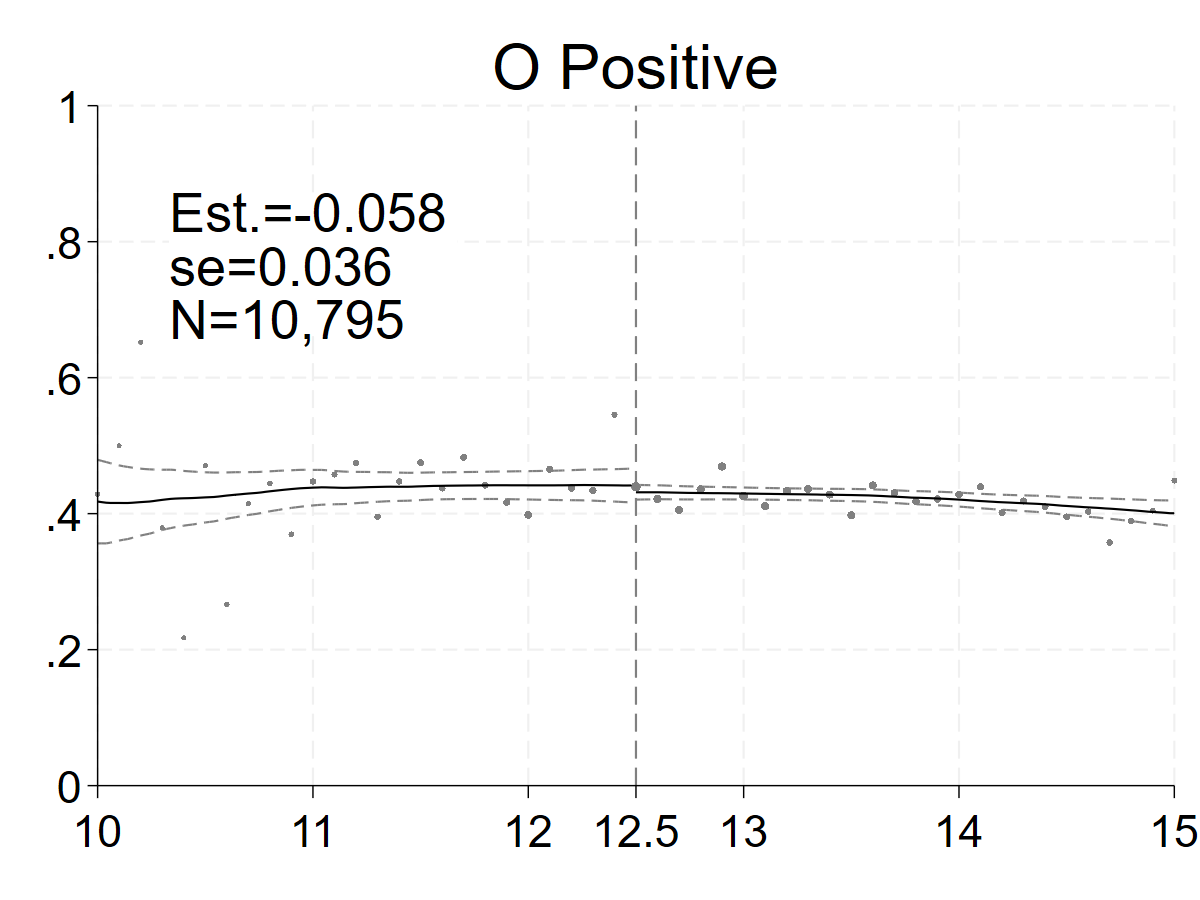}
    \includegraphics[width = 0.32\textwidth]{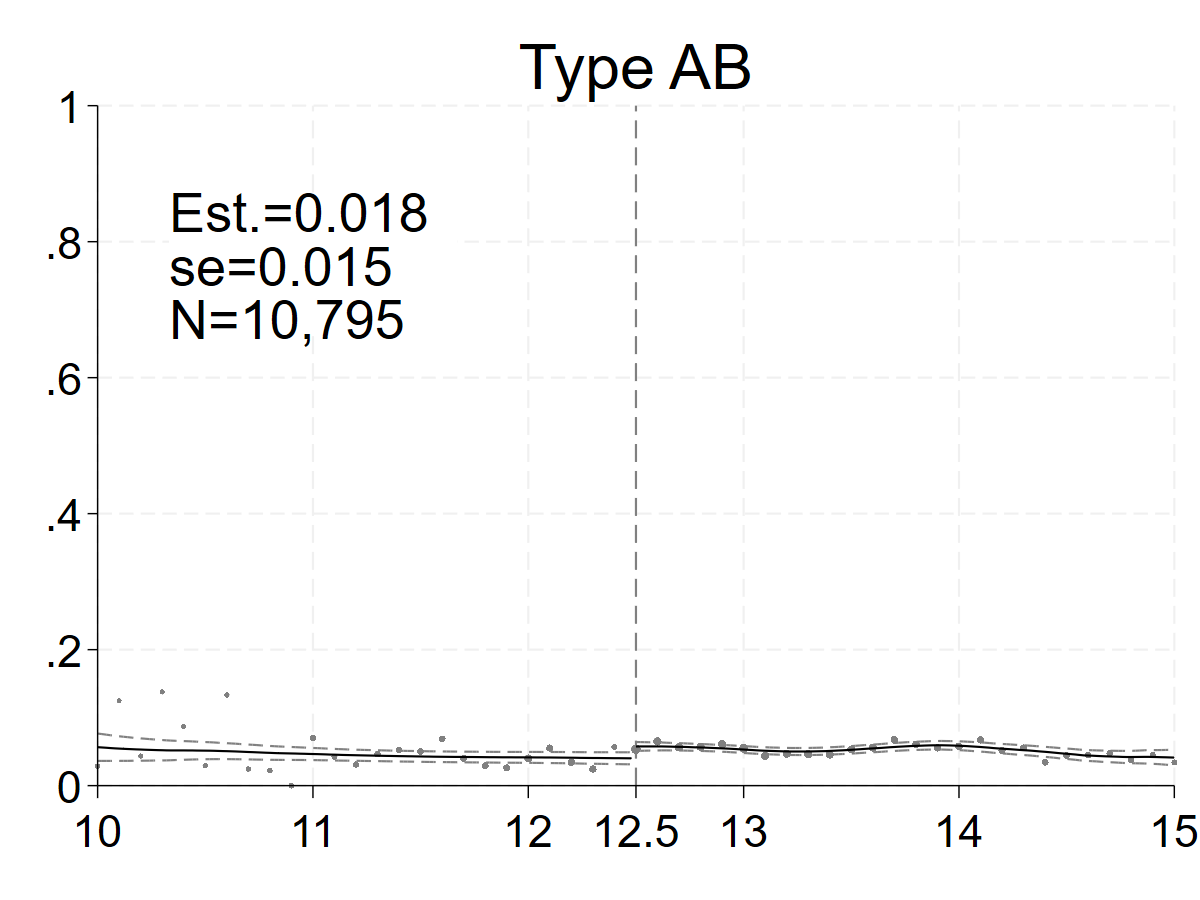}
    \includegraphics[width = 0.32\textwidth]{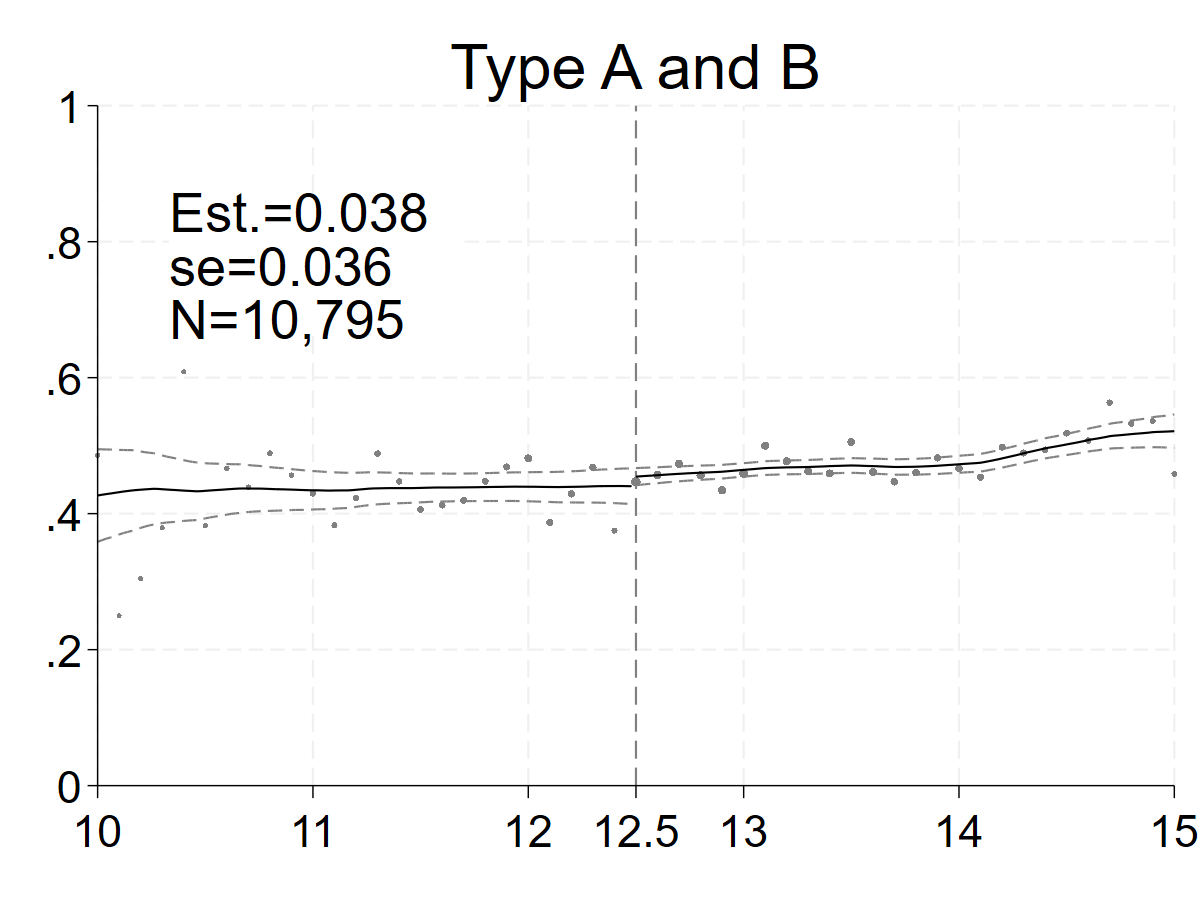}
\caption{Difference in probability to be a first donor (first row, left), number of attempted donation previously (first row, middle), age (first row, right), weight (second row, left), height (second row, middle) and probability to have O-Negative blood type (i.e., Universal Whole Blood donor) (second row, right), O-Positive blood type (i.e., the most common blood type) (third row, right), AB blood type (i.e., Universal Plasma donor) (third row, middle) and A or B blood type (third row, left)  around the eligibility threshold for women. Numbers shows estimate, standard error and sample size of RD estimates with the MSE-optimal bandwidth of 0.85. The x-axis represents the measured h-level.}
    \label{fig:discontinuitycovariates_F}
\end{figure}

Figure \ref{fig:discontinuitycovariates_F} is the analogous of Figure \ref{fig:discontinuitycovariates_M} for women. Covariates also exhibits discontinuities at the threshold for women. In line with our first theoretical framework, women are less experienced and older above the threshold. While we observe that women are less likely to be O negative above the threshold, this difference is not statistically significant.


As explained in Section \ref{sec:biasinRDDframework1}, the donors who are more likely to be bumped above the thresholds are also the ones who are more likely to come back (i.e., more experienced, older with more in-demand blood types). Therefore, we can not claim that donors just above and below the threshold are similar. This will bias our RDD estimates of a deferral on return behaviour downward, making it appear more negative than it truly is.


\FloatBarrier
\subsection{Bruhin et al. (2020) \label{sec:bruhinmanipulation}}

As mentioned earlier, we are not the first to exploit this structure as an RD design. \cite{bruhin2020sting} cleverly utilize the same discontinuity in h-level on a different dataset of blood donors to estimate the causal relationship between deferral and return behavior. \cite{bruhin2020sting} use data from Switzerland, which, similar to our dataset, exhibits a large amount of bunching in the running variable (see Figure \ref{fig:Histogram_Bruhin} in Appendix). However, in their setting, when the measured h-level falls below the threshold, it is remeasured twice more, and the average of the three measurements is reported. As \cite{bruhin2020sting} convincingly explains, this re-measurement protocol, which applies only when h-levels are below the threshold, creates bunching in the running variable. In contrast, in Abu Dhabi, the h-level is measured only once, and the observed bunching cannot be attributed to the measurement protocol.



\FloatBarrier
\section{Regression Discontinuity Design: Background and Manipulation\label{sec:litReview}}

The regression discontinuity (RD) design is a vital analytic tool for social scientists that can provide experiment-quality inference \citep{thistlethwaite1960regression}. It has gained popularity due to its applicability in diverse contexts where controlled experiments are difficult to run. For instance, RD designs have been used in medical settings \citep[e.g.][]{almond2011after}, education \citep[e.g.][]{cohodes2014merit}, and political science \citep[e.g.][]{lee2008randomized}. 

The RD design exploits scenarios in which each person or unit has an associated score, called the ``running variable," and treatments are assigned based on whether the score\footnote{We use the terms ``score'' and ``running variable'' interchangeably throughout this article.} falls above or below a threshold. Under the assumption that the score cannot be precisely manipulated near the threshold, the assignment to treatment is as good as random in a narrow window around the threshold. This quasi-randomization is exploited in the RD design to infer local average treatment effects.

Inference critically relies on the assumption that the score cannot be precisely manipulated 
\citep{cattaneo2020practical}, but this assumption is often problematic in practice. 
As mentioned earlier, recent studies have shown that teachers inflate students' grades to push them above grade thresholds \citep{diamond2016long}, police officers discount the recorded speed in roadside speed tests to pull it below fine thresholds \citep{goncalves2022should,goncalves2021few}, and close union certification elections are manipulated \citep{frandsen2017party}, or as in our paper, health care workers bump hemoglobin test results to allow blood donations to take place. In the many situations where the running variable is manipulated, ignoring the manipulation can yield biased inferences \citep{mccrary2008manipulation,frandsen2017party}.


To address violations of the RD design assumptions, \cite{mccrary2008manipulation} introduced an intuitive test for identifying the presence of running variable manipulation by examining its density. Given manipulation, we would expect to see a discontinuity in the running variable, resulting from individuals moving across the cutoff to secure a favorable treatment. McCrary suggests an empirical hypothesis test for this density discontinuity, making use of a density estimator originally proposed by \cite{cheng1997automatic}. Alternative tests for continuity of the running variable in the RD design have been proposed in \cite{arai2022testing} and \cite{bugni2021testing}. Other density estimators have subsequently been proposed by both \cite{otsu2013estimation} and \cite{cattaneo2020simple}. \cite{frandsen2017party} extended this approach by developing an alternative test for manipulation for the case of a discrete running variable. These methods have been widely adopted in the applied literature as a falsification test for checking the assumption of no manipulation; see \cite{jales2017identification} for a full review of them in the context of regression discontinuities and \cite{kleven2016bunching} for a related but distinct treatment from the point of view of bunching.

The applied literature suggests manipulation in RD designs occurs frequently \citep[e.g.][] {angrist2019maimonides, davis2013bounding, dee2019causes}. Yet, there is comparatively little clarity on how to proceed when these tests indicate manipulation. One method used in empirical work is the donut hole RD design \citep[such as in][] {almond2011after, bajari2017estimating, castleman2018intensive, kirdar2018effects}. The donut RD design deletes data within a window around the cutoff, with the goal being to remove all manipulated observations. Then RD estimates may be obtained from a parametric model fit on data outside the window and extrapolated to the cutoff. As yet, donut hole designs do not have a solid theoretical foundation and are therefore subject to ad-hoc estimation specifications. More fundamentally, by deleting \textit{all} the data around the cutoff, they weaken the RD design itself, which relies on exactly the data around the cutoff the most. 

One paper that does not resort to donut-hole style deletion is \cite{diamond2016long}. They consider Swedish math test data in which there is evidence of teachers inflating students' grades. They develop an estimator to determine the causal effect of the score manipulation on future educational attainment and earnings. While their focus is on a different causal effect than the one we consider, the paper develops several useful methods that we will incorporate here. We pursue the ``partial identification" approach, popularized by Manski and later by Tamer \citep[see e.g.,][]  {manski1990nonparametric, manski2002inference, haile2003inference}. The core idea is that, in scenarios in which a treatment effect cannot be point-identified (even with an infinite sample size), it can sometimes still be bounded. These bounds might be very informative in practice---for example, allowing us to rule out negative or positive treatment effects. \cite{davis2013bounding} show one way such a bound may be derived. However, they only provide a one-sided bound.

\cite{gerard2020bounds} take a two-sided partial identification approach to analyzing RD designs in the presence of manipulation. Their method posits the existence of subpopulations of manipulators and non-manipulators and defines the causal effect on the non-manipulators as the inferential target. We adopt the same framework but differ in our estimation technique. 
We compare our and \cite{gerard2020bounds}'s approach in Section \ref{sec:comparaisongerard2020}.

\section{RDD with manipulated running variable: Methodology}
\label{sec:methodology}

Manipulation of the running variable --- and the resultant selection bias --- have been a seemingly insurmountable roadblock to inference. To address this issue, we pursue the ``partial identification" approach, popularized by Manski and later by Tamer \citep{manski1990nonparametric, manski2002inference, haile2003inference}. Under partial identification, we obtain informative bounds on the treatment effect even when point estimation is not possible due to the manipulation.\footnote{An R package implementing this method is available at https://github.com/rajkumarkarthik/rdpartial, and the authors are available to assist with any user questions.}

\subsection{Intuition}

Our method addresses manipulation of the running variable by exploiting a simple fact: while we cannot identify which specific observations are manipulated, we can -- under reasonable assumptions -- estimate how many individuals are likely to be manipulators at each value of the running variable.

To estimate the proportion of manipulated observations, we proceed in three steps. First, we use a data-driven algorithm to select a manipulation region around the threshold (for an example, see the gray area in Figure \ref{fig:unmaniphist}). Next, using data outside this region, we estimate the counterfactual distribution: what the data would have looked like without the manipulation (see the red line in Figure \ref{fig:unmaniphist}). Third, we compare this estimated distribution against the observed histogram to infer the proportion of manipulated observations above the threshold. In Figure \ref{fig:unmaniphist}, this is visualized via the gray area above the red line, among bars exceeding the threshold h-level of 12.5 g/dL.

Using these estimated manipulator proportions, we estimate a fuzzy RDD under two extreme scenarios. In the first, we obtain the largest possible causal estimate by designating as manipulators the most ``pessimistic" cohort and excluding them from the analysis. In the second, we obtain the smallest possible causal estimate by designating as manipulators the most ``optimistic" cohort and again excluding them. This yields partial identification bounds: a range within which the true causal effect must lie, given the extent of manipulation. 

\subsection{Notation and Preliminaries}

Suppose we have a regression discontinuity design with running variable $X_i$ for units $i = 1, \dots, N$. These units assumed to be sampled from a larger super-population. $X_i$ is assumed discrete, with possible values enumerated in $\mathcal{X}$ such that $|\mathcal{X}| < \infty$. Suppose also we have cutoff $c \in \mathcal{X}$, with $N_{\ell}$ units falling to the left of the cutoff and $N_r$ units falling at or above the cutoff, where $N_{\ell} + N_r = N$. In the discussion to follow, we will assume for simplicity that the entries of $\mathcal{X}$ are integers lying within a given range.

We associate with each unit $i$ a pair of unseen potential outcomes $(Y_i(0), Y_i(1)) \in \mathbb{R}^2$, corresponding to the value of the outcome if unit $i$ does not or does receive the treatment, respectively. We also associate with each unit an observed value of the running variable $\tilde X_i$ and a true, unobserved value of the running variable $X_i$. As a shorthand, define $H_i \in \{0, 1\}$ as an indicator of whether a unit is an ``honest" subject, as opposed to a manipulator:
\[ H_i = \left\{ \begin{array}{ll} 1 & \text{ if } X_i = \tilde X_i \\ 0 & \text{ if } X_i \neq \tilde X_i. \end{array}\right. \] 

The researcher observes a  treatment assignment $W_i \in \{0, 1\}$. We will begin by considering the case of a sharp RD design, such that $\tilde X_i < c \implies W_i = 0$ and $\tilde X_i \geq c \implies W_i = 1$. We will later extend results to the fuzzy RD design, where --- as is the case with our blood donation setting --- the probability probability of treatment is discontinuous at $\tilde X_i \geq c$. In either case, note that treatment assignments are based on the observed running variable rather than the true running variable. The researcher observes 
\[ Y_i = W_i Y_i(1) + (1-W_i)Y_i(0)\, \] 
the outcome for each unit. Our estimand of interest is 
\[ \tau(c) = E(Y_i(1) - Y_i(0) \mid X_i = c, H_i = 1) \] 
where the expectation is with respect to the super-population from which our data is sampled.

\subsection{Causal Estimation}

Many standard sharp RDD causal effect estimators can be written in the following form: 
\begin{equation}\label{eq:stDef}
\hat \tau(c) = c^{\star T} \underbrace{\left(\boldsymbol{\tilde X}_r^T \boldsymbol{M}_r \boldsymbol{\tilde X}_r \right)^{-1} \boldsymbol{\tilde X}_r^T  \boldsymbol{M}_r Y_r}_{\hat \beta_r} -
c^{\star T} \underbrace{ \left(\boldsymbol{\tilde X}_{\ell}^T  \boldsymbol{M}_{\ell} \boldsymbol{\tilde X}_{\ell} \right)^{-1} \boldsymbol{\tilde X}_{\ell}^T  \boldsymbol{M}_{\ell} Y_{\ell}}_{\hat \beta_{\ell}} .
\end{equation}
Per the under-bracketed quantities, these estimators separately calculate two coefficient vectors: one from a regression relating outcomes to the running variable below the cutoff, the other above the cutoff. The causal estimate is given by the difference in these two regression predictions at the cutoff $c$.

Specifically, $\boldsymbol{\tilde X}_{\ell} \in \mathbb{R}^{N_{\ell} \times p}$ and $\boldsymbol{\tilde X}_r \in \mathbb{R}^{N_r \times p}$ are concatenated basis expansion of the observed variables $\tilde X_i$ for units to the left and right of the cutoff, respectively; $c^{\star} \in \mathbb{R}^p$ is an analogous basis expansion of the cutoff $c$; and $\boldsymbol{M}_{\ell} \in \mathbb{R}^{N_{\ell} \times N_{\ell}}$ and $\boldsymbol{M}_r \in \mathbb{R}^{N_r \times N_r}$ are diagonal matrices representing unit-level weights. We require that the unit-level weights depend solely on the distance between the observed running variable value $\tilde X_i$, and the cutoff, $c$. The popular local polynomial regression approach \cite{hahn2001identification} can be expressed in this form, as can spline formulations \cite{lemieux2008incentive} and simpler unweighted regressions. 

In the ideal case, there is no manipulation, such that $\tilde X_i = X_i$ for all units. If the researcher also correctly posits the functional form of the estimator, then $\hat \beta_r$ and $\hat \beta_{\ell}$ are unbiased estimators of the true coefficient vectors $\beta_r$ and $\beta_{\ell}$ relating $Y_i(1)$ and $Y_i(0)$ to $X_i$, respectively. As a consequence, $\hat \tau(c)$ is be an unbiased estimator of $\tau(c)$.\footnote{In practice, one uses a local polynomial estimator with an ex-ante unknown bandwidth and chooses the bandwidth to minimize mean squared error of the RDD estimator. The underlying idea is to be flexible with the functional form, while picking the specification with the least error. This ``nonparametric'' specification brings an asymptotic bias term that can be explicitly corrected for. See Sections 4.2 and 4.3 of \cite{cattaneo2020practical} for more on this. The method we introduce in this paper is agnostic to the exact design matrix used and is flexible to many common specifications.}

However, in the presence of manipulation, these estimators would be ``polluted" by the presence of individuals who had manipulated their running variable scores. That is, the RDD estimator is no longer a consistent estimator for the true causal effect of interest, but is biased by the selection of manipulators into treatment. 

We will consider the case of unidirectional manipulation. In most applications, assuming the manipulation to be unidirectional is well-grounded and fits empirical evidence. If being part of the treatment group is beneficial, one faces incentives to manipulate the running variable to be eligible but not to be ineligible. Similarly, if being in the treatment group is detrimental, people face incentives to manipulate the running variable to be ineligible but not to be eligible. Furthermore, unidirectional manipulation of the running variables leads to bunching at the threshold, which is observed in many empirical applications. For instance, taxpayers benefit by misreporting income below kink points but do not have any reason to misreport income above those kink points \citep{saez2010taxpayers}. In \cite{diamond2016long} teachers have incentives to inflate students' scores but have no incentives to reduce students' scores (see section 2.2 where teachers' incentives are discussed).

With unidirectional manipulation, only the estimate $\beta_r$ is polluted by the presence of manipulators. A valid point estimate of $\beta_r$ \emph{could} be recovered if we had access to the values $H_i$ for each unit. Denote as
$ \boldsymbol{H}_r = \text{diag}(H_1, \dots, H_{N_r}) \in \{0, 1\}^{N_r \times N_r}$ the diagonal matrix containing the honesty indicators for units above the cutoff. That is, $H_i$ is 0 if $i$ is a manipulator and $1$ otherwise. Then the estimator 
\begin{equation*}
    \hat \tau(c) = c^{\star T} \left(\boldsymbol{\tilde X}_r^T \boldsymbol{M}_r \boldsymbol{H}_r \boldsymbol{\tilde X}_r \right)^{-1} \boldsymbol{\tilde X}_r^T  \boldsymbol{M}_r  \boldsymbol{H}_r Y_r - c^{\star T} \hat \beta_{\ell},
\end{equation*}
with $\hat \beta_{\ell}$ as defined in (\ref{eq:stDef}), would ``ignore" all manipulators. Thus $\hat \tau^{\star}(c)$ cannot be computed because $\boldsymbol{H}_r$ is not known in practice, which is the reason we motivate partial identification. Nevertheless, we use its definition to provide the basis for our procedure. 

\subsection{An Optimization Problem}

Though we cannot directly obtain the $H_i$ values that populate the diagonal of $\boldsymbol{H}_r$, we can obtain information about the count of manipulators at each unique value of $\tilde X_i$ at or above $c$. This is a central tenet of our approach: we use this information to linearly constrain $\boldsymbol{H}_r$, and then pose the problem as an optimization.

We obtain the counts of manipulators by estimating the un-manipulated density of the forcing variable from the observed density, as in \cite{diamond2016long}. We assume the un-manipulated density is log-concave, except for rounding effects at integers. The log-concavity allows our density to be unimodal. We postulate the existence of a ``manipulation region,'' within (and only within) which there are deviations in the observed density from the un-manipulated density. Further, we enforce a constraint that the ``missing'' mass to the left of the cutoff $c$ in the manipulation region (i.e. the un-manipulated density minus the observed density) equals the ``excess'' mass to the right of the cutoff (i.e. the observed density minus the un-manipulated density). We estimate this model for the un-manipulated density using a Poisson regression, which gives us the un-manipulated density along with the counts of manipulators (i.e. the excess mass to the right of the threshold). 

Note that, in the ideal case, we could identify the manipulation region and simply solve for the log-concave distribution closest to our observed distribution outside of the manipulation region. In turn, this approach would yield identifiability on the un-manipulated region. However, the manipulation region itself is not identified, nor are we aware of an efficient algorithm to compute this closest distribution for known manipulation regions, under our constraints. Our approach, heavily influenced by \cite{diamond2016long}, allows us to heuristically approximate the ideal procedure. This approach is highly scalable, and yields un-manipulated distributions that are smooth, well-behaved, and appear very reasonable to the naked eye.

With our estimated un-manipulated distribution in hand, we can proceed to obtaining the partial identification bounds. Suppose that $\mathcal{X}$ contains $n_r$ unique values greater than or equal to $c$. The honest subject counts can be represented as a vector $\nu \in \mathbb{R}^{n_r}$ whose $j^{th}$ entry $\nu_j$ corresponds to the number of honest subjects for whom $X_i = \tilde X_i = c + (j - 1)$. Then, defining a matrix 
\[ \mathbb{Z} = \left(\begin{array}{llll}I(\tilde X_i = c) & I(\tilde X_i = c) & \dots & I(\tilde X_i = c) \\
I(\tilde X_i = c + 1) & I(\tilde X_i = c + 1) & \dots & I(\tilde X_i = c + 1) \\
\vdots & \vdots & \ddots & \vdots \\ 
I(\tilde X_i = c + n_r - 1) & I(\tilde X_i = c + n_r - 1) & \dots & I(\tilde X_i = c + n_r - 1) \end{array} \right) \,, \] 
we can capture our knowledge of the honest subject counts via the constraint $\mathcal{Z} H_r = \nu$ where $\text{diag}(\textbf{H}_r) = H_r$.

Now, we can compute the largest and smallest values of our estimated causal effect $\hat \tau^{\star}(c)$ that is consistent with this constraint. We will consider only the upper bound, as the lower bound is computed identically except using minimization rather than maximization. For simplicity of notation, we also drop most of the ``r" subscripts in defining Optimization Problem \ref{optProb1}: 

\setcounter{equation}{0}

\begin{equation}\label{optProb1}
\begin{aligned}
\text{maximize} & \hspace{5mm} c^{\star T} \left(\boldsymbol{\tilde X}^T \boldsymbol{M} \boldsymbol{H} \boldsymbol{\tilde X} \right)^{-1} \boldsymbol{\tilde X}^T \boldsymbol{M} \boldsymbol{H} Y - c^{\star T} \hat \beta_{\ell} \\
\text{subject to} & \hspace{5mm} H =  \text{diag}\left(\boldsymbol{H}\right), \\
& \hspace{5mm} H \in \{0, 1\}^{N_r},\\
& \hspace{5mm} \mathbb{Z}H = \nu \,. 
\end{aligned}
\end{equation}

The Boolean constraint on $H$ will yield an intractable problem. Observe that relaxing the constraint can only widen the bounds, so we instead solve the simplified problem:

\begin{equation}\label{optProb2}
\begin{aligned}
\text{maximize} & \hspace{5mm} c^{\star T} \left(\boldsymbol{\tilde X}^T \boldsymbol{M} \boldsymbol{H} \boldsymbol{\tilde X} \right)^{-1} \boldsymbol{\tilde X}^T \boldsymbol{M} \boldsymbol{H} Y - c^{\star T} \hat \beta_{\ell} \\
\text{subject to} & \hspace{5mm} H =  \text{diag}\left(\boldsymbol{H}\right), \\
& \hspace{5mm} 0 \leq H \leq 1,\\
& \hspace{5mm} \mathbb{Z}H = \nu \,. 
\end{aligned}
\end{equation}

\subsection{Reduction to a Linear Program}

Optimization Problem \ref{optProb2} can be reduced to a linear program through several observations. Define $\boldsymbol{\tilde X_i}$ as the $i^{th}$ row of $\boldsymbol{\tilde X}$ and $M_{ij}$ the $(i, j)^{th}$ entry of $\boldsymbol{M}$. Then
\begin{align*}
\left(\boldsymbol{\tilde X}^T \boldsymbol{M} \boldsymbol{H} \boldsymbol{\tilde X} \right) = \sum_{i} M_{ii} H_i  \boldsymbol{\tilde X_i} \boldsymbol{\tilde X_i}^T 
\end{align*}
Recall that $\boldsymbol{\tilde X_i} = \boldsymbol{\tilde X_j} \iff \tilde X_i = \tilde X_j$, since $\boldsymbol{\tilde X_i}$ is simply a basis expansion of $\tilde X_i$. Moreover, since the regression weights are based on the distance between $\tilde X_i$ and $c$, we will also have that $\tilde X_i = \tilde X_j \iff M_{ii} = M_{jj}$. We will define 
\begin{align*}
\boldsymbol{\tilde X_{(\ell)}} &:= \boldsymbol{\tilde X_i} \text{ for $i$ such that $X_i = \ell$} \\
M_{(\ell)} &:= M_{ii} \text{ for $i$ such that $X_i = \ell$} 
\end{align*}
We also denote as $\nu_{(\ell)}$ the entry in $\nu$ corresponding to the number of honest subjects for whom $\tilde X_i = \ell$. We assume the problem is feasible and hence a solution must exist that satisfies $\mathbb{Z} H = \nu$. Imposing this constraint, we can now write
\begin{align*}
\left(\boldsymbol{\tilde X}^T \boldsymbol{M} \boldsymbol{H} \boldsymbol{\tilde X} \right) &= \sum_{\ell = c}^{c + n_r - 1} \left( \sum_{i = 1}^N H_i I(\tilde X_i = \ell) \right)  M_{(\ell)} \left( \boldsymbol{\tilde X_{(\ell)}} \boldsymbol{\tilde X_{(\ell)}^T} \right) \\
&= \sum_{\ell = c}^{c + n_r - 1} \nu_{(\ell)}  M_{(\ell)} M_{(\ell)} \left( \boldsymbol{\tilde X_{(\ell)}} \boldsymbol{\tilde X_{(\ell)}^T} \right) \,,
\end{align*}
where the second line follows from satisfaction of the $\mathbb{Z} H = \nu$ constraints. Note that the expression on the right-hand side is not a function of the $H_i$. Hence, we can define a constant matrix,
\[ \boldsymbol{\Phi} = \left(\boldsymbol{\tilde X}^T \boldsymbol{M} \boldsymbol{H} \boldsymbol{\tilde X} \right) \] 
and rewrite our optimization problem as 
\begin{equation}\label{optProb3}
\begin{aligned}
\text{maximize} & \hspace{5mm} c^{\star T}  \boldsymbol{\Phi}^{-1} \boldsymbol{\tilde X}^T \boldsymbol{M} \boldsymbol{H} Y - c^{\star T} \hat \beta_{\ell} \\
\text{subject to} & \hspace{5mm} H =  \text{diag}\left(\boldsymbol{H}\right), \\
& \hspace{5mm} 0 \leq H \leq 1,\\
& \hspace{5mm} \mathbb{Z}H = \nu \,. 
\end{aligned}
\end{equation}

Now, recall that the trace of a constant is simply the constant itself. Hence, we can rewrite the first term of the objective as 
\begin{align*}
c^{\star T}  \boldsymbol{\Phi}^{-1} \boldsymbol{\tilde X}^T \boldsymbol{M} \boldsymbol{H} Y &= \Tr \left(c^{\star T}  \boldsymbol{\Phi}^{-1} \boldsymbol{\tilde X}^T \boldsymbol{M} \boldsymbol{H} Y \right) \\
&= \Tr \left(  Y c^{\star T}  \boldsymbol{\Phi}^{-1} \boldsymbol{\tilde X}^T \boldsymbol{M} \boldsymbol{H} \right)\\
&= \text{diag} \left( Y c^{\star T}  \boldsymbol{\Phi}^{-1} \boldsymbol{\tilde X}^T \boldsymbol{M}  \right) H
\end{align*}
where the second line follows from the cyclic property of the trace; and the third line from matrix algebra rules and the fact that $H = \text{diag}(\boldsymbol{H})$.

Hence, we can reduce our optimization problem to a simple linear program, given by 
\begin{equation}\label{linearizedOptProblem}
\begin{aligned}
\text{maximize} & \hspace{5mm} \text{diag} \left( Y c^{\star T}  \boldsymbol{\Phi}^{-1} \boldsymbol{\tilde X}^T \boldsymbol{M}  \right) H - c^{\star T} \hat \beta_{\ell} \\
\text{subject to} & \hspace{5mm} 0 \leq H \leq 1,\\
& \hspace{5mm} \mathbb{Z}H = \nu \,. 
\end{aligned}
\end{equation}
This problem can be efficiently solved by conventional solvers. 

\subsection{Extension to Fuzzy RDD Case}

Our blood donation setting involves a fuzzy, rather than sharp, RDD --- meaning that the probability of a donation increases sharply at the threshold, but not all the way from zero to one. 

Our approach can be extended easily to the case of the fuzzy RDD. In this case, we suppose the estimate of the causal effect is obtained via an instrumental variable approach. The numerator is the difference of the mean treated outcomes just above and just below the cutoff, and the denominator is the difference of the treatment probabilities just above and just below the cutoff.

We require that the the linear probability model be used to estimate the change in treatment probabilities at the cutoff. Further, we require the same basis expansion be used as in estimating the discontinuity in the outcome. Then, we can define an alternative estimator for the causal effect,
\[ \hat \tau_f^{\star}(c) = \frac{c^{\star T} \left(\boldsymbol{\tilde X}_r^T \boldsymbol{M}_r \boldsymbol{H}_r \boldsymbol{\tilde X}_r \right)^{-1} \boldsymbol{\tilde X}_r^T  \boldsymbol{M}_r  \boldsymbol{H}_r Y_r - c^{\star T} \hat \beta_{\ell}}{c^{\star T} \left(\boldsymbol{\tilde X}_r^T \boldsymbol{M}_r \boldsymbol{H}_r \boldsymbol{\tilde X}_r \right)^{-1} \boldsymbol{\tilde X}_r^T  \boldsymbol{M}_r  \boldsymbol{H}_r Z_r - c^{\star T} \hat \alpha_{\ell}} \] 
where $Z_r$ is the vector of treatment indicators to the right of the cutoff and $\hat \alpha_{\ell}$ is the fitted coefficient corresponding to the regression of $Z$ on $X$ to the left of the cutoff. 

The upper bound on the causal effect can be obtained via an optimization problem in which we maximize $\tau_f^{\star}(c)$ subject to constraints on the histogram of honest subjects. As in the prior section, we can substitute $\boldsymbol{\Phi}$ to obtain the optimization problem: 
\begin{equation}\label{fuzzyOptProb}
\begin{aligned}
\text{maximize} & \hspace{5mm} \frac{c^{\star T} \boldsymbol{\Phi}^{-1} \boldsymbol{\tilde X}^T \boldsymbol{M} \boldsymbol{H} Y - c^{\star T} \hat \beta_{\ell}}{c^{\star T} \boldsymbol{\Phi}^{-1} \boldsymbol{\tilde X}^T \boldsymbol{M} \boldsymbol{H} Z - c^{\star T} \hat \alpha_{\ell}} \\
\text{subject to} & \hspace{5mm} H = \text{diag}(\boldsymbol{H}), \\
& \hspace{5mm} 0 \leq H \leq 1, \\
& \hspace{5mm} \mathbb{Z} H = \nu \,,
\end{aligned} 
\end{equation}
where we have again dropped ``r" subscripts for simplicity. Finally, using the same trace trick as in the prior section, we can reduce the problem to a linear fractional programming problem,
\begin{equation}\label{fuzzyOptLinearizedProb}
\begin{aligned}
\text{maximize} &\hspace{5mm} \frac{\text{diag} \left(Y c^{\star T} \boldsymbol{\Phi}^{-1} \boldsymbol{\tilde X}^T \boldsymbol{M} \right)^T H - c^{\star T} \hat \beta_{\ell} }{\text{diag} \left(Z c^{\star T} \boldsymbol{\Phi}^{-1} \boldsymbol{\tilde X}^T \boldsymbol{M} \right)^T H - c^{\star T}\hat \alpha_{\ell}}\\
\text{subject to} & \hspace{5mm} 0 \leq H \leq 1\,,\\
& \hspace{5mm} \mathbb{Z} H = \nu \,. 
\end{aligned}
\end{equation}

Using results from \cite{boyd2004convex}, this problem can be transformed into an equivalent linear program and solved via conventional solvers. 

\subsection{Bootstrap Inference}\label{subsec:boot}

A natural question is how to do inference when using our estimation procedure. We suggest the use of a percentile bootstrap procedure which conditions upon the observed histogram of the running variable. 

Without loss of generality, suppose the entries of $\boldsymbol{\tilde X}$ are ordered row-wise by the value of $\tilde X_i$ (with any ordering applied to ties). The key steps for each bootstrap replicate $b$ are given below. 
\begin{itemize}
    \item For each value in $x \in \mathcal{X}$, sample $\sum_i I(\tilde X_i = x)$ units with replacement. Order the units based on the observed value of $\tilde X_i$. 
    \begin{itemize}
        \item Because we are conditioning on the observed histogram of the running variable, the estimated un-manipulated density will be unchanged from the original sample. Hence, any matrices or vectors based on the running variable will be unchanged from the original problem.
        \item For units to the left of the cutoff $c$, we can compute coefficients $\hat \beta_{\ell}^{(b)}$ and $\hat \alpha_{\ell}^{(b)}$ based on the resampled data in replicate $b$. 
        \item For units to the right of the cutoff, the values of $\boldsymbol{ \Phi}, \boldsymbol{\tilde X}$, and $\boldsymbol{ M}$ corresponding to the bootstrap sample will be unchanged from those in the original sample. However, we will now have bootstrap samples $Z^{(b)}$ and $Y^{(b)}$ corresponding to replicate $b$. 
\end{itemize}
\item The bootstrap estimate  $\hat \tau_f^{(b)}(c)$ is defined as the solution to 
\begin{equation*}
\begin{aligned}
\text{maximize} &\hspace{5mm} \frac{\text{diag} \left( Y^{(b)} c^{\star T} (\boldsymbol{ \Phi})^{-1} \boldsymbol{ X}^T \boldsymbol{ M} \right)^T H - c^{\star T} \hat \beta_{\ell}^{(b)} }{\text{diag} \left( Z^{(b)} c^{\star T} (\boldsymbol{ \Phi})^{-1} \boldsymbol{ X}^T \boldsymbol{ M} \right)^T H - c^{\star T}\hat \alpha_{\ell}^{(b)}}\\
\text{subject to} & \hspace{5mm} 0 \leq H \leq 1\,,\\
& \hspace{5mm} \mathbb{Z} H = \nu \,. 
\end{aligned}
\end{equation*}
\end{itemize}

Suppose we repeat this process for $b = 1, \dots, B$, where we typically assume $B$ is a few hundred samples. Denote as $Q_{u}(v)$ the function which takes in a vector $v$ and returns the $u^{th}$ quantile of $v$. Then the upper bound of our interval at level $\alpha$ is given by 
\[ U = Q_{1-\alpha/2} \left( \{ \hat \tau_f^{(b)}(c) \}_{b = 1}^B \right)\,.\]
The lower bound, $L$, is computed analogously, as the $\alpha/2$ quantile of the lower bounds across our bootstrap replicates. 

A few remarks are in order. First, a more recent strain of literature considers bootstrapping methods for the regression discontinuity design without making parametric assumptions about the potential outcome surfaces or the treatment probability model \citep{calonico2014robust, chiang2019robust, he2020wild}. 
We make no such claims here, and under-coverage is plausible for the bootstrap intervals derived for $T(\cdot)$ for a fixed value of $H$ if the underlying parametric model is incorrect. However, several steps in the algorithm --- the convex relaxation, and the union over all possible intervals --- may widen our bounds more than is strictly necessary for coverage. Hence, in cases where the chosen basis expansion is reasonably close to the true model, we still expect our intervals to achieve close-to-nominal coverage. 

Second, the proposed procedure accounts for uncertainty in the estimation of the causal effect, but not for uncertainty in the estimation of the un-manipulated histogram. If there is relatively little data near the threshold $c$, there may be substantial variability in the histogram estimation step. An alternative procedure would involve bootstrapping from the entire dataset and re-estimating the un-manipulated histogram within each replicate. Such a procedure would yield wider bounds than the one we propose here. 

\section{Results}\label{sec:application}

We now use our method to estimate the effect of deferral on return behavior, beginning with female donors (Section \ref{sec:resultwomen}), and following with male donors (Section \ref{sec:resultmen}). The estimation is more straightforward for women because there is less manipulation in the running variable and women only donate whole blood, simplifying the threshold considerations.

\FloatBarrier
\subsection{Results for women \label{sec:resultwomen}}

\begin{figure}[ht]
	\centering
	\includegraphics[width=0.49\textwidth]{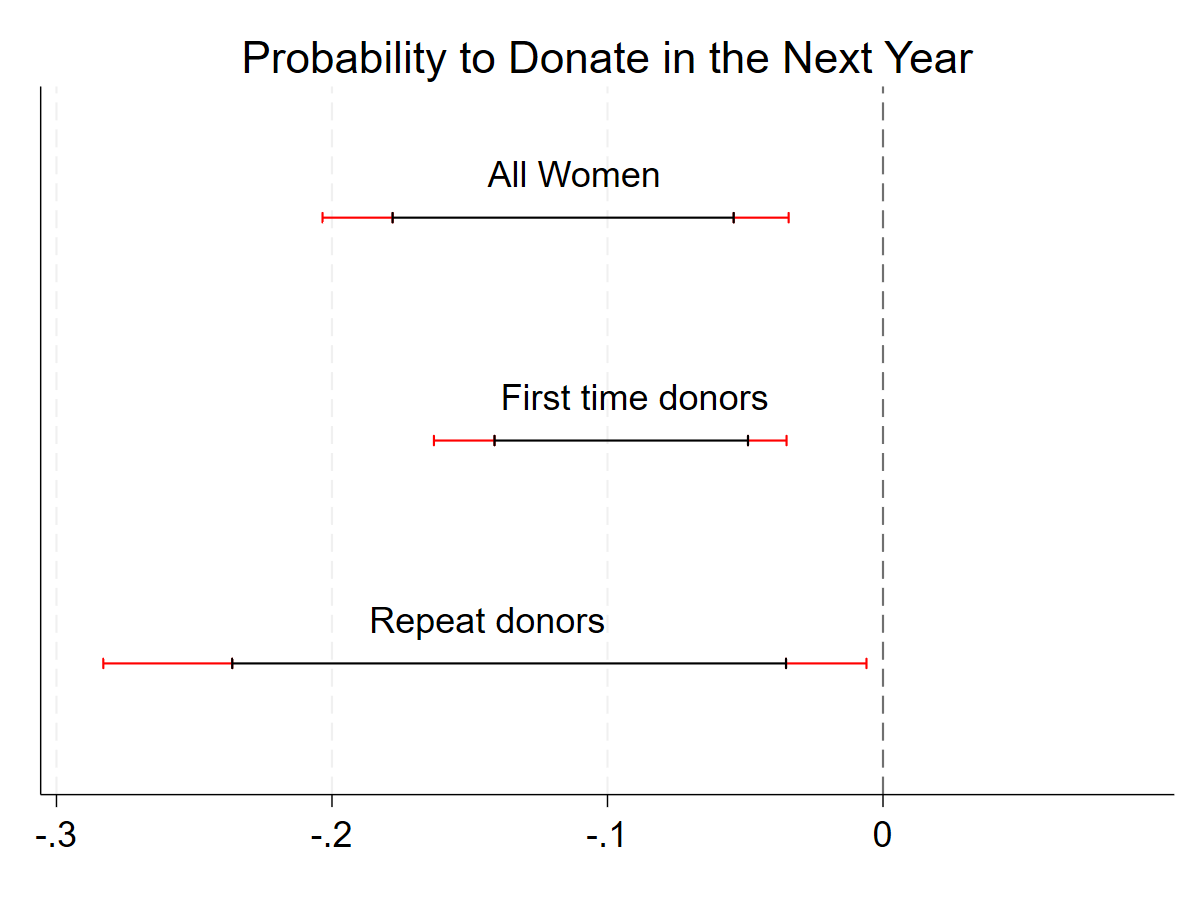}
        \includegraphics[width=0.49\textwidth]{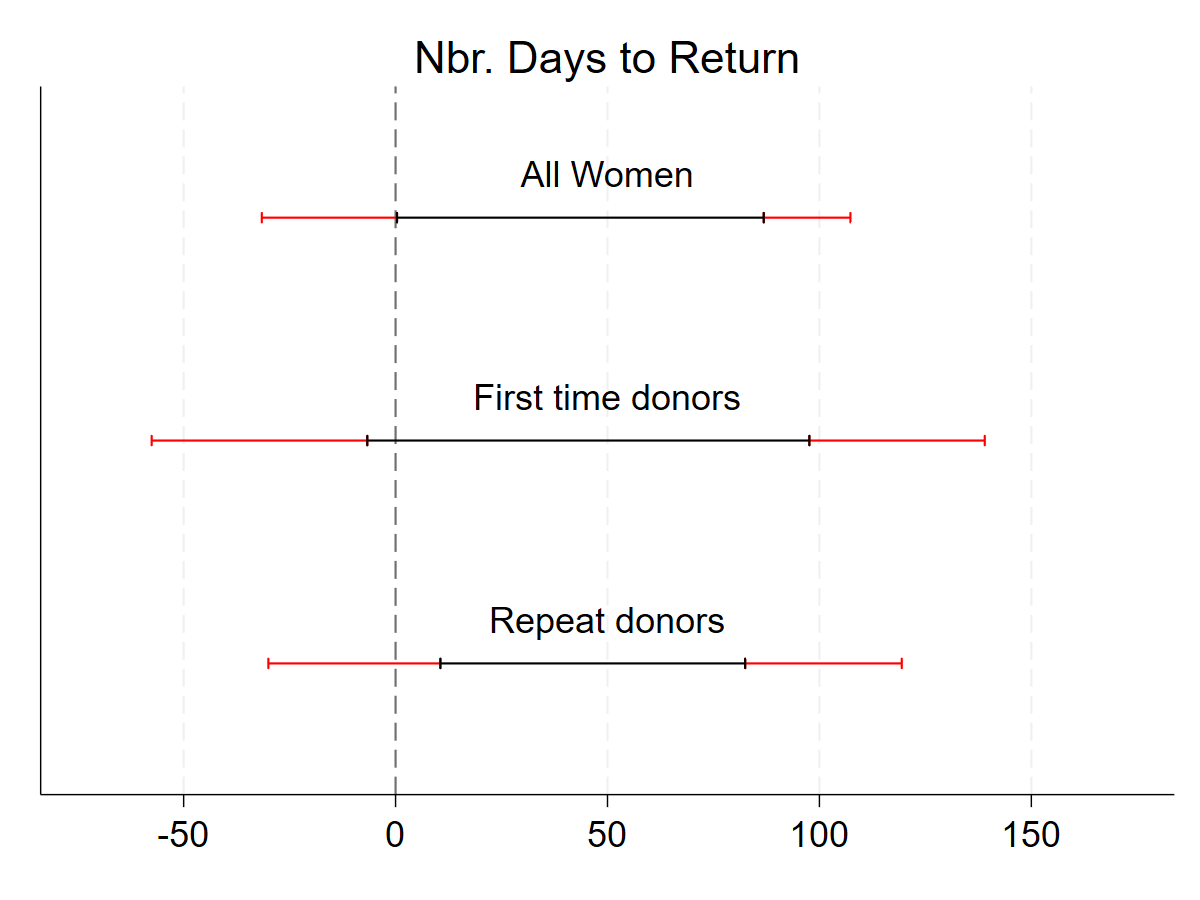}
\caption{Estimation results for female donors. Causal effect of a deferral on the probability of attempting to donate within the next year
(Left) and on the number of days donors take to return if they return (Right). In black the lower and upper bounds and in red the 95\% confidence intervals for those bounds. \label{fig:resultwomen}}
\end{figure}

Figure \ref{fig:resultwomen} reports the results for female donors for first and repeat donors. On the left is the probability to return in the next 12 months, and on the right, the number of days a donor takes to return if she returns. Overall, we find that a deferral reduces the probability of returning in the next year by 0.054 to 0.18 percentage points, and for donors who return, it increases the number of days they take to do so by 0 to 87 days.\footnote{When estimating the effect of a deferrals on the number of days donors take to return if they return, the sample size is smaller as we concentrate on donors who do return. Hence, the larger confidence intervals.} We explain below how we obtain those results.  

\subsubsection{Recovering the un-manipulated distribution}
\label{sec:unmanip-distr}

We do not know the precise range of hemoglobin levels for which manipulation occurred. Hence, we take an approach similar to that of \cite{diamond2016long} in seeking to automatically identify the manipulation region and obtain the un-manipulated histogram. We will discuss the process used to estimate the effect of the probability to come back in the next year for of all female donors. Other estimations are done similarly. 

We consider twelve possible manipulation windows. The narrowest window contains only hemoglobin levels 12.4 (just below the donation cutoff) and 12.5 (the cutoff exactly). The widest window extends from hemoglobin level 12.1 to 12.7. Our modeling strategy makes use of a constrained Poisson model to estimate the density. The model is fit to the histogram points outside the manipulation window, and the un-manipulated histogram is recovered by obtaining the model's predictions for points within the manipulation window. 

The Poisson models are fit by predicting the density from a spline expansion of the hemoglobin level. The models are constrained so that they meet the following intuitive requirements: 
\begin{itemize}
    \item The predicted counts must be greater than or equal to the observed counts for hemoglobin levels within the manipulation window but below the cutoff; and lesser than or equal to the observed counts for hemoglobin levels within the manipulation window but above the cutoff.
    \item The total number of predicted units within the manipulation window must equal the total number of observed units within the window. 
\end{itemize}

For predictors, we consider a cubic spline expansion of the hemoglobin levels, with anywhere from 3 to 15 degrees of freedom. For each possible manipulation window, we perform a five-fold cross-validation in order to select the optimal spline degree. This means hemoglobin levels outside the manipulation window are randomly split into five equally sized folds, and the constrained Poisson model of a given spline order is iteratively fit to all but one fold. Predictions are made on the held-out fold, and the total sum of squared errors is computed.

Visually, we observe that the raw data histograms exhibit substantial bunching at hemoglobin levels equal to whole numbers (e.g. 13.0). This makes intuitive sense, as donation center employees may be inclined to round reported hemoglobin levels. To better fit the observed data, we include a term in our model to account for the whole number bunching, yielding spikes in the fitted histogram. Denote as $\#k$ the observed count of individuals for whom $X_i = k$. The specification for our Poisson model is 

\[ \log(\# k) = \beta_0 \mathds{1}_{k \in \mathbb{N}} + \sum_{s = 1}^S \beta_s b(k),\]
where $b(k)$ is the spline basis expansion for the running variable and $S$ is the number of knots. As mentioned above, we allow the number of knots to vary over 3 to 15 and choose the model with the best cross-validated fit to the data. The values $\beta_0$ and $\beta_s$ are coefficients fit to the data.

We interpret our un-manipulated histogram as estimating what the distribution would have been, had there not been manipulation to explicitly authorize low-hemoglobin individuals to donate, but had the usual rounding effects remained. 

\begin{figure}[ht]
    \centering
    \includegraphics[width = 0.99\textwidth]{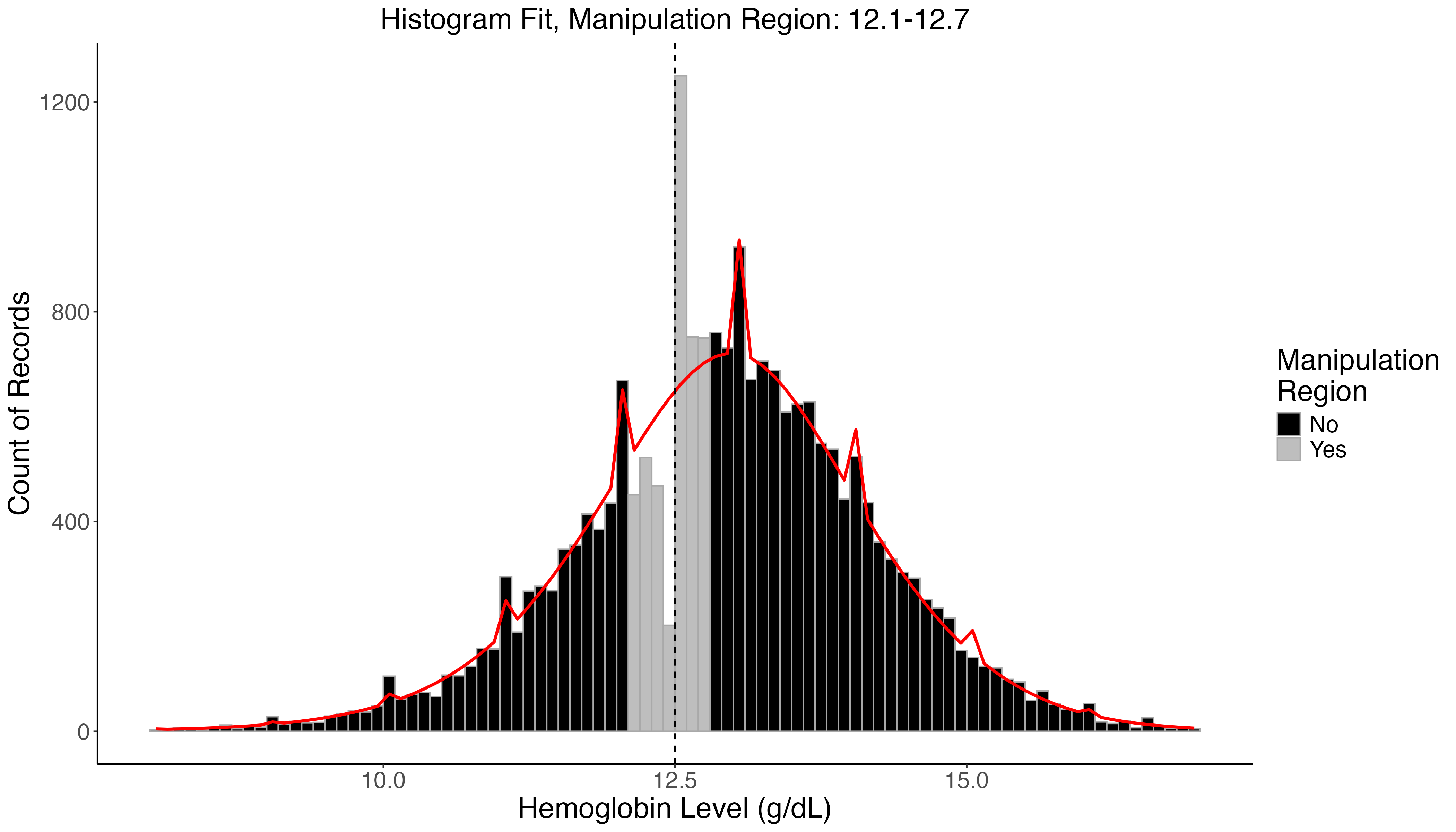}
    \caption{Histogram of hemoglobin levels for female donors. The bars in gray represent our manipulation window, while the bars in black represent levels outside the window. In red, we provide the estimated un-manipulated histogram.}
    \label{fig:unmaniphist}
\end{figure}

We select the (manipulation region-degrees of freedom) pair that has the minimum SSE. This turns out to be achieved with manipulation region $12.1$-$12.7$ and 10 degrees of freedom. The fitted histogram can be seen in Figure \ref{fig:unmaniphist}. The plot largely matches our visual intuition for what the un-manipulated histogram should look like.\footnote{Appendix \ref{sec:robustCheck} reports the results for different manipulation windows and knots. Results are insensitive to those choices.}

Crucially, these results imply that there are 588 manipulators for whom the observed hemoglobin level is 12.5 (662 honest donors vs. 1,250 observed donors), 67 manipulators for whom the observed hemoglobin level is 12.6 (685 honest donors vs. 752 observed donors), and 47 manipulators for whom the observed hemoglobin level is 12.7 (703 honest donors vs. 750 observed donors). We assume this is a full set of manipulators.

\subsubsection{Partial Identification Bounds}

We proceed with the estimation, using a local polynomial regression of order one to estimate both the numerator and denominator of $\hat \tau_f^{\star}(c)$ at the cutoff. Local weighting is implemented using tricubic weights. 

Converting this problem to the form of Optimization Problem \ref{fuzzyOptLinearizedProb}, we obtain the partial identification region of (-0.178, -0.054) for the causal effect of attempting a donation on the probability of attempting another donation within one year. We obtain slightly wider bounds of (-0.205, -0.051) for the probability of attempting another donation within the period of our data. 

\begin{figure}
  \begin{adjustwidth}{0cm}{}
    \centering
    \includegraphics[width = 0.49\textwidth]{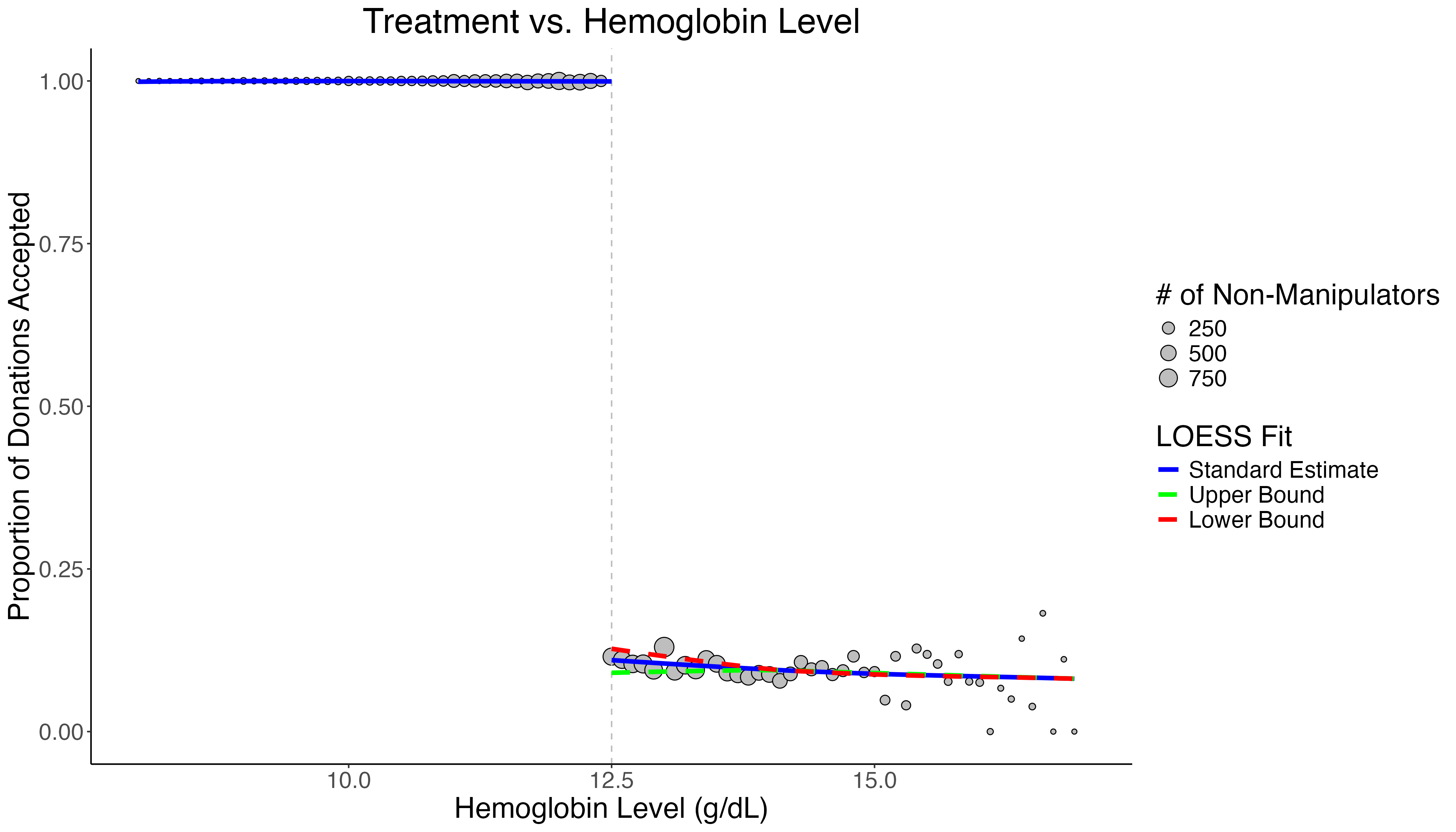}
    \includegraphics[width = 0.49\textwidth]{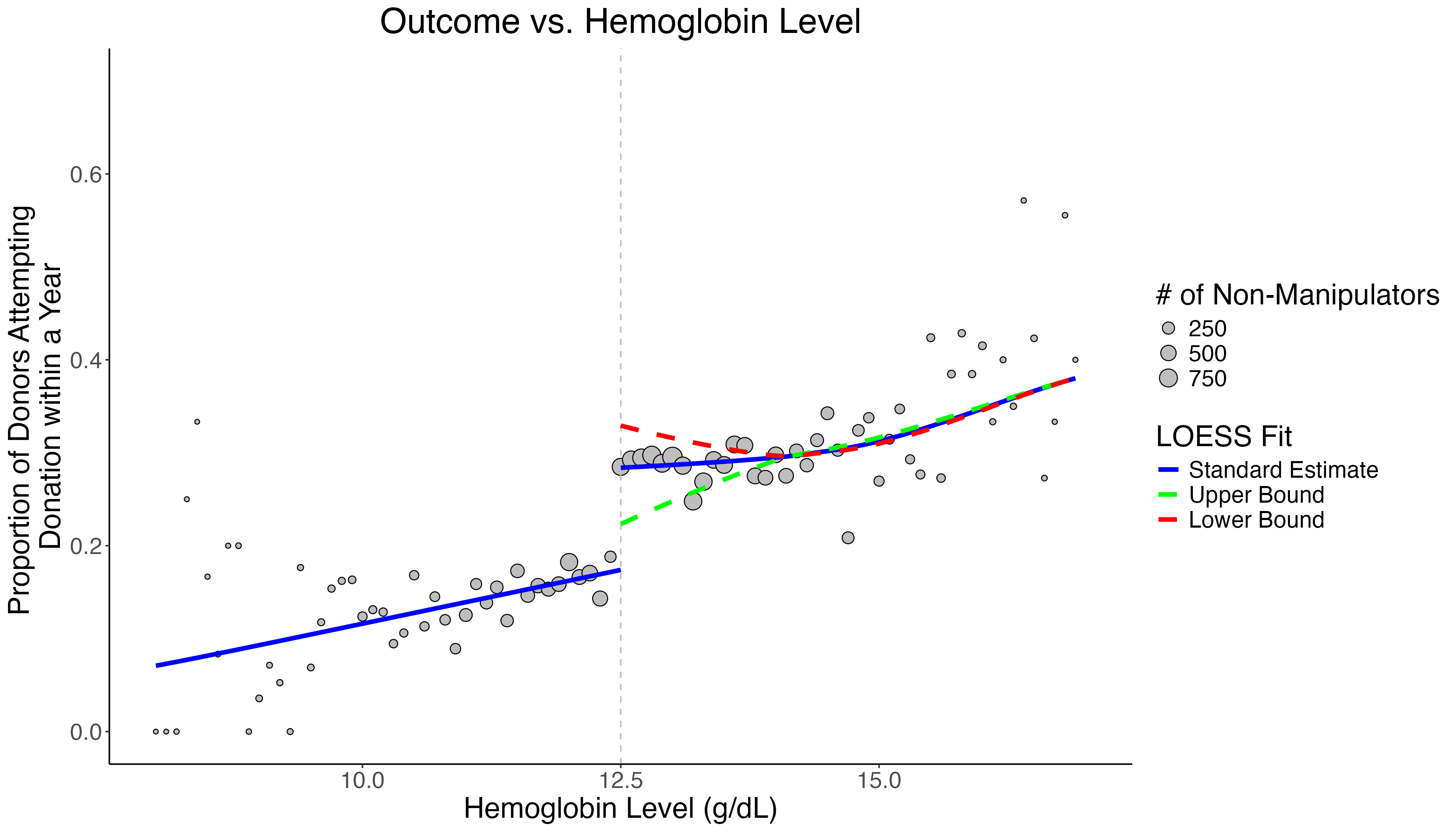}
    \caption{On the left, estimation of the denominator (probability to have a successful donation) and on the right the numerator (probability to come back in the next 12 months) of the causal effect estimator.}
    \label{fig:effectPlot}
 \end{adjustwidth}
\end{figure}

The effect of our method is visualized in Figure \ref{fig:effectPlot}, where we focus on the causal effect on attempting a donation within one year. The left panel displays the LOESS curve fits for the treatment as a function of the running variable (the denominator of our causal effect estimate); the right panel displays the LOESS curve fits for the outcome as a function of the running variable (the numerator of our causal effect estimate). 

The impact of the exclusions induced by our optimization problem can be seen most clearly in the right panel. The upper bound on the causal effect is obtained primarily by tagging as manipulators those women for whom the hemoglobin level is 12.5 and who did not attempt to donate again in one year. These women are then excluded from the estimation, resulting in the downward curvature of the green line as it approaches the cutoff from the right. Similarly, the lower bound is obtained by tagging as manipulators those women for whom the hemoglobin level is 12.5 and who \emph{did} attempt to donate again in one year. Their exclusion yields the upward curvature of the red line as it approaches the cutoff from the right. 

In the left panel, we see that the exclusions do not dramatically affect the LOESS fit for the treatment indicator as a function of the running variable. The LOESS curves diverge only slightly as they approach the cutoff from the right. 

We perform the analysis separately for first time and repeat donors and for the number of days donors take to return if they return. Figure \ref{fig:resultwomen} reports the results.\footnote{Appendix \ref{sec:figNdaytonextdonations} show the relevant figures when the outcome variable is the number of days donors take to return if they return.}

\subsubsection{Inference}
\label{sec:inference}

We can also use the bootstrap procedure from Section 4.6 to obtain a 95$\%$ confidence interval for each of the causal effects. The 95$\%$ CI for the effect of donation deferral on attempting another donation within one year is $(-0.203,-0.034)$. The corresponding interval for the probability of attempting another donation at all is $(-0.230,-0.028)$. Notably, the right ends of these intervals still do not include zero. For the number of days one takes to return, the 95\% confidence interval is $(-31.56,107.32)$ days.

In Appendix \ref{sec:robustCheck}, we confirm that the results are stable across different choices of manipulation region and number of knots. In Appendix Section \ref{sec:placeboCheck} we rerun our procedure with three choices of placebo variables: height, weight, and age. Because none of these variables could plausibly be affected by donation deferral, we expect to see the partial identification regions and confidence intervals contain zero --- and they do, across all choices of the manipulation region.

Taken together, these results underscore the robustness of the causal effect of deferral on donation probability: even under worst-case bounds, we still estimate an effect lower than zero. Moreover, we do not see the same pattern when considering placebo variables. 

\subsection{Results for men \label{sec:resultmen}}

We now estimate the effect of deferral on the probability to return in the next year for male donors.\footnote{For conciseness, we focus our estimation on the intensive margin for male donors.} Male donors face two cut-offs: (i) with an h-level above 13g/dL, they can donate plasma\footnote{In this Section, when referring to plasma donations, we refer to plasma/platelet and red-cell aspheris donations. We pool them together, as they all have the same eligibility threshold.} and (ii) with an h-level above 13.5g/dL, they can donate whole blood.

We estimate the effect of being deferred at the threshold of 13.5g/dL, that is, the effect of being allowed to donate whole blood vs. being deferred. Figure \ref{fig:histhlevel} shows that almost all successful donations are plasma donations for reported h-levels between 13 and 13.4g/dL. Plasma donors are, on average, more experienced than whole blood donors; they have more previous donations and are less likely to be first-time donors  (see Figure \ref{fig:discontinuitycovariates_M}). Therefore, on average, donors falling below the 13.5g/dL threshold are more experienced and more likely to return than donors falling above this threshold. Ignoring this in our estimation would be problematic and bias our estimates downward. 

\begin{figure}[ht]
	\centering
        \includegraphics[width=0.49\textwidth]{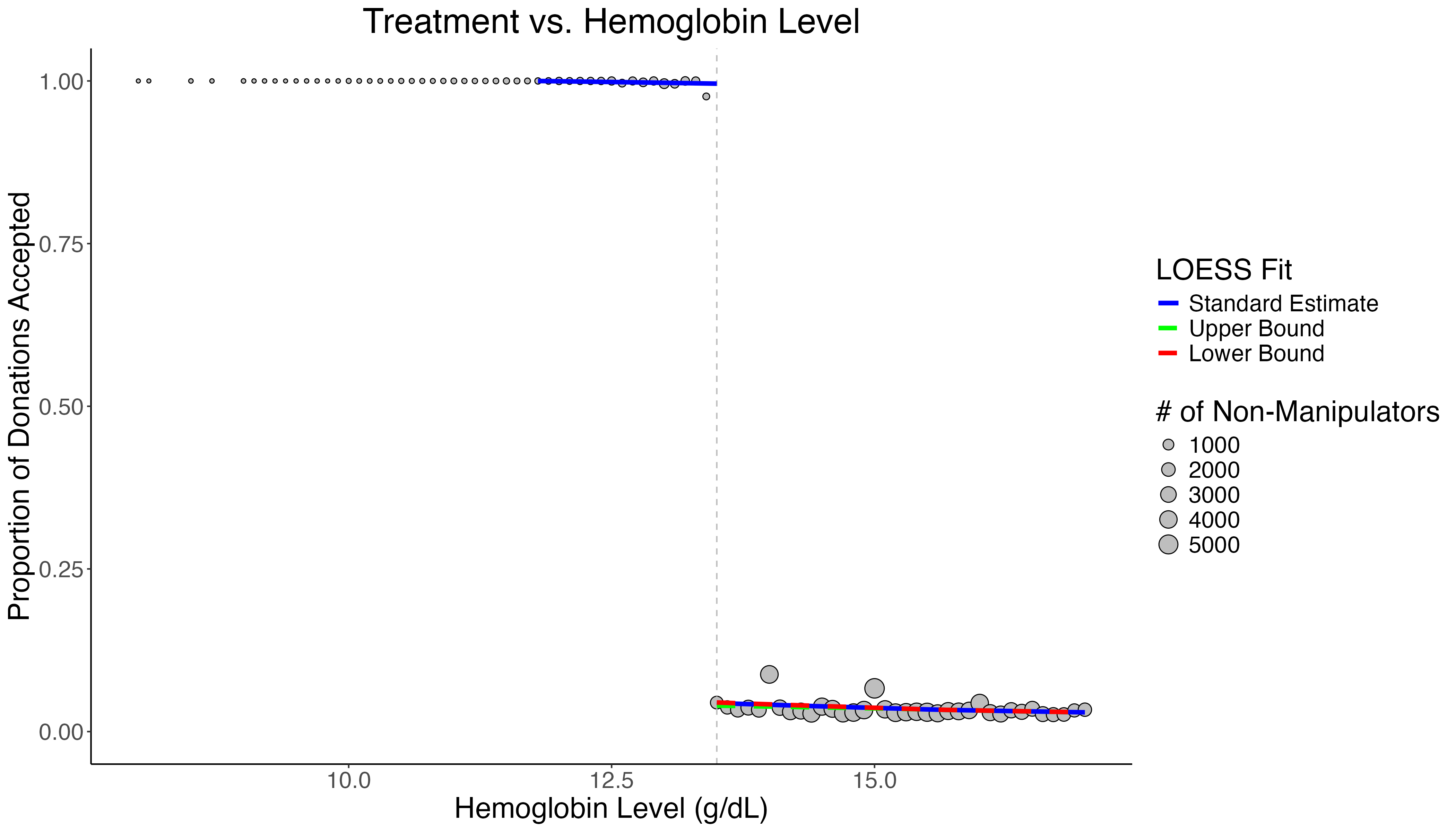}
        \includegraphics[width=0.49\textwidth]{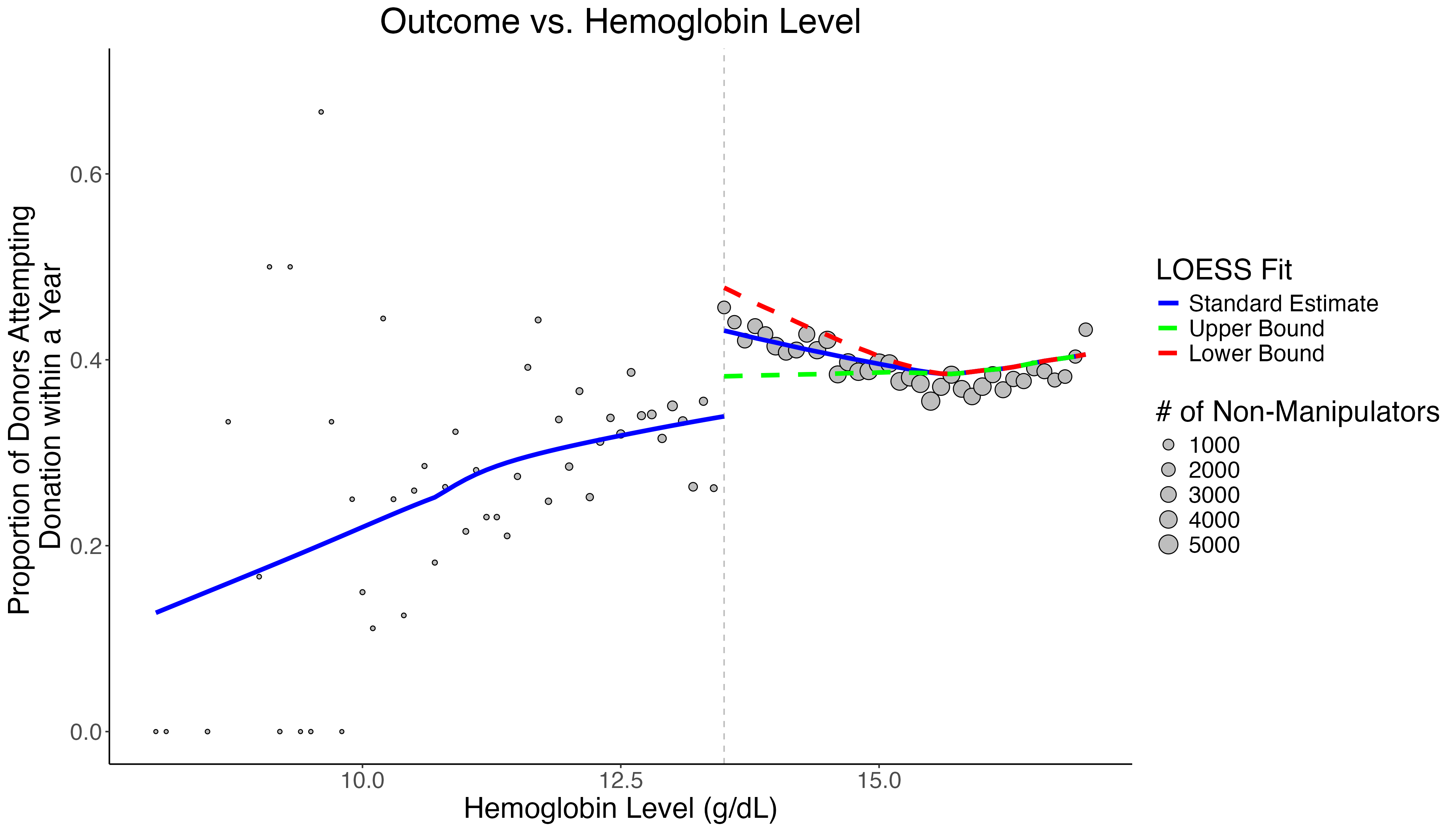}
\caption{Effect of being deferred for whole blood donors. Estimation made at the 13.5g/DL threshold excluding plasma donations. On the left, estimation of the denominator (probability of a successful donation), and on the right the numerator (probability to return in the next 12 months) of the causal effect estimator. The reweighted curves are shown in the green and red dashed lines. \label{fig:Men_WB}}
\end{figure}

To circumvent this issue, we disregard plasma donations from our sample and run the estimations as we did for female donors. Figure \ref{fig:Men_WB} reports the different steps of the estimation using a manipulation window of (12.9, 13.6). We estimate the effect to be between $-0.145$ and $-0.045$. Hence, the effect is similar in direction and size to the effect we found for women. All the placebo analyses yield intervals containing zero: $(-2.078, 1.316)$ for weight, $(-0.731, 0.816)$ for height, and $(-0.549, 1.543)$ for age.

\begin{figure}[ht]
    \centering
	\includegraphics[width=0.8\textwidth]{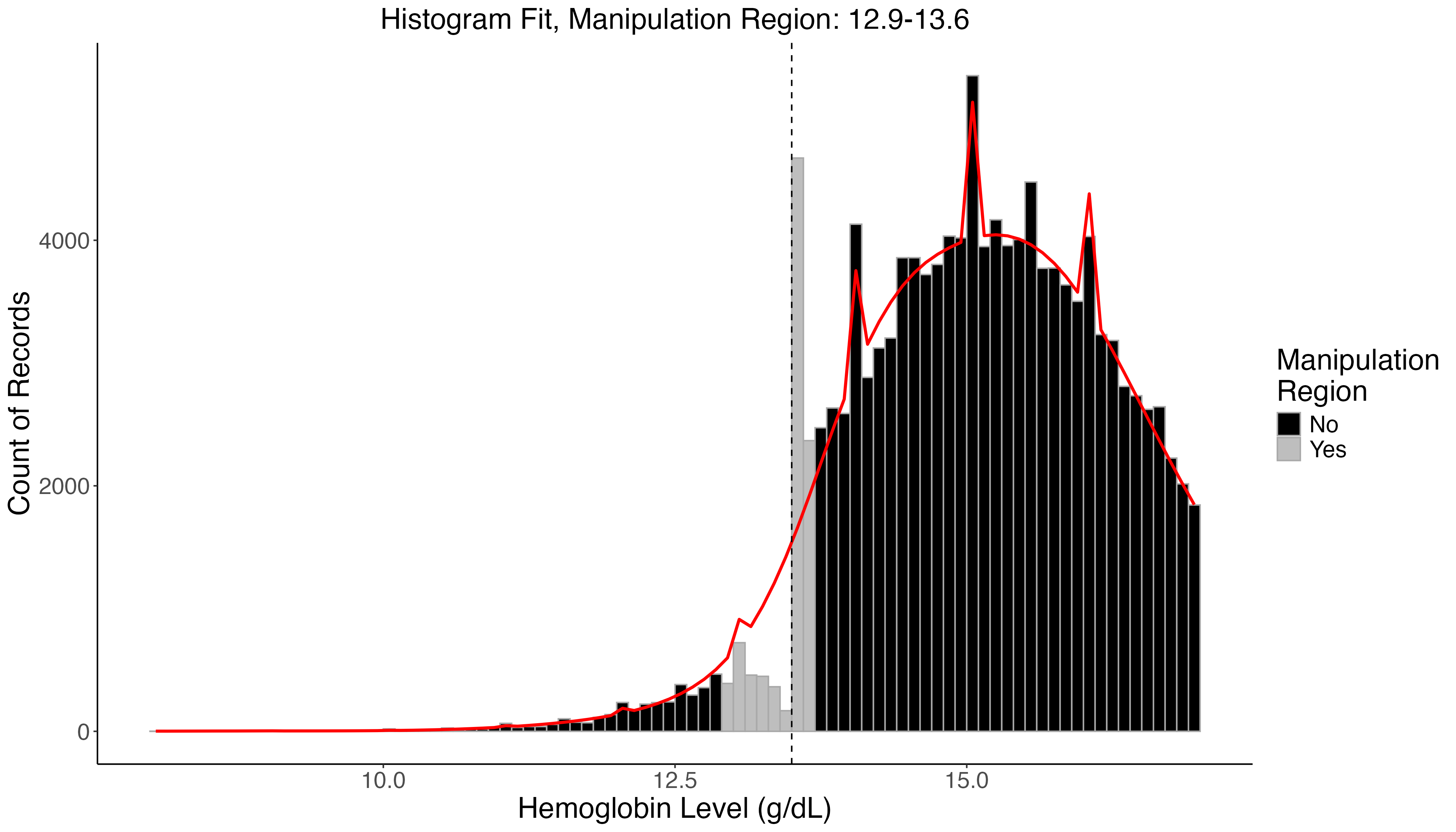}
\caption{Histogram of the distribution of h-level with the estimated un-manipulated histogram (in red). Exclude plasma donations. \label{fig:Men_WB_hist}}
\end{figure}

In our dataset, we know the outcome of successful donations, but do not know the intended donations for unsuccessful donations. Therefore, when analyzing whole blood donations, we can remove the successful plasma donations but cannot remove the attempted plasma donations that were unsuccessful. Those results should, therefore, be taken with caution. 

Appendix \ref{sec:resultsmen_plasma} investigates the effect of being deferred at the threshold of 13g/dL, that is, the effect of being allowed to donate plasma vs being deferred. To do so, we disregard the successful whole blood donations from our sample. 



\subsection{Alternative Approaches}

\begin{figure}[ht]
	\centering
	\includegraphics[width=0.6\textwidth]{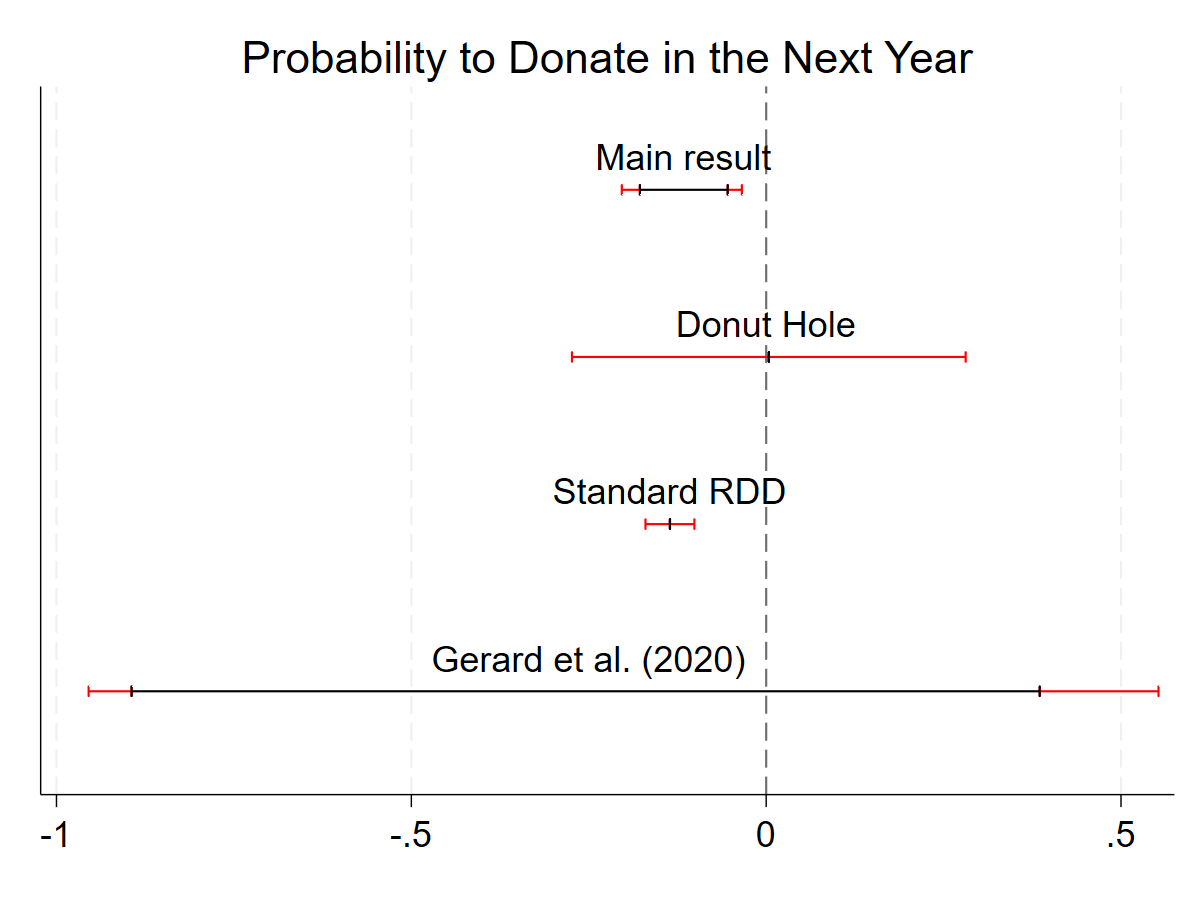}
\caption{Comparing alternative approaches. In black the lower and upper bounds and in red the 95\% confidence intervals for those bounds. \label{fig:comparingalternativemethods}}
\end{figure}

In this section we compare our estimations with alternatives approaches used in the literature in case of RDD with a manipulated running variable: the Donut Hole approach and Gerard et al's (2020) partial identification approach.\footnote{We also present the results from a standard RDD that does not account for potential manipulation of the running variable.} We focus on the estimation on the probability to return in the next year for women donors as it's a cleaner setting with only one threshold and less manipulation than for men. Figure \ref{fig:comparingalternativemethods} report the results from these estimation. Our method provides tighter bounds that the alternative approaches. We explain why below.

\subsubsection{Donut Hole Approach\label{subsubsec:donuthole}}

In RDDs with manipulation, a common approach used in the literature is the donut hole design.\footnote{\cite{noack2023donut} discusses the statistical properties of donut regression discontinuity designs.} Here, data within a window around the RDD threshold, known as the donut hole, is deleted and estimates are obtained at the threshold by extrapolation of the fitted functional form from outside the donut hole. We use our optimal manipulation region computed in Section \ref{sec:unmanip-distr}, i.e. $[12.1, 12.6]$ as the donut hole and delete all observations inside it. We then compute our RD point estimates and confidence intervals using the \texttt{rdrobust} R package \citep{calonico2014robust, calonico2015rdrobust}. 

Due to the nature of donut hole designs, which compute effects at the threshold where no data actually exists (by virtue of all data in its neighborhood being removed), the estimates can be very noisy due to a combination of extrapolation and sampling errors. Indeed, for our donut hole specification, for the causal effect of donation acceptance on the probability that the donor will attempt a return donation within a year, we obtain a 95\% confidence interval of $(-0.274,0.281)$. This interval is also significantly wider than the corresponding partial bounds interval we obtained in Section \ref{sec:inference}. We also perform a sensitivity analysis of the estimates with respect to the specification of the donut hole, which we reproduce in Appendix \ref{sec:donut-robust}.

These results show how our partial identification methodology for RDDs, even if not point identified, may provide \textit{tighter} bounds for the causal effects of interest, and thereby more power to detect both signs of treatment effects and magnitude ranges. Given that the RDD is a data-intensive procedure, since it primarily looks at observations in a very narrow window around a cutoff, we believe this improvement in power as well as design transparency would be important to researchers.

\subsubsection{Standard RDD \label{subsubsec:StandardRDD}}

In addition, we report the results of a standard RDD that ignores the manipulation of the running variable. This approach yields an estimated effect of -0.13 with a 95\% confidence interval of (-0.201, -0.061).\footnote{As for the donut hole, we estimate it with the \texttt{rdrobust} R package with the optimal bandwidth. The estimate is robust to the choice of bandwidth.} Notably, this point estimate lies within the partial identification bounds derived with our methodology. However, while ignoring the manipulation yields conclusions similar to those obtained from our partial identification bounds, this estimate cannot be considered a reliable causal effect due to the violation of a key assumption of the regression discontinuity design—that the running variable is not subject to manipulation. Consequently, this point estimate should be interpreted as a correlation, similar to the estimates discussed in Section 2.2, rather than as evidence of a causal relationship.

\subsubsection{Gerard et al. (2020) \label{sec:comparaisongerard2020}}

We are also interested in comparing the bounds obtained by our method versus those obtained from the method of \cite{gerard2020bounds} (henceforth GRR2020). Fortunately, their approach is provided via the R package \texttt{rdbounds} \citep{rdbounds}.

GRR2020's method does not take as input an explicit manipulation region. Rather, users provide a bandwidth for kernel density estimation of the density to the left and right of the cutoff. The paper does not provide theoretical results on the optimal bandwidth for estimation. The  Silverman ``rule of thumb" \citep{silverman1986density}, implemented as the default in the R function \texttt{density()}, suggests bandwidths of approximately 0.10 for density estimation to the left and right of the cutoff. However, values this small cause the \texttt{rdbounds} algorithm to estimate a proportion of manipulators larger than 100\% among those with a reported h-level of 12.5, yielding an error. Hence, we consider a grid of possible bandwidth values ranging from 0.4 (the smallest value the algorithm succeeds) to 1.0 for the bandwidth.\footnote{By default, if a single value is provided, it is used as the bandwidth for both density estimation and for estimation of the conditional expectation functions to the left and right of the cutoff.} Results are provided in Figure \ref{fig:gerard} below, and Table \ref{tab:gerard} in Appendix.

\begin{figure}[ht]
	\centering
	\includegraphics[width=0.6\textwidth]{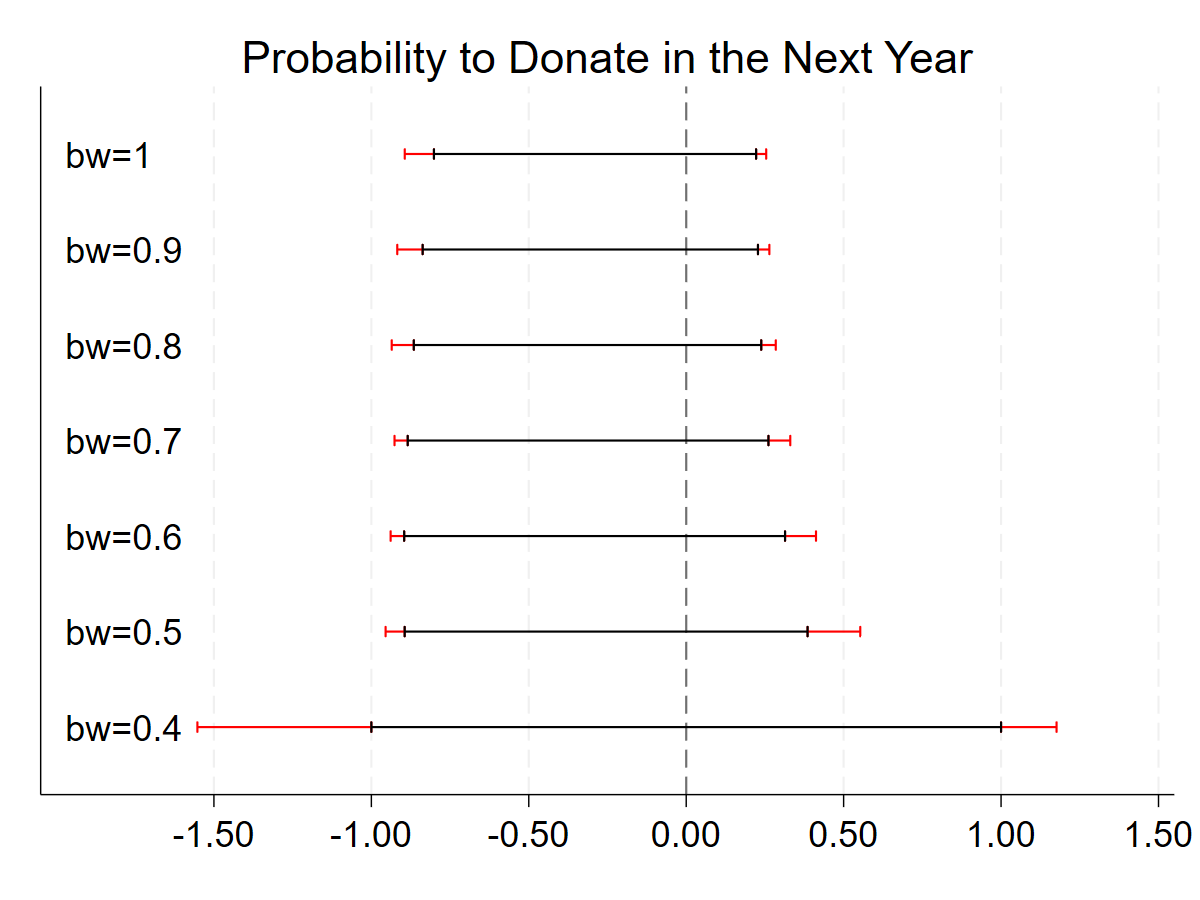}
\caption{Causal effect of donation acceptance on probability of attempting to donate within the next year, calculated under various bandwidths using GRR2020's method. In black the estimated bounds and in red the 95\% confidence intervals. The numbers on the left of each estimated bounds is the bandwidth used. \label{fig:gerard}}
\end{figure}

Figure \ref{fig:gerard} shows that GRR2020's bounds are significantly wider than ours. The partial identification bounds include zero at all values of the bandwidth. There are two central reasons for this discrepancy in the bounds.

First, GRR2020's algorithm estimates a significantly higher proportion of manipulators among those with a reported h-level of 12.5. Recall that we estimate that 588 out of 1,250 individuals with h-level 12.5 are manipulators, or 47\%. The proportion estimated by GRR2020's algorithm ranges from 91\% at the narrowest bandwidth to 61\% at the widest. Larger proportions of manipulators naturally lead to wider partial identification bounds.

The disparity is caused by the different assumptions in our underlying manipulation model and GRR2020's. We assume the un-manipulated density -- that is, the density of all individuals, including manipulators and honest participants -- is smooth (except spikes at whole numbers due to rounding). This means that, under our model, the aggregate density would not violate a McCrary test if no manipulation had occurred. By contrast, GRR2020 assumes that the density of the honest participants \emph{only} is smooth across the cut-off. It does not consider what the values of the running variable would be among manipulators had they not been able to manipulate their scores.

\begin{figure}[ht]
	\centering
	\includegraphics[width=0.7\textwidth]{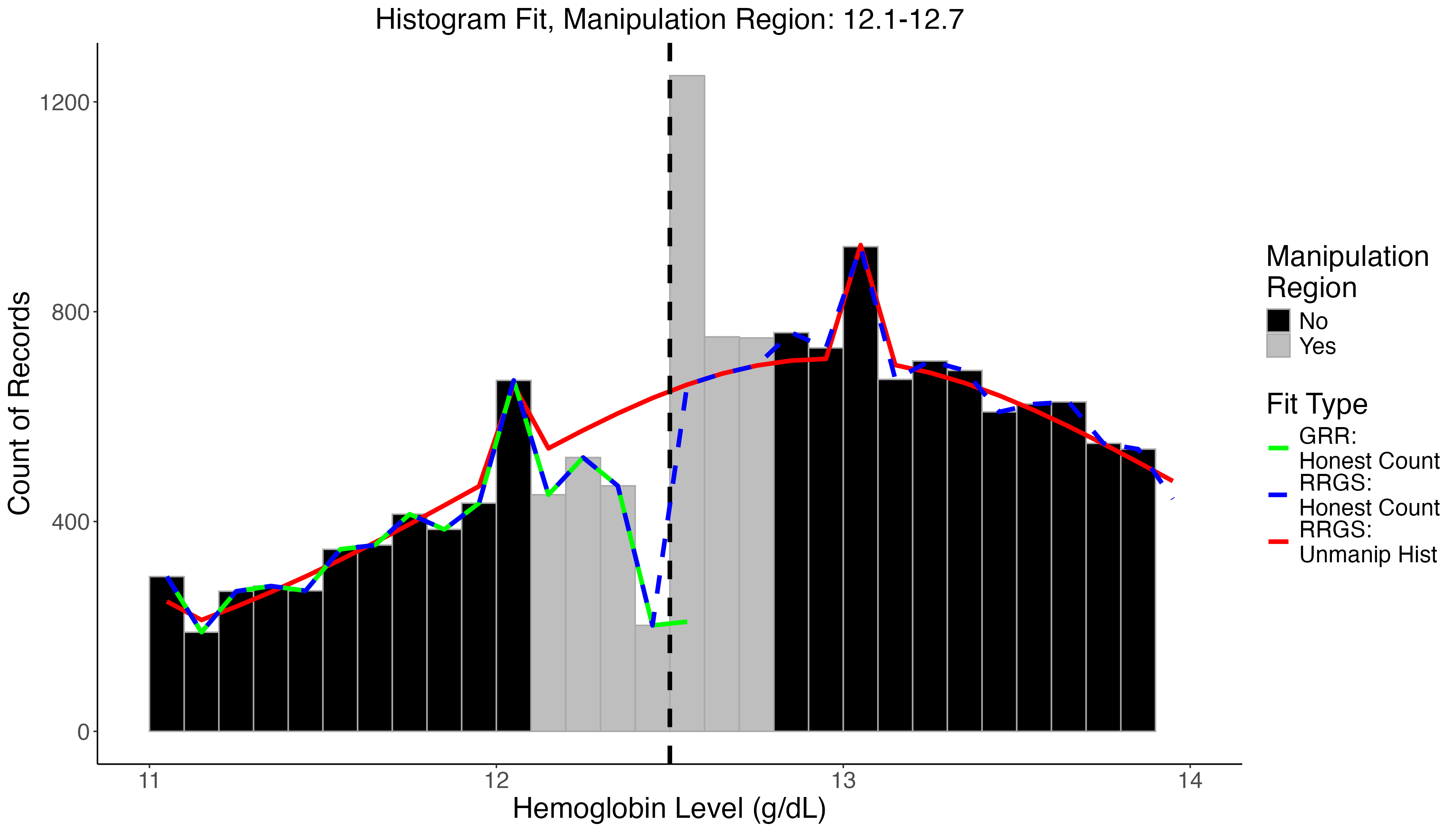}
\caption{Comparing our and GRR2020's approaches for identifying honest and manipulators. In red, the "unmanipulated histogram", and in blue, the "honest counts" using our method.  In green, the "honest counts" using GRR2020's method with an bandwidth of 0.5.\label{fig:GRR_Comparison}}
\end{figure}

Figure \ref{fig:GRR_Comparison} illustrates those different assumptions using our data for female donors. In red, the "unmanipulated histogram", and in blue, the "honest counts" using our method. For the "honest count", we take the raw histogram values everywhere except inside the manipulation region at and above the cut-off, where we default to the unmanipulated histogram counts. In green, the "honest counts" using GRR2020's method (note that it has no notion of a manipulation region nor an unmanipulated histogram). It also assumes all observations to the left of the cut-off are honest participants, hence why the blue and green lines coincide. However, it assumes smoothness of the manipulator counts over the cut-off, such that at h-level 12.5, the estimated honest count is very small (209). Moreover, it does not estimate the honest counts above the cut-off, as the honest counts are not identifiable at these points in their framework.

\FloatBarrier

These different assumptions matter in our empirical application because, as seen in Figure \ref{fig:GRR_Comparison}, the density of the running variable dips down just below the cut-off of 12.5. Under our assumptions, there is a significant increase in the count of honest subjects (blue line) from an h-level of 12.4 (202 subjects) to an h-level of 12.5 (653 subjects). This is not problematic because we assume that the density would be ``smoothed out" where the many manipulators with reported h-levels of 12.5 and 12.6 returned to their true hemoglobin levels just below the cut-off. However, the assumptions of GRR2020 (green line) imply that there cannot be a large uptick between the count of honest subjects just below the cut-off and the cut-off itself. Hence, it assumes many more manipulators than we do. 

A second subtle distinction is the order in which computations are done for the two approaches. For our algorithm, we first identify the ``best-case" and ``worst-case" manipulators, exclude them from our sample, and then estimate the conditional mean functions to obtain our upper and lower bounds. The code in \texttt{rdbounds} proceeds in the opposite order. The CDF of units with a reported h-level of 12.5 is computed only once and is then adjusted post-hoc to account for the manipulation.  

Other parameters could plausibly affect our bounds and those of GRR2020, such as the choice of kernel and the order of the local polynomial used to estimate the conditional mean functions. We have proceeded with the defaults used in \texttt{rdbounds}, and exploratory analysis indicates that other parameter choices make little difference in estimating the bounds. 

This exercise demonstrates that different assumptions and computational choices underlie our method and that of GRR2020, yielding different estimates of the frequency of manipulators and, hence, different widths of the partial identification bounds. In the blood-donation data, our log-concavity assumption yields larger estimates of the count of honest subjects near the cutoff than does the smoothness assumption of GRR2020, so we obtain narrower intervals. We believe the log-concavity assumption is well-justified by the observed data histogram. We also note that a very similar histogram shape was observed for blood donation in Switzerland by \cite{bruhin2020sting}. However, the viability of this assumption will rely on subject matter knowledge. 

As the preceding discussion illustrates, our method and that of GRR2020 both provide partial identification bounds in the presence of manipulation of the running variable, but rely on different modeling assumptions. Our approach assumes that the un-manipulated running variable is log-concave, a plausible assumption in many empirical settings. When this assumption holds and manipulation is substantial (e.g., clear bunching near the threshold), our method can yield tighter and more informative bounds by explicitly modeling the entire unmanipulated distribution. By contrast, GRR2020 only assumes that the density of honest participants is smooth across the cutoff and does not model the un-manipulated distribution of manipulators. This weaker, local assumption may be preferable when manipulation is limited or when the shape of the un-manipulated distribution is uncertain. In many settings, we believe it is prudent to report both sets of bounds, as we do here, to provide a robustness check and to illustrate the sensitivity of conclusions to different identifying assumptions.

\FloatBarrier

\section{Discussion}\label{sec:discusson}

The present paper makes two contributions. First, it contributes to the literature on charitable giving by providing causal evidence that deferring donors negatively impacts their return behavior. Second, it adds to the literature on regression discontinuity designs by providing a general method for deriving partial identification bounds of causal effects in RDDs with manipulation. To motivate our approach, we develop two conceptual frameworks that formalize how agents adjust the running variable around the threshold, highlighting when RD designs may be biased.

Future research could investigate the mechanisms explaining why deferred donors are less likely to return. 

A first possible mechanism is that rejections might provide information to donors that would affect their beliefs about the value of their volunteer effort or the likelihood of being rejected again. For instance, a rejection could signal that their offerings are not very valuable to the organization or that the benefit of their volunteer service is less than the volunteer had expected. If so, volunteers would rationally update their belief in the benefits to costs and consequently be less likely to donate. There is some empirical evidence to support this plausible causal effect. \cite{craig2017waiting} shows that the time until donors return to make a subsequent donation is adversely affected by how long they had to wait when attempting to make donations in the past. In their model, donors rationally update their beliefs on their expected wait time based on the past wait times they have experienced, and the higher these wait time beliefs are, the longer it will be until they donate again. They present empirical evidence consistent with this theoretical framework.

A second possible mechanism is that temporary deferral may be perceived as a form of social exclusion, which can diminish intrinsic motivation to engage in prosocial behavior \citep{hutchison2007social}. For example, \cite{twenge2007social} find that individuals who experience social rejection are significantly less likely to engage in prosocial behavior in subsequent tasks.

Finally, there is growing evidence that prosocial behavior exhibits habit formation \citep{meer2013habit}. In particular, \cite{bruhin2025understanding}, using a large-scale field experiment, find that habit formation is the primary driver of persistent blood donation. Relatedly, \cite{heger2022giving} show in laboratory experiments that choosing to give "now" increases the likelihood of future donations.

 Our empirical investigation was made difficult by an issue often faced by researchers attempting to use a regression discontinuity design: the running variable was manipulated. Traditional approaches to RDDs with manipulation either ignore manipulation or use a donut-hole method \cite[as in][]{almond2011after}. These approaches can be improved upon. Ignoring manipulation does not address the selection bias that emerges from manipulation, undermining the very causal nature that motivates an RD design. On the other hand, donut-hole approaches to RDDs have fragile statistical properties. While a traditional RDD estimator is a careful tradeoff between bias and variance in a local polynomial estimator, the donut-hole estimator potentially incorporates both large bias from the misspecification of functional form and large variance from the extrapolation of the functional form (even if correctly specified) outside the support of the data used to fit the model. 

 Instead, partial bounds approaches, such as GRR2020's and ours, allow to maintain a causal interpretation of the RDD and to reduce variance because they do not discard \textit{all} the data in the manipulation region, only those that might plausibly be associated with manipulation. Further, by providing a range of possible causal estimates in the face of manipulation, these approaches add to the transparency and credibility of the RDD by not claiming point identification with debatable assumptions. 

More generally, we derive partial bounds that account for the uncertainty in identifying compliers and sampling uncertainty, conditional on correctly identifying the amount of manipulation. We further show the robustness of our empirical results to various specifications and degrees of manipulation. It would be fruitful to develop bounds that incorporate uncertainty in estimating the manipulation mass directly and produce a theory of optimal (narrowest) bounds. We leave these exercises to future work.

Outside of our blood donation setting, we believe researchers could take this analytical approach when dealing with RD designs with evidence of manipulation. 

Running our optimization procedure, the analyst can confirm that the effect is robust even in the presence of manipulation. Such an approach is significantly more credible than a donut hole approach, which excludes potentially informative data. We suggest reporting the partial identification bounds and confidence intervals as part of the robustness check.

\bibliographystyle{apalike}
\bibliography{biblio,biblio2}
\newpage
\appendix

\bigskip
\begin{center}
{\large\bf SUPPLEMENTARY MATERIAL}
\end{center}

\section{Correlational Evidences}

\begin{table}[H]
\centering
\footnotesize
\begin{tabular}{l ccc ccc}
\toprule
& \multicolumn{6}{c}{All Donors} \\
& \multicolumn{3}{c}{P(Return next 12 months)} & \multicolumn{3}{c}{Nbr days to return}  \\
\cmidrule(lr){2-4} \cmidrule(lr){5-7} 
Constant & $0.256^{***} $  & $-0.391^{***} $  & $-0.304^{***} $  & $356.8^{***} $  & $611.3^{***} $  & $555.1^{***} $ \\
 & \footnotesize{($0.001$) }  & \footnotesize{($0.039$) }  & \footnotesize{($0.043$) }  & \footnotesize{($ 1.7$) }  & \footnotesize{($27.2$) }  & \footnotesize{($28.2$) } \\
Female & $-0.085^{***} $  & $-0.001 $  & $0.002 $  & $35.7^{***} $  & $ 1.3 $  & $ 2.1 $ \\
 & \footnotesize{($0.003$) }  & \footnotesize{($0.005$) }  & \footnotesize{($0.005$) }  & \footnotesize{($ 3.7$) }  & \footnotesize{($ 4.5$) }  & \footnotesize{($ 4.5$) } \\
Repeat Donation & $0.385^{***} $  & $0.362^{***} $  & $0.334^{***} $  & $-163.5^{***} $  & $-161.5^{***} $  & $-155.2^{***} $ \\
 & \footnotesize{($0.002$) }  & \footnotesize{($0.003$) }  & \footnotesize{($0.003$) }  & \footnotesize{($ 1.9$) }  & \footnotesize{($ 2.4$) }  & \footnotesize{($ 2.4$) } \\
 \textit{Successful Donations} \\
Plasma/Platelet & $0.332^{***} $  & $0.352^{***} $  & $0.385^{***} $  & $-175.2^{***} $  & $-177.5^{***} $  & $-192.2^{***} $ \\
 & \footnotesize{($0.003$) }  & \footnotesize{($0.004$) }  & \footnotesize{($0.004$) }  & \footnotesize{($ 1.4$) }  & \footnotesize{($ 1.9$) }  & \footnotesize{($ 2.1$) } \\
Apheresis & $0.058 $  & $0.042 $  & $0.037 $  & $-7.9 $  & $ 3.3 $  & $ 7.0 $ \\
 & \footnotesize{($0.069$) }  & \footnotesize{($0.068$) }  & \footnotesize{($0.069$) }  & \footnotesize{($44.7$) }  & \footnotesize{($43.8$) }  & \footnotesize{($45.1$) } \\
\textit{Unsuccessful Donations}  \\
Failed Phlebotomy & $-0.135^{***} $  & $-0.125^{***} $  & $-0.108^{***} $  & $63.0^{***} $  & $65.1^{***} $  & $56.3^{***} $ \\
 & \footnotesize{($0.010$) }  & \footnotesize{($0.010$) }  & \footnotesize{($0.012$) }  & \footnotesize{($14.5$) }  & \footnotesize{($14.5$) }  & \footnotesize{($14.7$) } \\
Change of mind & $0.013 $  & $0.048^{***} $  & $0.172^{***} $  & $-40.4^{***} $  & $-44.4^{***} $  & $-50.0^{***} $ \\
 & \footnotesize{($0.008$) }  & \footnotesize{($0.008$) }  & \footnotesize{($0.011$) }  & \footnotesize{($ 7.1$) }  & \footnotesize{($ 7.3$) }  & \footnotesize{($ 7.4$) } \\
\ \ Low h-level  & $-0.124^{***} $  & $-0.124^{***} $  & $-0.044^{***} $  & $44.0^{***} $  & $44.8^{***} $  & $40.2^{***} $ \\
 & \footnotesize{($0.004$) }  & \footnotesize{($0.004$) }  & \footnotesize{($0.008$) }  & \footnotesize{($ 5.2$) }  & \footnotesize{($ 5.8$) }  & \footnotesize{($ 6.1$) } \\
\ \ Other temporary  & $-0.095^{***} $  & $-0.102^{***} $  & $0.025 $  & $-5.8 $  & $ 5.3 $  & $-6.3 $ \\
 & \footnotesize{($0.007$) }  & \footnotesize{($0.007$) }  & \footnotesize{($0.014$) }  & \footnotesize{($ 9.1$) }  & \footnotesize{($10.4$) }  & \footnotesize{($10.6$) } \\
Weight (in kg) & $0.002^{***} $  & $0.002^{***} $  & $-0.5^{***} $  & $-0.5^{***} $ \\
 & \footnotesize{($0.000$) }  & \footnotesize{($0.000$) }  & \footnotesize{($ 0.1$) }  & \footnotesize{($ 0.1$) } \\
Height (in cm) & $0.002^{***} $  & $0.002^{***} $  & $-0.9^{***} $  & $-0.9^{***} $ \\
 & \footnotesize{($0.000$) }  & \footnotesize{($0.000$) }  & \footnotesize{($ 0.2$) }  & \footnotesize{($ 0.2$) } \\
Age (in years) & $0.003^{***} $  & $0.003^{***} $  & $-1.5^{***} $  & $-1.6^{***} $ \\
 & \footnotesize{($0.000$) }  & \footnotesize{($0.000$) }  & \footnotesize{($ 0.1$) }  & \footnotesize{($ 0.1$) } \\
A+ & $-0.093^{***} $  & $63.1^{***} $ \\
 & \footnotesize{($0.009$) }  & \footnotesize{($ 6.5$) } \\
AB- & $-0.127^{***} $  & $69.3^{***} $ \\
 & \footnotesize{($0.022$) }  & \footnotesize{($18.9$) } \\
AB+ & $-0.129^{***} $  & $75.2^{***} $ \\
 & \footnotesize{($0.010$) }  & \footnotesize{($ 7.3$) } \\
B- & $-0.027 $  & $27.7^{**} $ \\
 & \footnotesize{($0.014$) }  & \footnotesize{($ 9.4$) } \\
B+ & $-0.124^{***} $  & $70.4^{***} $ \\
 & \footnotesize{($0.010$) }  & \footnotesize{($ 6.6$) } \\
O- & $0.101^{***} $  & $-36.0^{***} $ \\
 & \footnotesize{($0.011$) }  & \footnotesize{($ 6.9$) } \\
O+ & $-0.056^{***} $  & $46.0^{***} $ \\
 & \footnotesize{($0.009$) }  & \footnotesize{($ 6.3$) } \\
\midrule
Mean  & $0.442 $  & $0.445 $  & $0.469 $  & $228.6 $  & $214.4 $  & $213.1 $ \\
Std. Dev.  & $0.497 $  & $0.497 $  & $0.499 $  & $291.8 $  & $293.2 $  & $292.3 $ \\
N  & $242,699 $  & $176,343 $  & $166,613 $  & $134,117 $  & $96,367 $  & $95,764 $ \\
\bottomrule
\end{tabular}
             \caption{Donor's return behavior depending on the attempted donations' outcome and the donors' characteristics for all donors. OLS regression. On the left, the dependent variable is a dummy variable equal to 1 if the donors return in the next 12 months. On the right, the dependent variable is the number of days the donor takes to donate again if he donated again. In parentheses, std errors clustered by donors. Only donations made a year before the end of the dataset. $^{***}$ significance at the 0.1\% level, $^{**}$ significance at the 1\%, $^{*}$ significance at the 5\% level.\label{tab:ols_return_full_all} }
\end{table}

\begin{table}[H]
\centering
\footnotesize
\begin{tabular}{l ccc ccc}
\toprule
& \multicolumn{6}{c}{Women} \\
& \multicolumn{3}{c}{P(Return next 12 months)} & \multicolumn{3}{c}{Nbr days to return}  \\
\cmidrule(lr){2-4} \cmidrule(lr){5-7} 
Constant & $0.199^{***} $  & $-0.147$  & $-0.058 $  & $385.3^{***} $  & $612.4^{***} $  & $523.8^{***} $ \\
 & \footnotesize{($0.004$) }  & \footnotesize{($0.079$) }  & \footnotesize{($0.109$) }  & \footnotesize{($ 5.1$) }  & \footnotesize{($93.3$) }  & \footnotesize{($99.0$) } \\
Repeat Donation & $0.290^{***} $  & $0.285^{***} $  & $0.230^{***} $  & $-143.5^{***} $  & $-131.6^{***} $  & $-122.4^{***} $ \\
 & \footnotesize{($0.008$) }  & \footnotesize{($0.009$) }  & \footnotesize{($0.010$) }  & \footnotesize{($ 6.7$) }  & \footnotesize{($ 7.5$) }  & \footnotesize{($ 7.7$) } \\
\textit{Unsuccessful Donations}  \\
Failed Phlebotomy & $-0.136^{***} $  & $-0.152^{***} $  & $-0.154^{***} $  & $85.9^{*} $  & $101.0^{**} $  & $92.1^{*} $ \\
 & \footnotesize{($0.019$) }  & \footnotesize{($0.020$) }  & \footnotesize{($0.022$) }  & \footnotesize{($38.2$) }  & \footnotesize{($38.1$) }  & \footnotesize{($40.3$) } \\
Change of mind & $-0.039^{**} $  & $-0.034^{*} $  & $0.175^{***} $  & $-73.1^{***} $  & $-61.2^{**} $  & $-59.8^{**} $ \\
 & \footnotesize{($0.014$) }  & \footnotesize{($0.016$) }  & \footnotesize{($0.030$) }  & \footnotesize{($18.5$) }  & \footnotesize{($19.0$) }  & \footnotesize{($20.5$) } \\
\ \ Low h-level  & $-0.124^{***} $  & $-0.143^{***} $  & $-0.021 $  & $38.0^{***} $  & $52.8^{***} $  & $43.2^{***} $ \\
 & \footnotesize{($0.006$) }  & \footnotesize{($0.006$) }  & \footnotesize{($0.013$) }  & \footnotesize{($ 9.1$) }  & \footnotesize{($ 9.7$) }  & \footnotesize{($10.9$) } \\
\ \ Other temporary  & $-0.128^{***} $  & $0.046 $  & $ 9.3 $  & $-15.2 $ \\
 & \footnotesize{($0.013$) }  & \footnotesize{($0.035$) }  & \footnotesize{($24.2$) }  & \footnotesize{($25.6$) } \\
Weight (in kg) & $0.001^{**} $  & $0.001^{**} $  & $ 0.1 $  & $ 0.1 $ \\
 & \footnotesize{($0.000$) }  & \footnotesize{($0.000$) }  & \footnotesize{($ 0.3$) }  & \footnotesize{($ 0.3$) } \\
Height (in cm) & $0.001^{**} $  & $0.001 $  & $-1.3^{*} $  & $-1.0 $ \\
 & \footnotesize{($0.001$) }  & \footnotesize{($0.001$) }  & \footnotesize{($ 0.6$) }  & \footnotesize{($ 0.6$) } \\
Age (in years) & $0.004^{***} $  & $-1.5^{***} $ \\
 & \footnotesize{($0.000$) }  & \footnotesize{($ 0.4$) } \\
A+ & $-0.097^{**} $  & $67.9^{**} $ \\
 & \footnotesize{($0.033$) }  & \footnotesize{($26.1$) } \\
AB- & $-0.123 $  & $ 5.2 $ \\
 & \footnotesize{($0.092$) }  & \footnotesize{($79.0$) } \\
AB+ & $-0.136^{***} $  & $76.7^{*} $ \\
 & \footnotesize{($0.036$) }  & \footnotesize{($30.9$) } \\
B- & $0.030 $  & $19.1 $ \\
 & \footnotesize{($0.050$) }  & \footnotesize{($35.9$) } \\
B+ & $-0.113^{***} $  & $57.9^{*} $ \\
 & \footnotesize{($0.033$) }  & \footnotesize{($26.7$) } \\
O- & $0.131^{***} $  & $-57.8^{*} $ \\
 & \footnotesize{($0.036$) }  & \footnotesize{($26.5$) } \\
O+ & $-0.050 $  & $40.2 $ \\
 & \footnotesize{($0.032$) }  & \footnotesize{($25.2$) } \\
\midrule
Mean  & $0.238 $  & $0.250 $  & $0.325 $  & $325.6 $  & $309.3 $  & $303.1 $ \\
Std. Dev.  & $0.426 $  & $0.433 $  & $0.468 $  & $315.3 $  & $312.9 $  & $310.2 $ \\
N  & $26,622 $  & $21,038 $  & $15,538 $  & $9,120 $  & $7,403 $  & $7,009 $ \\
\bottomrule
\end{tabular}
             \caption{Donor's return behavior depending on the attempted donations' outcome and the donors' characteristics for women. Note that women do not donate plasma and red-cell apheresis. OLS regression. On the left, the dependent variable is a dummy variable equal to 1 if the donors return in the next 12 months. On the right, the dependent variable is the number of days the donor takes to donate again if he donated again. In parentheses, std errors clustered by donors. Only donations made a year before the end of the dataset. $^{***}$ significance at the 0.1\% level, $^{**}$ significance at the 1\%, $^{*}$ significance at the 5\% level.\label{tab:ols_return_full_F} }
\end{table}

\begin{table}[H]
\centering
\footnotesize
\begin{tabular}{l ccc ccc}
\toprule
& \multicolumn{6}{c}{Men} \\
& \multicolumn{3}{c}{P(Return next 12 months)} & \multicolumn{3}{c}{Nbr days to return}  \\
\cmidrule(lr){2-4} \cmidrule(lr){5-7} 
Constant & $0.252^{***} $  & $-0.412^{***} $  & $-0.318^{***} $  & $357.7^{***} $  & $610.8^{***} $  & $556.3^{***} $ \\
 & \footnotesize{($0.001$) }  & \footnotesize{($0.043$) }  & \footnotesize{($0.046$) }  & \footnotesize{($ 1.7$) }  & \footnotesize{($28.4$) }  & \footnotesize{($29.5$) } \\
Repeat Donation & $0.395^{***} $  & $0.371^{***} $  & $0.345^{***} $  & $-165.2^{***} $  & $-164.6^{***} $  & $-158.5^{***} $ \\
 & \footnotesize{($0.003$) }  & \footnotesize{($0.003$) }  & \footnotesize{($0.003$) }  & \footnotesize{($ 2.0$) }  & \footnotesize{($ 2.5$) }  & \footnotesize{($ 2.5$) } \\
\\
\textit{Successful Donations} \\
Plasma/Platelet & $0.326^{***} $  & $0.348^{***} $  & $0.379^{***} $  & $-174.4^{***} $  & $-176.5^{***} $  & $-191.0^{***} $ \\
 & \footnotesize{($0.003$) }  & \footnotesize{($0.004$) }  & \footnotesize{($0.004$) }  & \footnotesize{($ 1.4$) }  & \footnotesize{($ 1.9$) }  & \footnotesize{($ 2.2$) } \\
Apheresis & $0.055 $  & $0.038 $  & $0.033 $  & $-7.3 $  & $ 4.3 $  & $ 8.0 $ \\
 & \footnotesize{($0.069$) }  & \footnotesize{($0.068$) }  & \footnotesize{($0.069$) }  & \footnotesize{($44.7$) }  & \footnotesize{($43.7$) }  & \footnotesize{($45.0$) } \\
\\
\textit{Unsuccessful Donations}  \\
Failed Phlebotomy & $-0.135^{***} $  & $-0.119^{***} $  & $-0.096^{***} $  & $58.7^{***} $  & $59.0^{***} $  & $50.7^{**} $ \\
 & \footnotesize{($0.012$) }  & \footnotesize{($0.012$) }  & \footnotesize{($0.014$) }  & \footnotesize{($15.7$) }  & \footnotesize{($15.6$) }  & \footnotesize{($15.8$) } \\
Change of mind & $0.026^{**} $  & $0.067^{***} $  & $0.172^{***} $  & $-35.4^{***} $  & $-41.4^{***} $  & $-48.4^{***} $ \\
 & \footnotesize{($0.009$) }  & \footnotesize{($0.010$) }  & \footnotesize{($0.012$) }  & \footnotesize{($ 7.7$) }  & \footnotesize{($ 7.8$) }  & \footnotesize{($ 7.9$) } \\
\ \ Low h-level  & $-0.129^{***} $  & $-0.115^{***} $  & $-0.040^{***} $  & $45.9^{***} $  & $40.0^{***} $  & $36.6^{***} $ \\
 & \footnotesize{($0.006$) }  & \footnotesize{($0.007$) }  & \footnotesize{($0.010$) }  & \footnotesize{($ 6.3$) }  & \footnotesize{($ 7.2$) }  & \footnotesize{($ 7.4$) } \\
\ \ Other temporary  & $-0.093^{***} $  & $-0.098^{***} $  & $0.023 $  & $-5.4 $  & $ 4.8 $  & $-4.7 $ \\
 & \footnotesize{($0.008$) }  & \footnotesize{($0.009$) }  & \footnotesize{($0.016$) }  & \footnotesize{($ 9.9$) }  & \footnotesize{($11.5$) }  & \footnotesize{($11.6$) } \\
Weight (in kg) & $0.002^{***} $  & $0.002^{***} $  & $-0.5^{***} $  & $-0.5^{***} $ \\
 & \footnotesize{($0.000$) }  & \footnotesize{($0.000$) }  & \footnotesize{($ 0.1$) }  & \footnotesize{($ 0.1$) } \\
Height (in cm) & $0.002^{***} $  & $0.002^{***} $  & $-0.9^{***} $  & $-0.9^{***} $ \\
 & \footnotesize{($0.000$) }  & \footnotesize{($0.000$) }  & \footnotesize{($ 0.2$) }  & \footnotesize{($ 0.2$) } \\
Age (in years) & $0.003^{***} $  & $0.003^{***} $  & $-1.6^{***} $  & $-1.6^{***} $ \\
 & \footnotesize{($0.000$) }  & \footnotesize{($0.000$) }  & \footnotesize{($ 0.1$) }  & \footnotesize{($ 0.1$) } \\
A+ & $-0.093^{***} $  & $62.8^{***} $ \\
 & \footnotesize{($0.010$) }  & \footnotesize{($ 6.7$) } \\
AB- & $-0.129^{***} $  & $71.5^{***} $ \\
 & \footnotesize{($0.023$) }  & \footnotesize{($19.3$) } \\
AB+ & $-0.129^{***} $  & $75.2^{***} $ \\
 & \footnotesize{($0.011$) }  & \footnotesize{($ 7.5$) } \\
B- & $-0.031^{*} $  & $28.2^{**} $ \\
 & \footnotesize{($0.014$) }  & \footnotesize{($ 9.7$) } \\
B+ & $-0.125^{***} $  & $71.2^{***} $ \\
 & \footnotesize{($0.010$) }  & \footnotesize{($ 6.8$) } \\
O- & $0.097^{***} $  & $-34.0^{***} $ \\
 & \footnotesize{($0.011$) }  & \footnotesize{($ 7.1$) } \\
O+ & $-0.057^{***} $  & $46.5^{***} $ \\
 & \footnotesize{($0.010$) }  & \footnotesize{($ 6.5$) } \\
\midrule
Mean  & $0.467 $  & $0.472 $  & $0.484 $  & $221.5 $  & $206.5 $  & $206.0 $ \\
Std. Dev.  & $0.499 $  & $0.499 $  & $0.500 $  & $288.7 $  & $290.2 $  & $289.6 $ \\
N  & $216,077 $  & $155,305 $  & $151,075 $  & $124,997 $  & $88,964 $  & $88,755 $ \\
\bottomrule
\end{tabular}
             \caption{Donor's return behavior depending on the attempted donations' outcome and the donors' characteristics for men. OLS regression. On the left, the dependent variable is a dummy variable equal to 1 if the donors return in the next 12 months. On the right, the dependent variable is the number of days the donor takes to donate again if he donated again. In parentheses, std errors clustered by donors. Only donations made a year before the end of the dataset. $^{***}$ significance at the 0.1\% level, $^{**}$ significance at the 1\%, $^{*}$ significance at the 5\% level.\label{tab:ols_return_full_M} }
\end{table}

\FloatBarrier
\section{Proof \label{sec:proofmeasurementerror}}

To compute \( P(h > c \mid h_2 = k) \), where \( h \sim N(x, \sigma^2) \) and \( \epsilon \sim N(0, \theta^2) \), with \( h_2 = h + \epsilon \), we proceed as follows:

\subsection*{Step 1: Conditional Distribution of \( h \mid h_2 = k \)}

From the properties of jointly normal random variables, the conditional distribution of \( h \) given \( h_2 = k \) is:

\[
h \mid (h_2 = k) \sim N\left(\mu_{h|h_2}, \sigma^2_{h|h_2}\right)
\]

\textbf{Mean (\( \mu_{h|h_2} \)):}
\[
\mu_{h|h_2} = E[h] + \text{Cov}(h, h_2) \cdot \text{Var}(h_2)^{-1} \cdot (k - E[h_2])
\]

\begin{itemize}
 \item \( E[h] = x \)
 \item \( E[h_2] = E[h] + E[\epsilon] = x + 0 = x \)
 \item \( \text{Cov}(h, h_2) = \text{Cov}(h, h + \epsilon) = \text{Var}(h) = \sigma^2 \)
 \item \( \text{Var}(h_2) = \text{Var}(h) + \text{Var}(\epsilon) = \sigma^2 + \theta^2 \)
\end{itemize}

Substitute:
\[
\mu_{h|h_2} = x + \frac{\sigma^2}{\sigma^2 + \theta^2} (k - x)
\]

\textbf{Variance (\( \sigma^2_{h|h_2} \)):}
\[
\sigma^2_{h|h_2} = \text{Var}(h) - \frac{\text{Cov}(h, h_2)^2}{\text{Var}(h_2)}
\]

\[
\sigma^2_{h|h_2} = \sigma^2 - \frac{\sigma^4}{\sigma^2 + \theta^2} = \frac{\sigma^2 \theta^2}{\sigma^2 + \theta^2}
\]

Thus:
\[
h \mid (h_2 = k) \sim N\left( \mu_{h|h_2}, \sigma^2_{h|h_2} \right)
\]

Where:
\[
\mu_{h|h_2} = x + \frac{\sigma^2}{\sigma^2 + \theta^2} (k - x), \quad \sigma^2_{h|h_2} = \frac{\sigma^2 \theta^2}{\sigma^2 + \theta^2}
\]

\subsection*{Step 2: Compute \( P(h > c \mid h_2 = k) \)}

Since \( h \mid (h_2 = k) \sim N(\mu_{h|h_2}, \sigma^2_{h|h_2}) \), we compute the probability:

\[
P(h > c \mid h_2 = k) = P\left(Z > \frac{c - \mu_{h|h_2}}{\sigma_{h|h_2}}\right)
\]

where \( Z \sim N(0, 1) \), \( \mu_{h|h_2} \) and \( \sigma_{h|h_2} \) are as derived above.\\

Substituting:
\[
P(h > c \mid h_2 = k) = 1 - \Phi\left(\frac{c - \left[x + \frac{\sigma^2}{\sigma^2 + \theta^2} (k - x)\right]}{\sqrt{\frac{\sigma^2 \theta^2}{\sigma^2 + \theta^2}}}\right)
\]

\textbf{Final Result:}
\[
P(h > c \mid h_2 = k) = 1 - \Phi\left(\frac{c - \left[x + \frac{\sigma^2}{\sigma^2 + \theta^2} (k - x)\right]}{\sqrt{\frac{\sigma^2 \theta^2}{\sigma^2 + \theta^2}}}\right)
\]

Which is increasing with $k$ and $x$.

Where:
\begin{itemize}
	\item \( \Phi(\cdot) \) is the standard normal cumulative distribution function.
	\item \( x \) is the mean of \( h \).
	\item \( \sigma^2 \) and \( \theta^2 \) are the variances of \( h \) and \( \epsilon \), respectively.
\end{itemize}

\section{Conceptual Framework Extension to two thresholds\label{sec:framework2threshold}}

In this section, we extend the conceptual framework developed in Section \ref{sec:framework1threshold} to the case with 2 thresholds, one for whole blood and one for plasma, as it is the case for male donors.

\subsection{Conceptual Framework 1: Maximizing donation and minimizing adverse events\label{sec:framework1_2threshold}}

Let $V_w$ denote the utility the blood bank derives from a successful whole blood donation, $V_p$ the utility from a successful plasma donation, $V_r$ the utility from a rejection and $V_a$ the utility from an adverse event. We assume $V_w, V_p>V_r>V_a$ and, without loss of generality, set $V_a=0$.

Whether $V_w>V_p$ depends on the blood type. We assume $V_w>V_p$ for O-Negative and O-Positive blood types. O-Negative is the red cell universal donor, and O-Positive is the most common blood type and is therefore in high demand. We assume $V_w<V_p$ for the AB blood type (negative or positive), which is the plasma universal donor. For types A and B, we assume $V_w \approx V_p$.\footnote{See here for an explanation of which blood types are in demand: \url{https://www.redcrossblood.org/donate-blood/blood-types.html}} 

Let's also denote $p(s|h,w)$ and $p(s|h,p)$, the probability that an accepted Whole Blood and Plasma donation is successful given an observed h-level $h$. Let's denote $c_p$ for plasma donation and $c_w$ the threshold for whole blood donations, $c_p<c_w$ ($c_p=13$ and $c_w=13.5$). 

We assume $p(s|h,w)>p(s|h,p)$. That is, a whole blood donation is safer than a plasma donation. We also assume $p(s|h,w)$, $p(s|h,p)$ and $\frac{p(s|h,p)}{p(s|h,w)}$ increases with experiences. As donors get more experienced, both whole blood and plasma donations become safer, but as experience increases the increased safety increases faster for plasma than whole blood donations.

$h$ is the observed h-level and $h^*$ the reported one. Given an observed h-level $h$, the expected utility the nurse gets is:

\begin{equation}
EU(h) =
\begin{cases} 
V_w*p(s|h,w) & \text{for an attempted whole blood donation} \\ 
V_p*p(s|h,p) & \text{if an plasma donation} \\
V_r & \text{for a deferral} \\ 
\end{cases}
\end{equation}

\subsubsection{Case 1: $c_p<c_w<h$ \label{sec:case1}}

If $c_p<c_w<h$, nurses report truthfully $h^*=h$ and collect whole blood if $V_w*p(s|h,w)>V_p*p(s|h,p)$ and plasma otherwise.

\subsubsection{Case 2: $c_p<h<c_w$ \label{sec:case2}}

The nurse can report truthfully $h^*=h$ and collect plasma donation or she can report $h^*=c_w$ and collect whole blood.

She reports $h^*=c_w$ if she prefers a whole blood donation to a deferral:
\begin{align}
V_w*p(s|h,w)&>V_r \\
 \Rightarrow p(s|h,w)&>\frac{V_r}{V_w}
\end{align}
And if she prefers a whole blood to a plasma donation,
\begin{align}
V_w*p(s|h,w)&>V_p*p(s|h,p) \\
 \Rightarrow \frac{V_w}{V_p}&>\frac{p(s|h,p)}{p(s|h,w)}
\end{align}

Therefore, the nurse is more likely to manipulate the reported h-level when:

\begin{itemize}
 \item $V_r$ decreases
 \item $V_w$ increases. 
 \item $V_p$ decreases.
 \item $p(s|h,w)$ increases.
 \item $\frac{p(s|h,p)}{p(s|h,w)}$ decreases. 
\end{itemize}

Which means donor O negative and O positive blood types are more likely to be be manipulated (high $V_w$) and donors with AB blood types are less likely to be manipulated (high $V_p$). 

When it comes to donors' experience, the predictions are unclear. On the one hand, they should be more likely to manipulate for experienced donors because $p(s|h,w)$ increases with experience. On the other hand, they should be less likely to manipulate experienced donors because as one gets more experienced $\frac{p(s|h,p)}{p(s|h,w)}$ increases (in other words, they prefer to have the experienced donors giving plasma than whole blood).

\subsubsection{Case 3: $h<c_p<c_w$}

If $h<c_p<c_w$, the nurse has 3 possibilities. She can report the observed h-level truthfully, bump it at the threshold $c_p$ allowing a plasma donation, or bump it at the threshold $c_w$ allowing for a whole blood donation. 

\subsubsection*{Manipulation to $c_p$}

The nurse reports $h^*=c_p$ if:

She prefers a plasma donation to a deferral:
\begin{align}
V_p*p(s|h,p)&>V_r \\
\Rightarrow p(s|h,p)&>\frac{V_r}{V_p}
\end{align}

And she prefers a plasma donation to a whole blood donation:
\begin{align}
V_p*p(s|h,p)&>V_w*p(s|h,w) \\
\Rightarrow \frac{p(s|h,p)}{p(s|h,w)}&>\frac{V_w}{V_p}
\end{align}

Therefore, the nurse is more likely to manipulate when:

\begin{itemize}
 \item $V_r$ decreases
 \item $V_w$ decreases. 
 \item $V_p$ increases.
 \item $p(s|h,p)$ increases.
 \item $\frac{p(s|h,p)}{p(s|h,w)}$ increases. 
\end{itemize}

Which means donors with AB blood types, the plasma universal donors, are more likely to be be manipulated (high $V_p$) and donors with O are less likely to be manipulated to $c_p$ (high $V_w$). 

They should also manipulate more experienced donors because both $p(s|h,p)$ and $\frac{p(s|h,p)}{p(s|h,w)}$ increases with experiences.

\subsubsection*{Manipulation to $c_w$}

The nurse reports $h^*=c_w$ if

She prefers a whole blood donation to a deferral:
\begin{align}
V_w*p(s|h,w) &>V_r \\
\Rightarrow p(s|h,w)&>\frac{V_r}{V_w}
\end{align}

She prefers a whole blood donation to a plasma donation:
\begin{align}
V_w \cdot p(s|h,w) &> V_p \cdot p(s|h,p) \\
\Rightarrow \frac{V_w}{V_p} &> \frac{p(s|h,p)}{p(s|h,w)}
\end{align}

This is the same as in Case 2 above (Section \ref{sec:case2})

\subsubsection{Conclusion}

Therefore, O-Negative and O-Positive donors will be more likely to be bumped at the $c_w$ threshold, allowing whole blood donations and AB donors will be more likely to be bumped at the $c_p$ threshold. This is what we observe in Section \ref{sec:discontinuitycovariates}.

More experienced donors are more likely to be bumped at the $c_p$ threshold. However, it's unclear whether they will be less or more likely to be bumped at the $c_w$ threshold.

\subsection{Conceptual Framework 2: Manipulation due to measurement error \label{sec:theory2cutoff}}

\begin{figure}[ht]
    \centering	
    \includegraphics[width=0.49\textwidth]{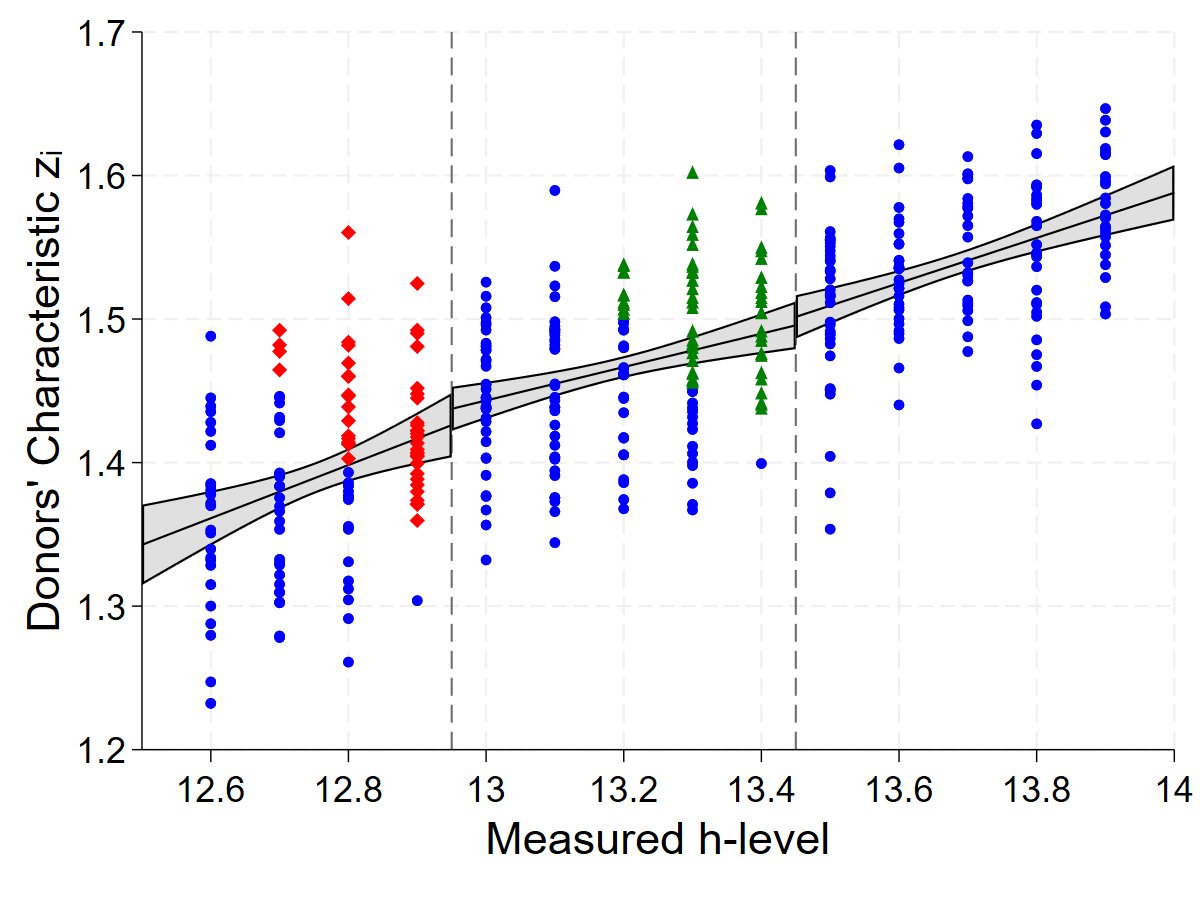}
    \includegraphics[width=0.49\textwidth]{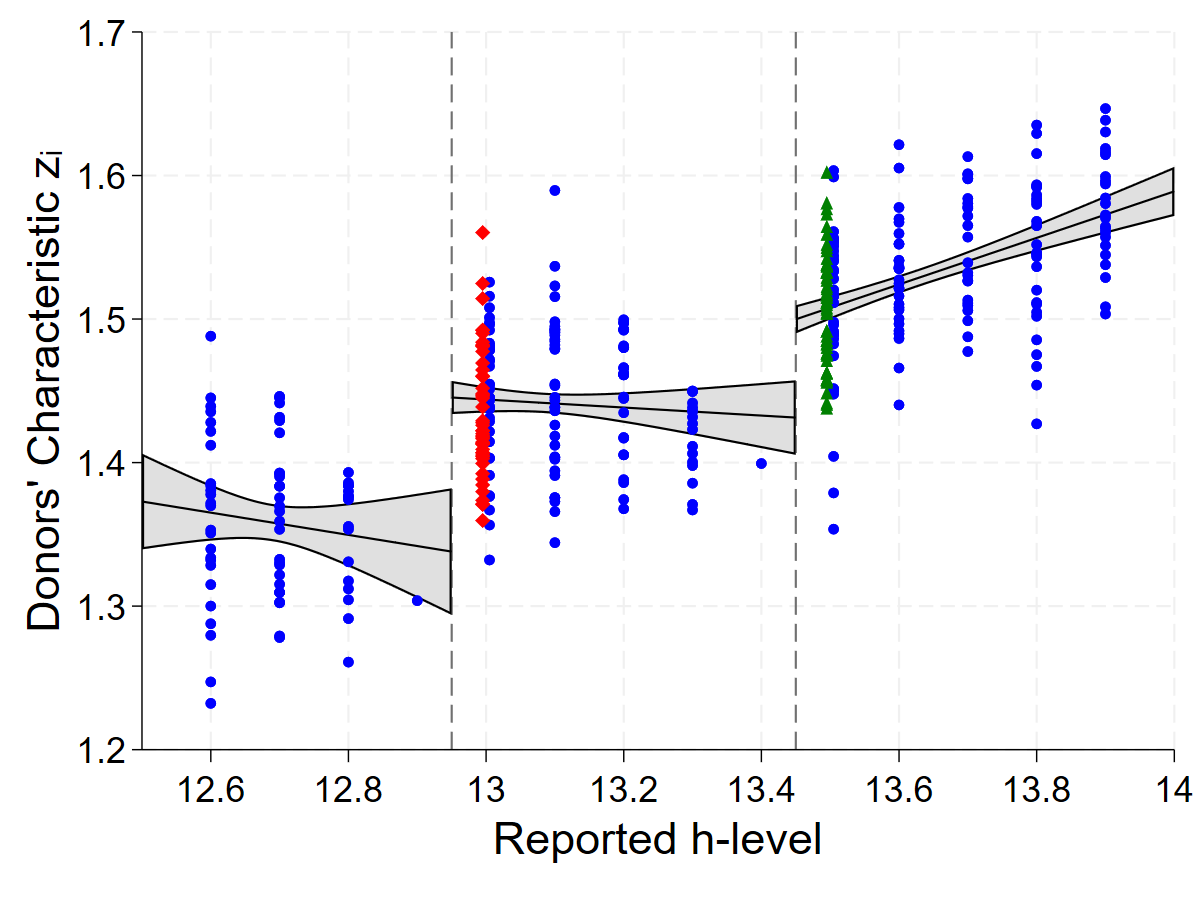}
\caption{Potential bias from misreporting. On the left, the x-axis is the measured h-level $\widetilde{h}$, on the right it is the reported h-level $h^*$. Red diamonds and green triangles represents misreported observations. \label{fig:theory_h2_2cutoff}}
\end{figure}

\begin{figure}[ht]
	\centering
		\includegraphics[width=0.49\textwidth]{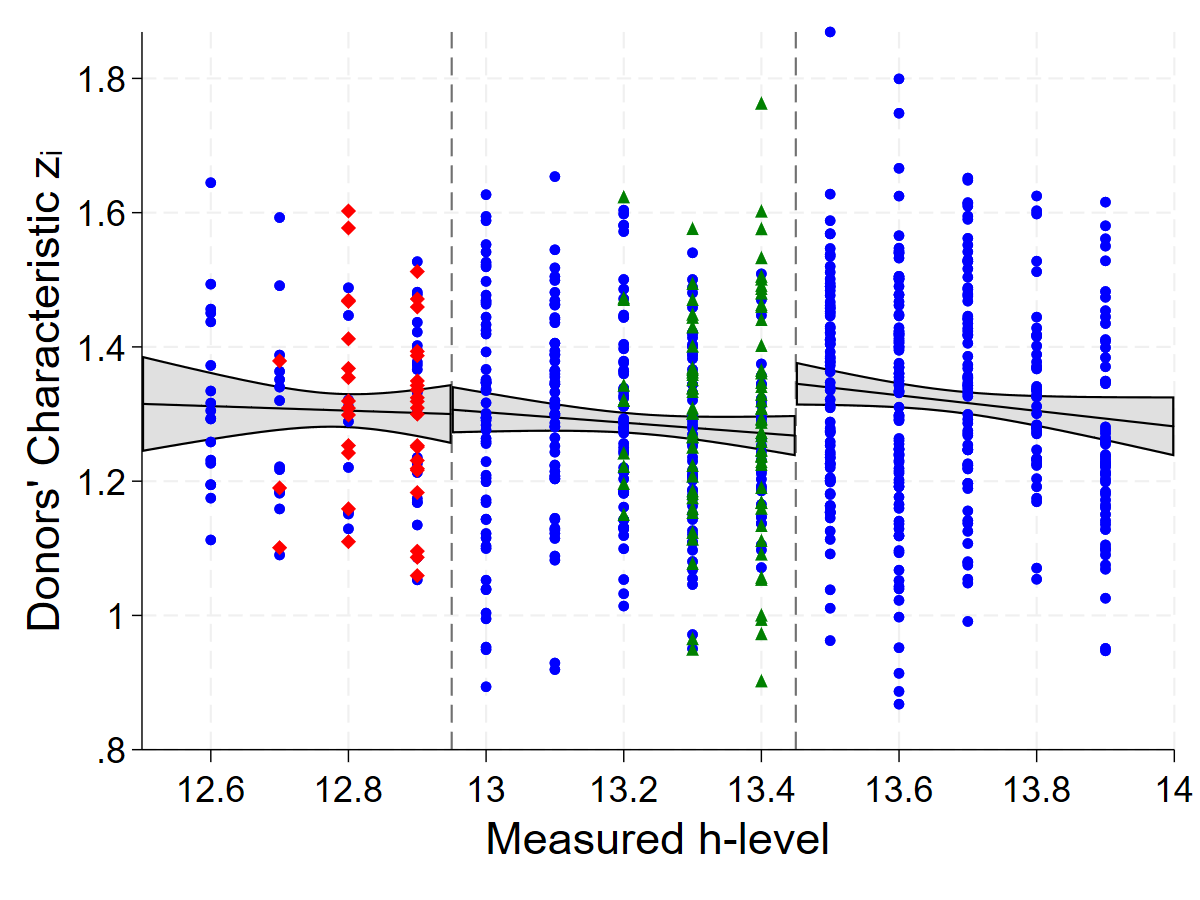}
          \includegraphics[width=0.49\textwidth]{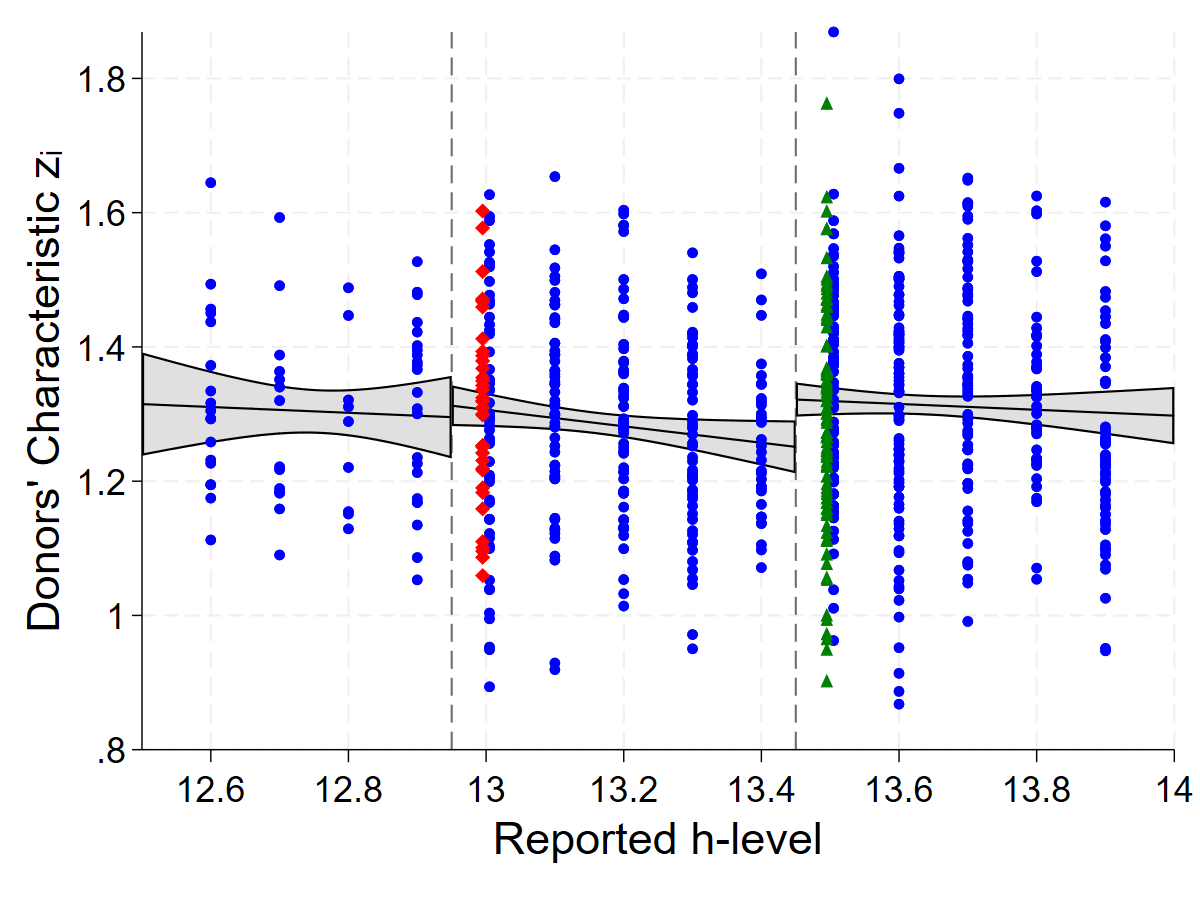}
\caption{Potential bias from misreporting. On the left, the x-axis is the measured h-level $\widetilde{h}$, on the right panel the reported h-level $h^*$. Red diamonds and green triangles represents misreported observations.\label{fig:h_manipulated_2cutoff_notincreasing}}
\end{figure}

Figure \ref{fig:theory_h2_2cutoff} and \ref{fig:h_manipulated_2cutoff_notincreasing} are the analogous to Figure \ref{fig:theory_h2} and \ref{fig:h_manipulated_notincreasing} in the case with two thresholds.

\section{Robustness Checks}\label{sec:robustCheck}

\subsection{Robustness to Different Manipulation Regions}\label{sec:robustCheck_manip}

\begin{figure}[ht]
	\centering
	\includegraphics[width=0.49\textwidth]{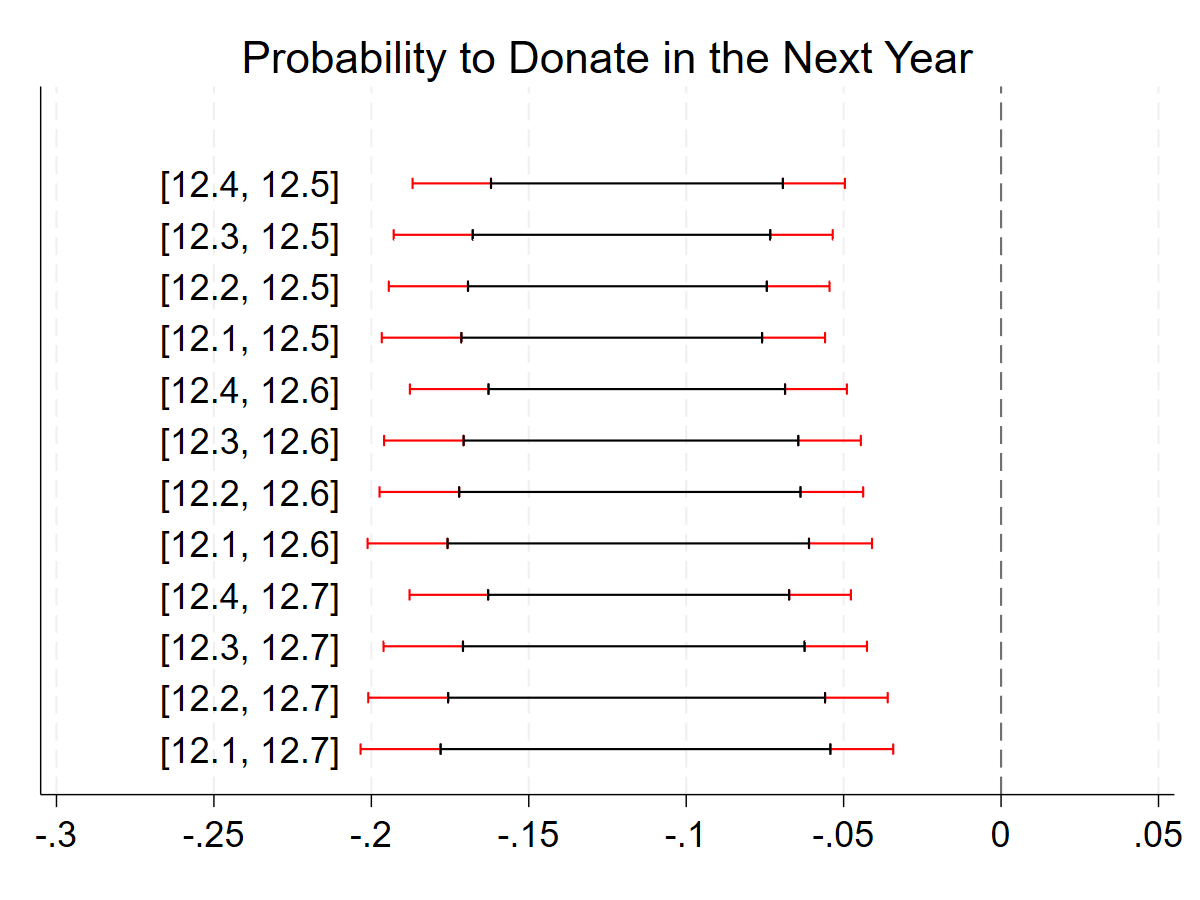}
        \includegraphics[width=0.49\textwidth]{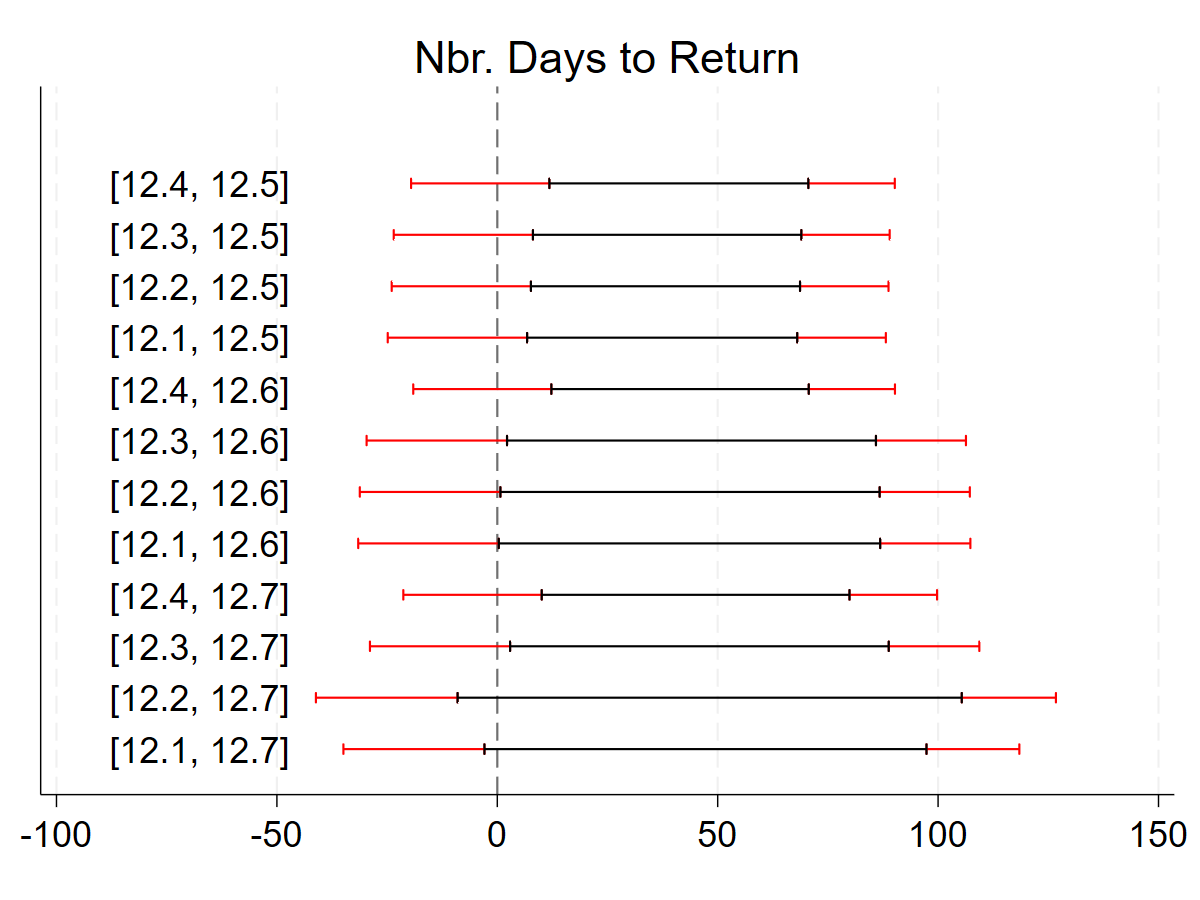}
\caption{Causal effect of a deferral on the probability of attempting to donate within the next year (Left) and on the number of days the donor takes to return if she returns (Right) for various manipulation regions. In black the lower and upper bounds and in red the 95\% confidence intervals for those bounds. \label{fig:robustnessmanipregionwomen}}
\end{figure}

We apply our procedure across various manipulation regions. Figure \ref{fig:robustnessmanipregionwomen}, Table \ref{tab:manipregion_women_preturn12months} and \ref{tab:manipregion_women_Ndaytonextdonation} show the results. Results are very stable. 

\begin{table}[ht]
\centering
\begin{tabular}{cccccc}
\toprule
\multicolumn{2}{c}{\textbf{\begin{tabular}[c]{@{}c@{}}Manipulation \\ Region\end{tabular}}}  & 
\multicolumn{2}{c}{\textbf{\begin{tabular}[c]{@{}c@{}}Partial\\ Identification\end{tabular}}}   & 
\multicolumn{2}{c}{\textbf{\begin{tabular}[c]{@{}c@{}}Confidence\\Interval \end{tabular}}}   \\ \midrule
\textbf{\begin{tabular}[c]{@{}l@{}}Start \end{tabular}} & \textbf{\begin{tabular}[c]{@{}l@{}}End\end{tabular}} & \textbf{\begin{tabular}[c]{@{}l@{}}Lower \\ Bound\end{tabular}} & \textbf{\begin{tabular}[c]{@{}l@{}}Upper \\ Bound\end{tabular}} & \textbf{\begin{tabular}[c]{@{}l@{}}Lower \\ Bound\end{tabular}} & \textbf{\begin{tabular}[c]{@{}l@{}}Upper \\ Bound\end{tabular}} \\ \midrule
12.4&	12.5&	-0.162	&-0.069&	-0.187&	-0.050 \\
12.3&	12.5&	-0.168&	-0.073&	-0.193	&-0.054 \\ 
12.2&	12.5&	-0.169&	-0.074	&-0.195	&-0.055 \\ 
12.1&	12.5&	-0.171&	-0.076&	-0.197&	-0.056 \\
12.4&	12.6&	-0.163&	-0.069&	-0.188&	-0.049 \\ 
12.3&	12.6&	-0.171&	-0.064&	-0.196	&-0.045 \\ 
12.2&	12.6&	-0.172&	-0.064	&-0.197&	-0.044 \\ 
12.1&	12.6&	-0.176&	-0.061	&-0.201&	-0.041  \\ 
12.4&	12.7&	-0.171&	-0.062&	-0.196&	-0.043\\
12.2&	12.7&	-0.176&	-0.056&	-0.201&	-0.036\\
\textit{12.1}&	\textit{12.7}&	\textit{-0.178}&	\textit{-0.054}&	\textit{-0.203}&	\textit{-0.034}\\
\bottomrule
\end{tabular}
\caption{\label{tab:manipregion_women_preturn12months} Causal effect of a deferral on the probability of attempting to donate within the next year for various manipulation regions. In italic the preferred specification. }
\end{table}

\begin{table}[ht]
\centering
\begin{tabular}{cccccc}
\toprule
\multicolumn{2}{c}{\textbf{\begin{tabular}[c]{@{}c@{}}Manipulation \\ Region\end{tabular}}}  & 
\multicolumn{2}{c}{\textbf{\begin{tabular}[c]{@{}c@{}}Partial\\ Identification\end{tabular}}}   & 
\multicolumn{2}{c}{\textbf{\begin{tabular}[c]{@{}c@{}}Confidence\\Interval \end{tabular}}}   \\ \midrule
\textbf{\begin{tabular}[c]{@{}l@{}}Start \end{tabular}} & \textbf{\begin{tabular}[c]{@{}l@{}}End\end{tabular}} & \textbf{\begin{tabular}[c]{@{}l@{}}Lower \\ Bound\end{tabular}} & \textbf{\begin{tabular}[c]{@{}l@{}}Upper \\ Bound\end{tabular}} & \textbf{\begin{tabular}[c]{@{}l@{}}Lower \\ Bound\end{tabular}} & \textbf{\begin{tabular}[c]{@{}l@{}}Upper \\ Bound
\end{tabular}} \\ \midrule
12.4&	12.5	&11.81	&70.57	&-19.56	&90.18 \\
12.3&	12.5	&8.08	&68.98	&-23.49	&89.02 \\
12.2&	12.5	&7.63	&68.69	&-23.96	&88.75 \\
12.1&	12.5	&6.77	&68.05	&-24.85	&88.15 \\ 
12.4&	12.6	&12.27	&70.65	&-19.06	&90.22 \\
12.3&	12.6	&2.21	&85.91	&-29.65	&106.32 \\
12.2&	12.6	&0.70	&86.74	&-31.19	&107.20 \\
\textit{12.1}&	\textit{12.6}	&\textit{0.34}	&\textit{86.87}	&\textit{-31.56}	&\textit{107.32} \\
12.4&	12.7	&10.08	&79.90	&-21.31	&99.77  \\
12.3&	12.7	&2.91	&88.80	&-28.89	&109.35 \\ 
12.2&	12.7	&-9.01	&105.37	&-41.15	&126.73 \\
12.1&	12.7	&-2.91	&97.37	&-34.91	&118.43 \\
\bottomrule
\end{tabular}
\caption{\label{tab:manipregion_women_Ndaytonextdonation} Causal effect of a deferral on the number of days the donor takes to return if she returns for various manipulation regions. In italic the preferred specification. }
\end{table}

\FloatBarrier
\subsection{Robustness to number of Knots}\label{sec:robustCheck_knots}

\begin{figure}[ht]
	\centering
	\includegraphics[width=0.49\textwidth]{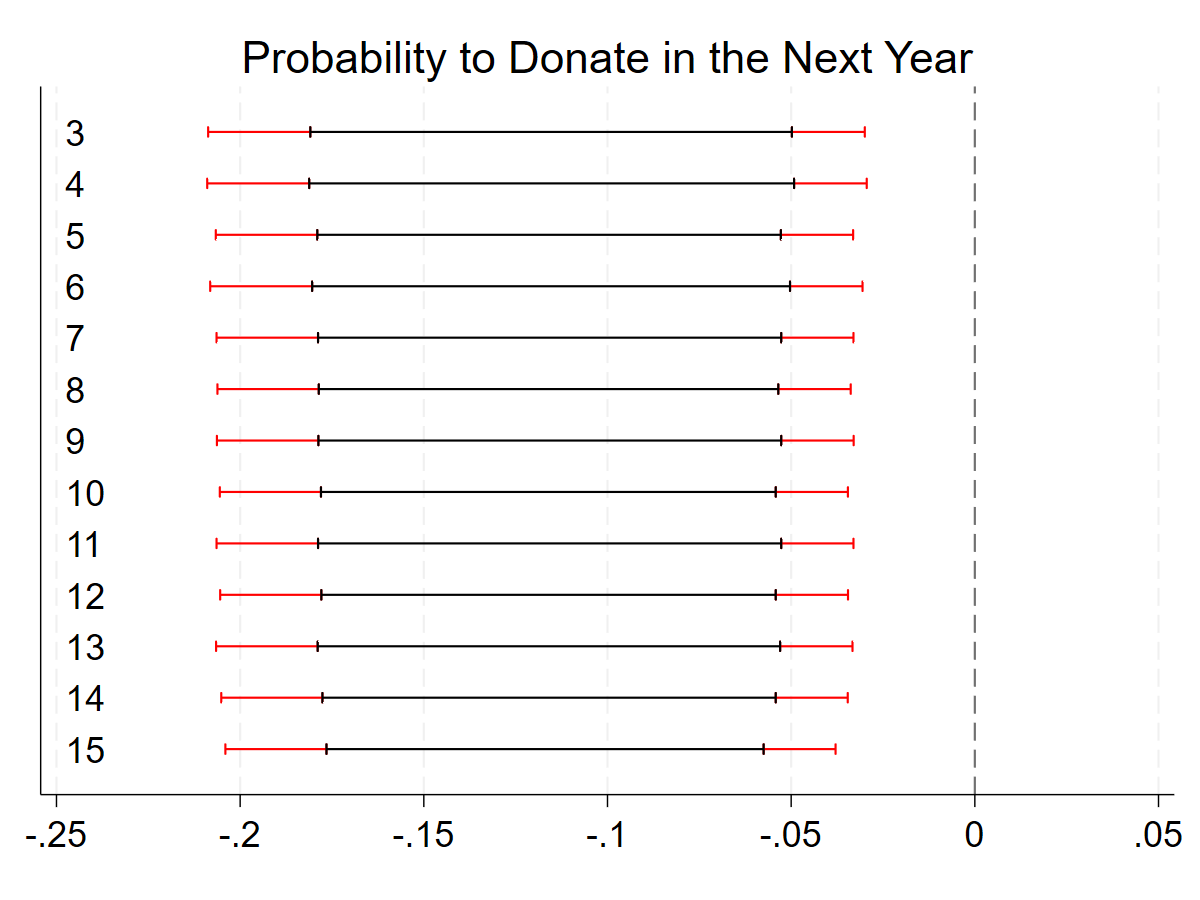}
\caption{Causal effect of a deferral on the probability of attempting to donate within the next year and on the number of days the donor takes to return if she returns for various knots. The number of knots is displayed on the left. In black the lower and upper bounds and in red the 95\% confidence intervals for those bounds. \label{fig:robustCheck_knots_women}}
\end{figure}

We apply our procedure across different number of knots. Figure \ref{fig:robustCheck_knots_women} and Table \ref{tab:robustCheck_knots_women} show the results. Results are insensitive to this choice.

\begin{table}[ht]
\centering
\begin{tabular}{c cc cc}
\toprule
\textbf{Knots}  & 
\multicolumn{2}{c}{\textbf{\begin{tabular}[c]{@{}c@{}}Partial\\ Identification\end{tabular}}}   & 
\multicolumn{2}{c}{\textbf{\begin{tabular}[c]{@{}c@{}}Confidence\\Interval \end{tabular}}}   \\ \midrule
& \textbf{\begin{tabular}[c]{@{}l@{}}Lower \\ Bound\end{tabular}} & \textbf{\begin{tabular}[c]{@{}l@{}}Upper \\ Bound\end{tabular}} & \textbf{\begin{tabular}[c]{@{}l@{}}Lower \\ Bound\end{tabular}} & \textbf{\begin{tabular}[c]{@{}l@{}}Upper \\ Bound
\end{tabular}} \\ \midrule
3	& -0.181	& -0.050	&-0.209	&-0.030 \\
4	& -0.181	& -0.049	&-0.209	&-0.029 \\
5	& -0.179	& -0.053	&-0.207	&-0.033 \\
6	& -0.180	& -0.050	&-0.208	&-0.031 \\
7	& -0.179	& -0.053	&-0.206	&-0.033 \\
8	& -0.179	& -0.054	&-0.206	&-0.034 \\
9	& -0.179	& -0.053	&-0.206	&-0.033 \\
\textit{10}	& \textit{-0.178}	& \textit{-0.054}	& \textit{-0.206}	& \textit{-0.035} \\
11	& -0.179	& -0.053	&-0.206	&-0.033 \\
12	& -0.178	& -0.054	&-0.205	&-0.035 \\
13	& -0.179	& -0.053	&-0.207	&-0.033 \\
14	& -0.178	& -0.054	&-0.205	&-0.035 \\
15	& -0.177	& -0.058	&-0.204	&-0.038 \\
\bottomrule
\end{tabular}
\caption{\label{tab:robustCheck_knots_women} Causal effect of a deferral on the probability of attempting to donate within the next year for various knots with a manipulation region of 12.1 to 12.6. In italic the preferred specification. }
\end{table}

\FloatBarrier
\section{Figures for the estimation of the extensive margin \label{sec:figNdaytonextdonations}}

Figures for the estimation of the effect of a successful donation on the number of days donors take to return if she returns. 

\begin{figure}[h]
    \centering
    \includegraphics[width = 0.7\textwidth]{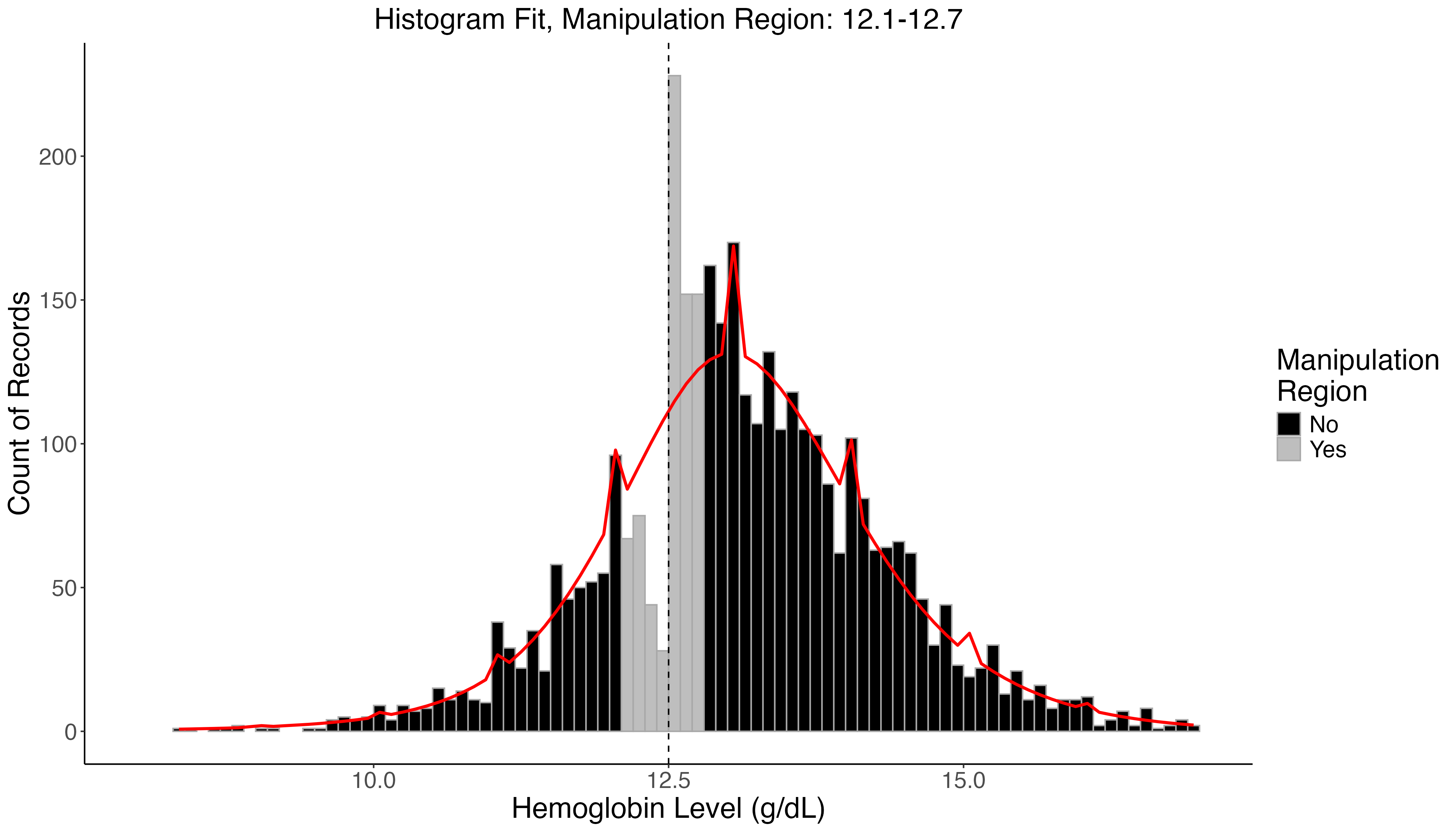}
    \caption{Histogram of hemoglobin levels for female donors who return. The bars in gray represent our manipulation window, while the bars in black represent levels outside the window. In red, we provide the estimated un-manipulated histogram.}
    \label{fig:unmaniphist_Ndaytonextdonations}
\end{figure}

\begin{figure}
  \begin{adjustwidth}{0cm}{}
    \centering
    \includegraphics[width = 0.49\textwidth]{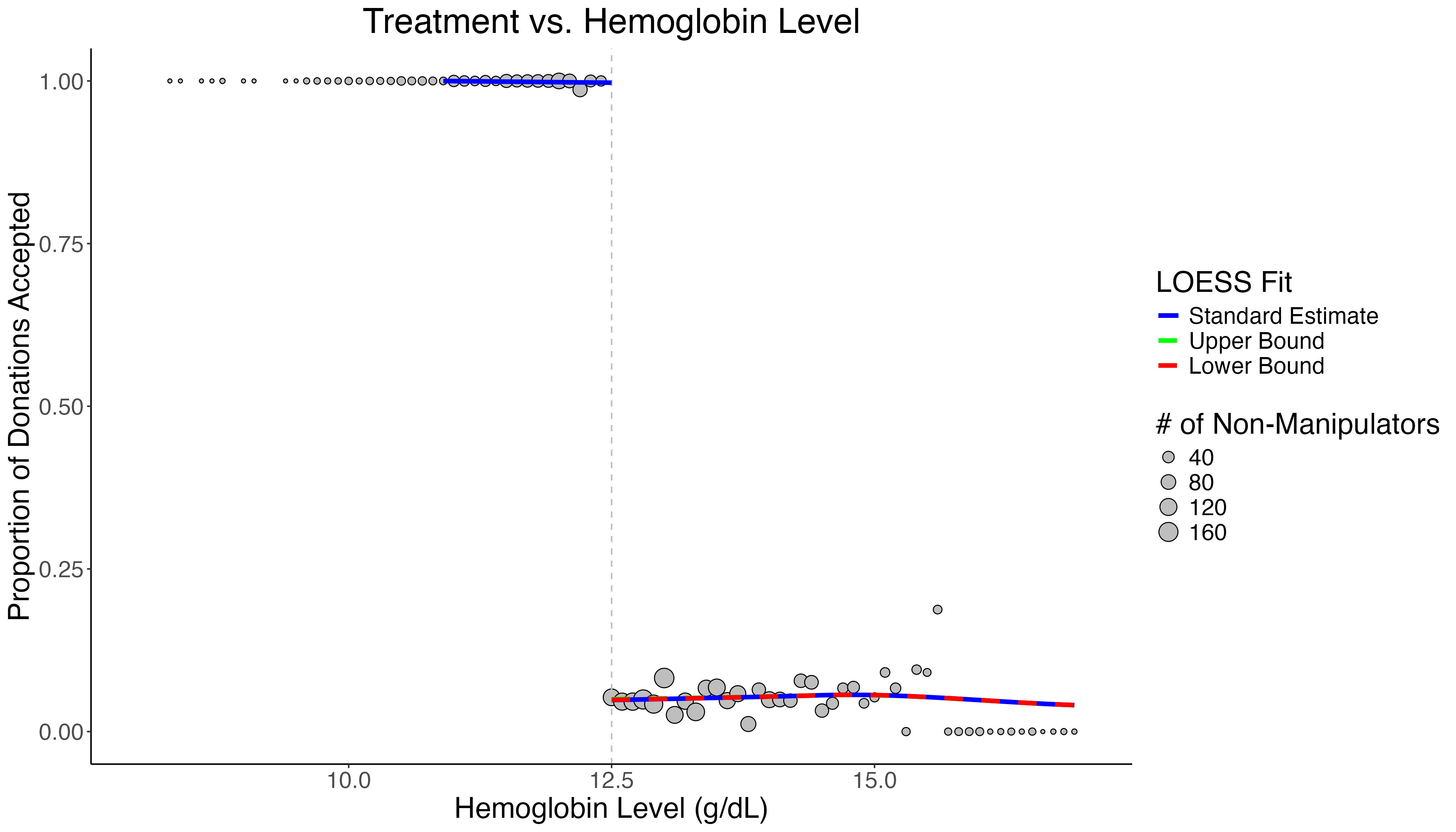}
    \includegraphics[width = 0.49\textwidth]{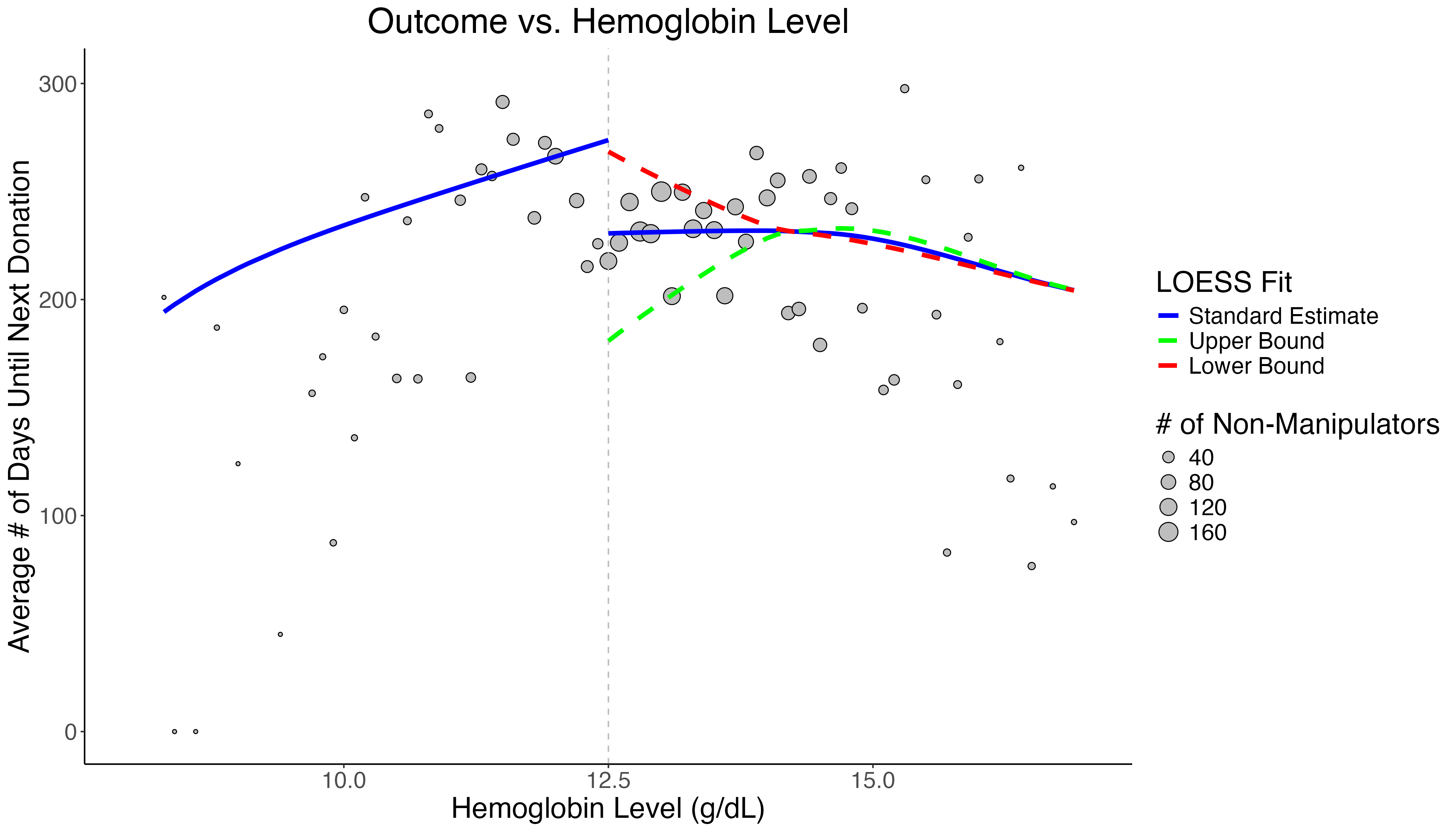}
    \caption{On the left, estimation of the denominator (probability to have a successful donation) and on the right the numerator (Number of days taken to return) of the causal effect estimator. Only female donors who return.}
    \label{fig:effectPlot_effectPlot}
 \end{adjustwidth}
\end{figure}

\FloatBarrier
\section{Placebo Tests}\label{sec:placeboCheck}

\begin{figure}[ht]
    \centering
	\includegraphics[width=0.49\textwidth]{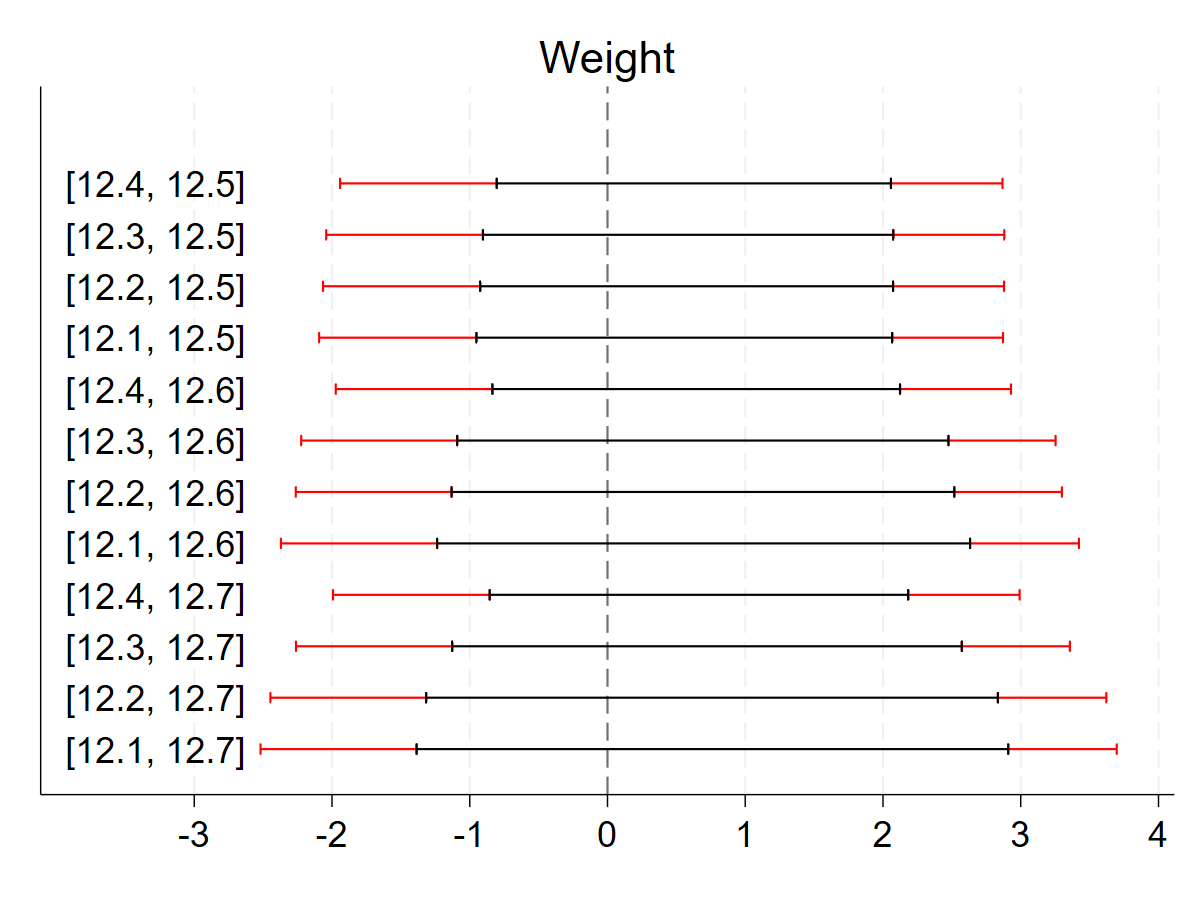}
        \includegraphics[width=0.49\textwidth]{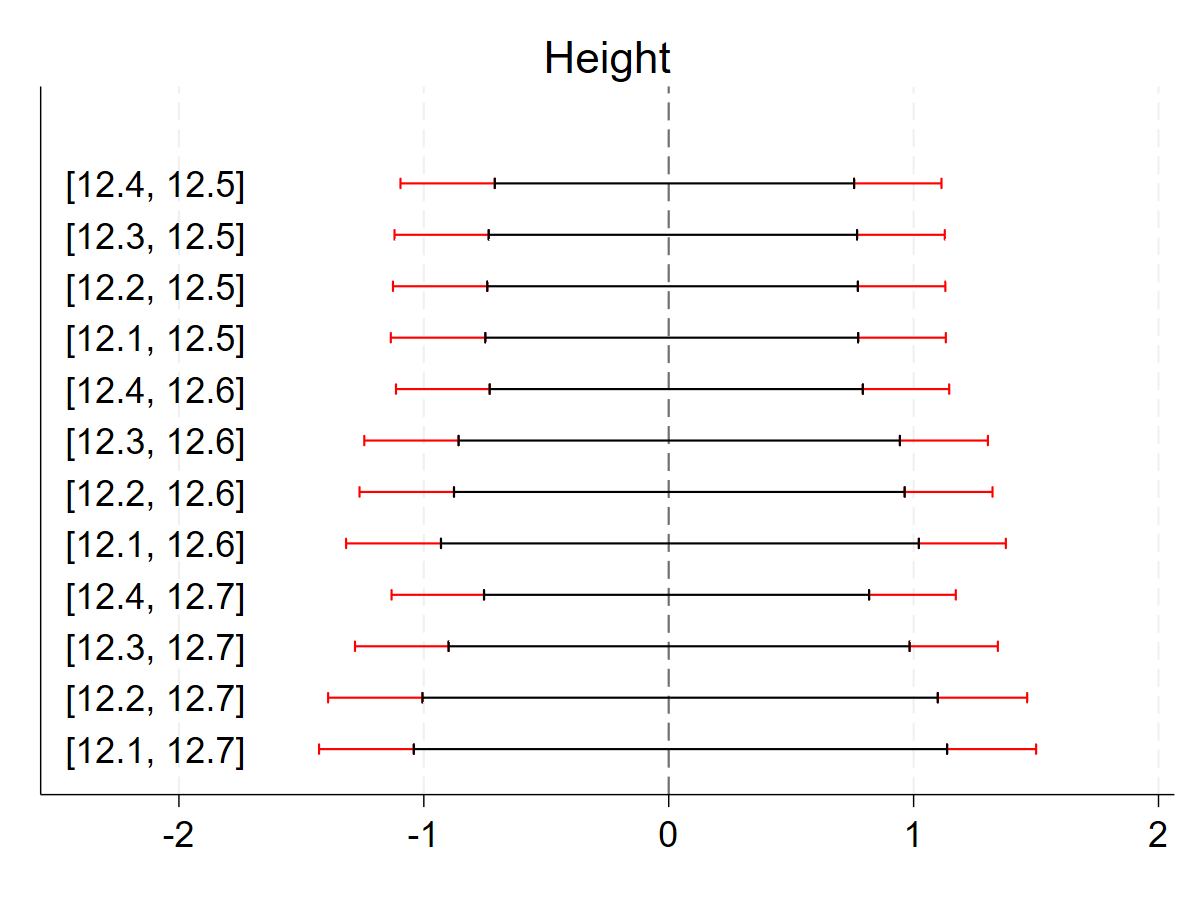}
        \includegraphics[width=0.49\textwidth]{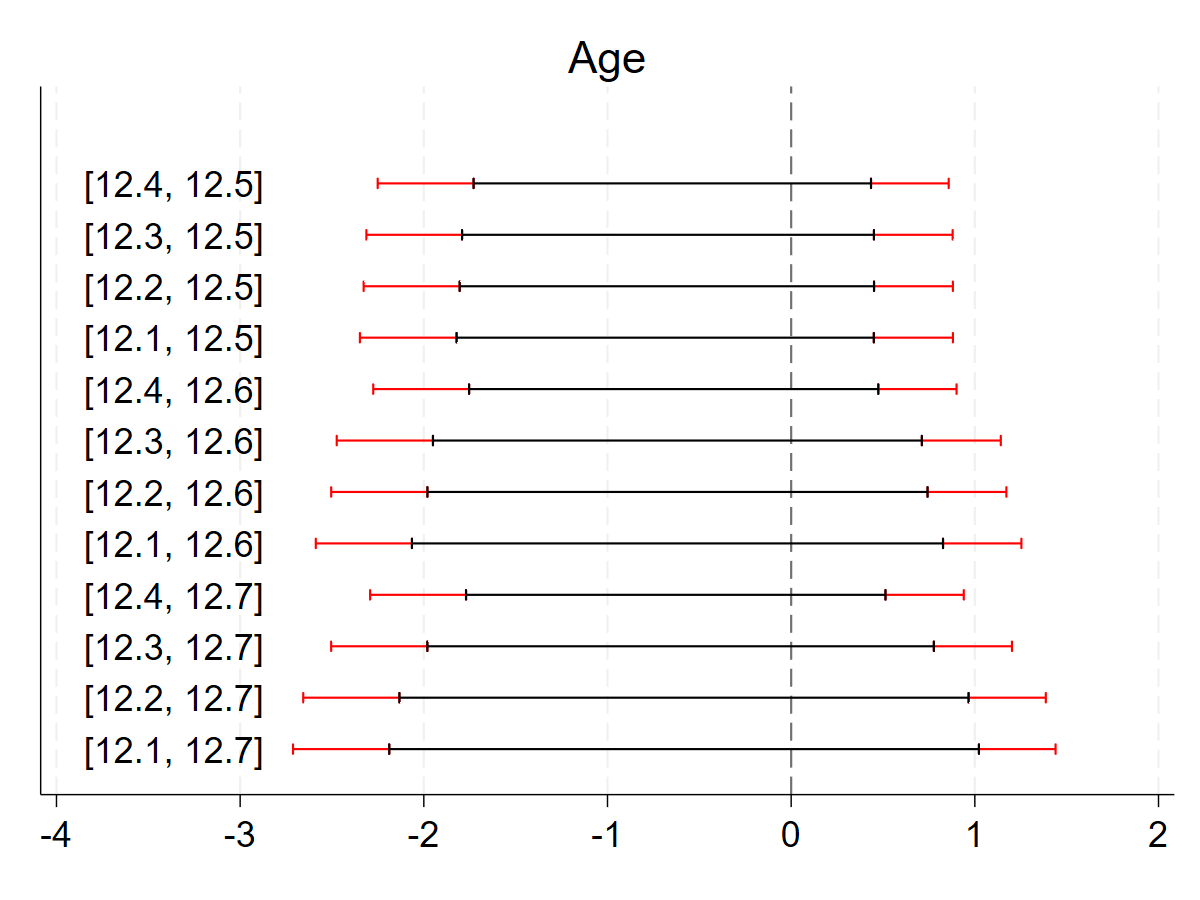}
\caption{Placebo test: causal effect of donation acceptance on a donor's weight (in kg), height (in cm) and age (in years) calculated for various manipulation regions. In black the lower and upper bounds and in red the 95\% confidence intervals for those bounds. \label{fig:placebo}}
\end{figure}

In Figure \ref{fig:placebo} and Tables \ref{tab:placeboWeight}, \ref{tab:placeboHeight}, and \ref{tab:placeboAge}, we repeat the analysis for three placebo variables: weight, height, and age. We again compute the partial identification bounds and confidence intervals for a range of manipulation regions.

Because there is no plausible relationship between deferral and these variables, we expect to see the intervals cover zero. For all three variables, the partial identification bounds indeed contain zero. 

\begin{table}[ht]
\centering
\begin{tabular}{cccccc}
\toprule
\multicolumn{2}{c}{\textbf{\begin{tabular}[c]{@{}c@{}}Manipulation \\ Region\end{tabular}}}  & 
\multicolumn{2}{c}{\textbf{\begin{tabular}[c]{@{}c@{}}Partial\\ Identification\end{tabular}}}   & 
\multicolumn{2}{c}{\textbf{\begin{tabular}[c]{@{}c@{}}Confidence\\Interval \end{tabular}}}   \\ \midrule
\textbf{\begin{tabular}[c]{@{}l@{}}Start \end{tabular}} & \textbf{\begin{tabular}[c]{@{}l@{}}End\end{tabular}} & \textbf{\begin{tabular}[c]{@{}l@{}}Lower \\ Bound\end{tabular}} & \textbf{\begin{tabular}[c]{@{}l@{}}Upper \\ Bound\end{tabular}} & \textbf{\begin{tabular}[c]{@{}l@{}}Lower \\ Bound\end{tabular}} & \textbf{\begin{tabular}[c]{@{}l@{}}Upper \\ Bound\end{tabular}} \\ \midrule
12.4	&12.5	&-0.804	&2.058	&-1.941	&2.868\\
12.3	&12.5	&-0.904	&2.075	&-2.042	&2.881\\
12.2	&12.5	&-0.924	&2.073	&-2.065	&2.879\\
12.1	&12.5	&-0.952	&2.067	&-2.093	&2.871\\
12.4	&12.6	&-0.835	&2.124	&-1.973	&2.930\\
12.3	&12.6	&-1.091	&2.475	&-2.224	&3.253\\
12.2	&12.6	&-1.132	&2.518	&-2.263	&3.299\\
12.1	&12.6	&-1.237	&2.632	&-2.371	&3.423\\
12.4	&12.7	&-0.855	&2.183	&-1.993	&2.992\\
12.3	&12.7	&-1.127	&2.572	&-2.261	&3.357\\
12.2	&12.7&	-1.316&	2.834	&-2.447	&3.621\\
\textit{12.1}	&\textit{12.7}	&\textit{-1.386}	&\textit{2.910}	&\textit{-2.519}	&\textit{3.697} \\
\bottomrule
\end{tabular}
\caption{\label{tab:placeboWeight} Placebo test: causal effect of donation acceptance on a donor's weight, calculated under various manipulation regions. In italic the preferred specification. }
\end{table}

\begin{table}[ht]
\centering
\begin{tabular}{cccccc}
\toprule
\multicolumn{2}{c}{\textbf{\begin{tabular}[c]{@{}c@{}}Manipulation \\ Region\end{tabular}}}  & 
\multicolumn{2}{c}{\textbf{\begin{tabular}[c]{@{}c@{}}Partial\\ Identification\end{tabular}}}   & 
\multicolumn{2}{c}{\textbf{\begin{tabular}[c]{@{}c@{}}Confidence\\Interval \end{tabular}}}   \\ \midrule
\textbf{\begin{tabular}[c]{@{}l@{}}Start \end{tabular}} & \textbf{\begin{tabular}[c]{@{}l@{}}End\end{tabular}} & \textbf{\begin{tabular}[c]{@{}l@{}}Lower \\ Bound\end{tabular}} & \textbf{\begin{tabular}[c]{@{}l@{}}Upper \\ Bound\end{tabular}} & \textbf{\begin{tabular}[c]{@{}l@{}}Lower \\ Bound\end{tabular}} & \textbf{\begin{tabular}[c]{@{}l@{}}Upper \\ Bound\end{tabular}} \\ \midrule
12.4&	12.5	&-0.710	&0.757	&-1.095	&1.114 \\
12.3&	12.5	&-0.735	&0.769	&-1.120	&1.127 \\
12.2&	12.5	&-0.741	&0.772	&-1.126	&1.130 \\
12.1&	12.5	&-0.749	&0.774	&-1.135	&1.132 \\
12.4&	12.6	&-0.731	&0.793	&-1.114	&1.145  \\
12.3&	12.6	&-0.858	&0.944	&-1.243	&1.303 \\
12.2&	12.6	&-0.877	&0.964	&-1.263	&1.322 \\
12.1&	12.6	&-0.930	&1.021	&-1.317	&1.377 \\
12.4&	12.7    &-0.754	&0.818	&-1.132	&1.172 \\
12.3&	12.7    &-0.899	&0.984	&-1.281	&1.344 \\
12.2&	12.7	&-1.006	&1.099	&-1.391	&1.464 \\
\textit{12.1}&	\textit{12.7}	&\textit{-1.041}	&\textit{1.137}	&\textit{-1.428}	&\textit{1.500} \\
\bottomrule
\end{tabular}
\caption{\label{tab:placeboHeight} Placebo test: causal effect of donation acceptance on a donor's height, calculated under various manipulation regions. In italic the preferred specification. }
\end{table}

\begin{table}[ht]
\centering
\begin{tabular}{cccccc}
\toprule
\multicolumn{2}{c}{\textbf{\begin{tabular}[c]{@{}c@{}}Manipulation \\ Region\end{tabular}}}  & 
\multicolumn{2}{c}{\textbf{\begin{tabular}[c]{@{}c@{}}Partial\\ Identification\end{tabular}}}   & 
\multicolumn{2}{c}{\textbf{\begin{tabular}[c]{@{}c@{}}Confidence\\Interval \end{tabular}}}   \\ \midrule
\textbf{\begin{tabular}[c]{@{}l@{}}Start \end{tabular}} & \textbf{\begin{tabular}[c]{@{}l@{}}End\end{tabular}} & \textbf{\begin{tabular}[c]{@{}l@{}}Lower \\ Bound\end{tabular}} & \textbf{\begin{tabular}[c]{@{}l@{}}Upper \\ Bound\end{tabular}} & \textbf{\begin{tabular}[c]{@{}l@{}}Lower \\ Bound\end{tabular}} & \textbf{\begin{tabular}[c]{@{}l@{}}Upper \\ Bound\end{tabular}} \\ \midrule
12.4&	12.5&	-1.729&	0.435&	-2.251&	0.858 \\
12.3&	12.5&	-1.792&	0.451&	-2.313&	0.879 \\
12.2&	12.5&	-1.805&	0.451&	-2.328&	0.881 \\
12.1&	12.5&	-1.822&	0.450&	-2.348&	0.881 \\
12.4&	12.6&	-1.753&	0.475&	-2.275&	0.900 \\
12.3&	12.6&	-1.950&	0.712&	-2.474&	1.141 \\
12.2&	12.6&	-1.981&	0.743&	-2.504&	1.172 \\
12.1&	12.6&	-2.065&	0.827&	-2.588&	1.254 \\
12.4&	12.7&	-1.770&	0.514&	-2.292&	0.940 \\
12.3&	12.7&	-1.981&	0.777&	-2.505&	1.202 \\
12.2&	12.7&	-2.133&	0.966&	-2.657&	1.386 \\
\textit{12.1}&	\textit{12.7}&	\textit{-2.188}&	\textit{1.022}&	\textit{-2.712}&	\textit{1.440} \\
\bottomrule
\end{tabular}
\caption{\label{tab:placeboAge} Placebo test: causal effect of donation acceptance on a donor's age, calculated under various manipulation regions. In italic the preferred specification. }
\end{table}

\section{Comparing Alternative Methods}

\FloatBarrier
\subsection{Donut hole estimates}
\label{sec:donut-robust}

\begin{figure}[ht]
	\centering
	\includegraphics[width=0.6\textwidth]{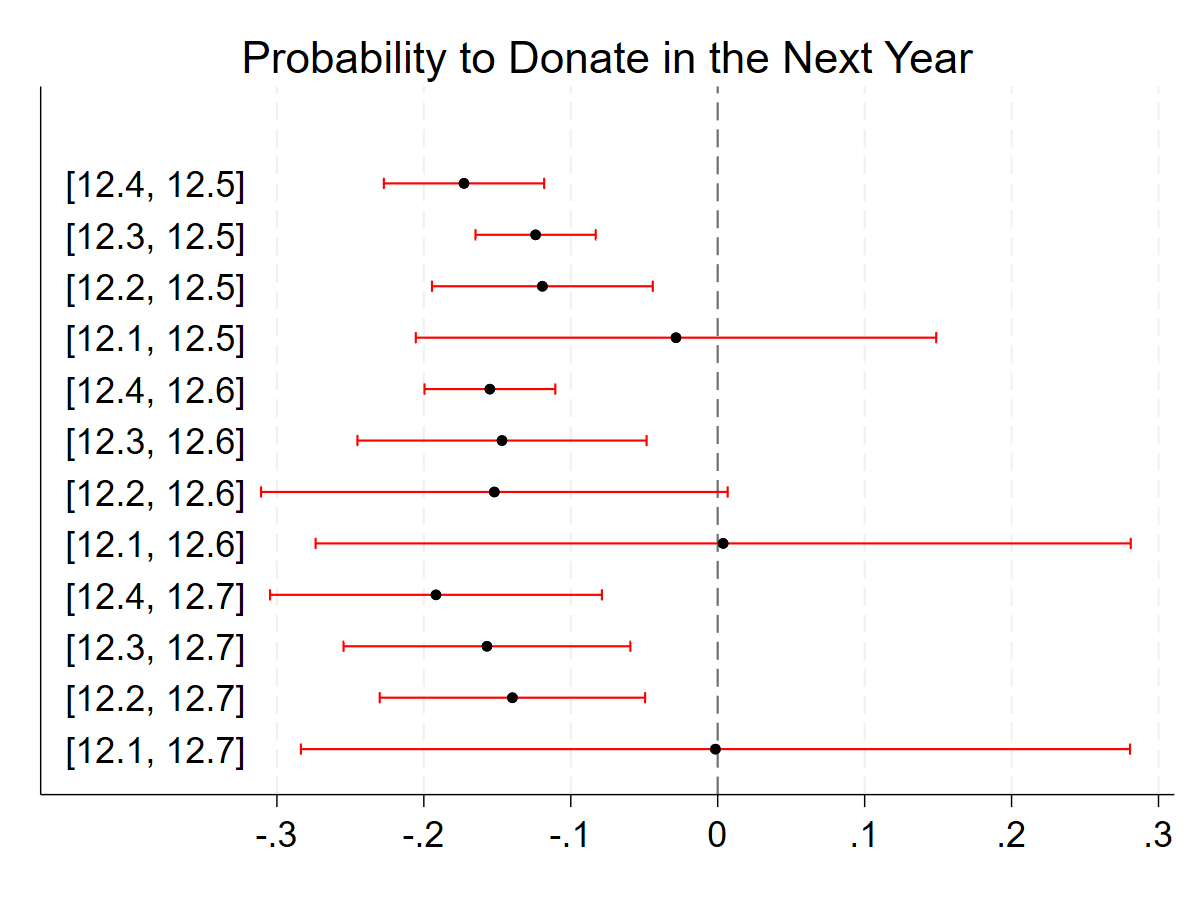}
\caption{Donut hole estimation for the causal effect of donation acceptance on probability of attempting to donate within the next year for various manipulation regions. In black the point estimates and in red the 95\% confidence intervals for those bounds. \label{fig:donuthole_manipregion_women}}
\end{figure}

In this section, we study the sensitivity of the donut hole estimates for the causal effect of donation acceptance (i.e. on the donor \textit{not} being deferred) on their probability of returning to the blood bank to attempt another donation within a year. For our candidate set of donut holes, we use the same specifications of the manipulation region as in Appendix \ref{sec:robustCheck}. For a given donut hole, we delete all observations within it and compute ``robust'' RD point estimates at the cutoff and 95\% confidence intervals using the rdrobust R package. The results are reproduced in Figure \ref{fig:donuthole_manipregion_women} and Table \ref{tab:donuthole_manipregion_women}, with the preferred donut hole specification in italics in Table \ref{tab:donuthole_manipregion_women}. Depending on the specification of the size and location of the donut hole, we may or may not reject the null hypothesis of no causal effects at the 95\% confidence level.

\begin{table}[ht]
\centering
\begin{tabular}{ccccrr}
\toprule
\multicolumn{2}{c}{\textbf{\begin{tabular}[c]{@{}c@{}}Donut\\Hole \end{tabular}}}  & \multicolumn{2}{c}{\textbf{\begin{tabular}[c]{@{}c@{}}Point\\Estimates \end{tabular}}}                                            & \multicolumn{2}{c}{\textbf{\begin{tabular}[c]{@{}c@{}}Confidence\\Interval \end{tabular}}}                                                                         \\ \midrule
\textbf{Start}          & \textbf{End}           & \textbf{Coefficient} & \textbf{\begin{tabular}[c]{@{}c@{}}Standard \\ Error\end{tabular}} & \textbf{\begin{tabular}[c]{@{}c@{}}Lower \\ Bound\end{tabular}} & \textbf{\begin{tabular}[c]{@{}c@{}}Upper \\ Bound\end{tabular}} \\ \midrule
12.4&	12.5&	$-0.173$&	0.028	&$-0.227$ & $-0.118$ \\
12.3&	12.5&	$-0.124$&	0.021	&$-0.165$ & $-0.083$ \\
12.2&	12.5&	$-0.119$&	0.038	&$-0.194$ &	$-0.044$ \\
12.1&	12.5&	$-0.028$&	0.090	&$-0.205$ &	$0.149$ \\
12.4&	12.6&	$-0.155$&	0.023   &$-0.200$ &	$-0.111$ \\
12.3&	12.6&	$-0.147$&	0.050	&$-0.245$ &	$-0.048$ \\
12.2&	12.6&	$-0.152$&	0.081	&$-0.311$ &	$0.007$ \\
\textit{12.1}&	\textit{12.6}&	\textit{0.004}&   \textit{0.142}   &\textit{$-$0.274} &	\textit{0.281} \\
12.4&	12.7&	$-0.192$&	0.058	&$-0.305$ &	$-0.07$ \\
12.3&	12.7&	$-0.157$&	0.050	&$-0.255$ &	$-0.059$ \\
12.2&	12.7&	$-0.140$&	0.046	&$-0.230$ &	$-0.049$ \\
12.1&	12.7&	$-0.002$&	0.144	&$-0.284$ &	$0.281$ \\
 \bottomrule                                   
\end{tabular}
\caption{\label{tab:donuthole_manipregion_women} The donut hole RD point estimates and confidence intervals of the causal effect of blood donation acceptance on the future return probability of donors within a year. In italic the preferred specification.}
\end{table}

\FloatBarrier
\subsection{GRR2020}

\begin{table}[ht]
\centering
\addtolength{\tabcolsep}{5pt}    
\begin{tabular}{ccrrrr}
\toprule
\textbf{Bandwidth} & \multirow{2}{*}{\textbf{\begin{tabular}[c]{@{}c@{}}Est. \\ Proportion of \\ Manipulators\end{tabular}}} & 
\multicolumn{2}{c}{\textbf{\begin{tabular}[c]{@{}c@{}}Partial\\ Identification\end{tabular}}}   & 
\multicolumn{2}{c}{\textbf{\begin{tabular}[c]{@{}c@{}}Confidence\\Interval \end{tabular}}}   \\ \cline{3-6}
 &  & \textbf{\begin{tabular}[c]{@{}l@{}}Lower \\ Bound\end{tabular}} & \textbf{\begin{tabular}[c]{@{}l@{}}Upper \\ Bound\end{tabular}} & \textbf{\begin{tabular}[c]{@{}l@{}}Lower \\ Bound\end{tabular}} & \textbf{\begin{tabular}[c]{@{}l@{}}Upper \\ Bound\end{tabular}} \\ \midrule
0.4 & 0.93	&$-1.00$	&1.00	&$-1.55$	&1.18  \\
0.5 &0.82	&$-0.89$	&0.39	&$-0.95$	&0.55 \\
0.6 &0.80	&$-0.90$	&0.31	&$-0.94$	&0.41 \\
0.7 &0.75	&$-0.88$	&0.26	&$-0.93$	&0.33 \\
0.8 &0.69   &$-0.87$    &0.24	&$-0.94$	&0.28 \\
0.9 &0.64   &$-0.84$    &0.23	&$-0.92$	&0.26 \\
1   &0.61   &$-0.80$	&0.22	&$-0.89$	&0.25 \\
\bottomrule
\end{tabular}
\caption{\label{tab:gerard} Results shown in Figure \ref{fig:gerard}. Causal effect of donation acceptance on probability of attempting to donate within the next year, calculated under various bandwidths using the GRR20202's method. }
\end{table}

\section{Additional Summary Statistics}

\begin{table}[ht]
    \centering
    \begin{tabular}{l c cccc}
         \toprule
         Type of  &  \multicolumn{4}{c}{Sample Size} &  N Days\\
         Donations  & \multicolumn{2}{c}{N} & \multicolumn{2}{c}{\%} & Deferred    \\
         \cmidrule(lr){2-3} \cmidrule(lr){4-5} \cmidrule(lr){6-6}  
           & Male & Female & Male & Female & All \\
         \cmidrule(lr){2-3} \cmidrule(lr){4-5} \cmidrule(lr){6-6} 
         & \multicolumn{5}{c}{Successful Donations} \\
         Whole Blood        & 181,989  & 17,364 & 84.22\% & 65.22\% & 56 \\
         Plasma/Platelet    & 20,751 &  1 & 9.60\% & 0\%  & 14 \\
         Red Cell Apheresis & 43 & 0 & 0.02\% & 0\%   & 112  \\
         \cmidrule(lr){2-3} \cmidrule(lr){4-5} \cmidrule(lr){6-6} 
         & \multicolumn{5}{c}{Unsuccessful Donations}\\
         Change of mind   & 2,706  & 842 & 1.25\% & 3.16\% & 0   \\
         Failed Phlebotomy & 1,115 & 314  & 0.50\% & 1.12\% & 1  \\
         Deferrals: \\
         \ \ Low h-level     & 6,779 & 7,331 & 3.14\%& 27.54\%  & 28  \\
         \ \ Other temporary  & 2,694 & 770 & 1.25\% & 2.89\%  & 96$^{\delta}$ \\
         \ \ Permanent       & - & - & 0\% & 0\% & -   \\ 
         \midrule
         Total         & 216,077  & 26,622 &   100\% &   100\% & -  \\
         \bottomrule
    \end{tabular}
    \caption{Sample size and number of days deferred for the different donations' outcomes. Only donations for which donors are eligible to come back one year before our dataset ends. $^{\delta}$ The reported number of days deferred for temporary deferrals is the average number of days deferred for such deferrals in our dataset.}
    \label{tab:sum_stat_oneyear}
\end{table}

\section{Causal effect of donating plasma for men \label{sec:resultsmen_plasma}}

\begin{figure}[ht]
	\centering
	\includegraphics[width=0.60\textwidth]{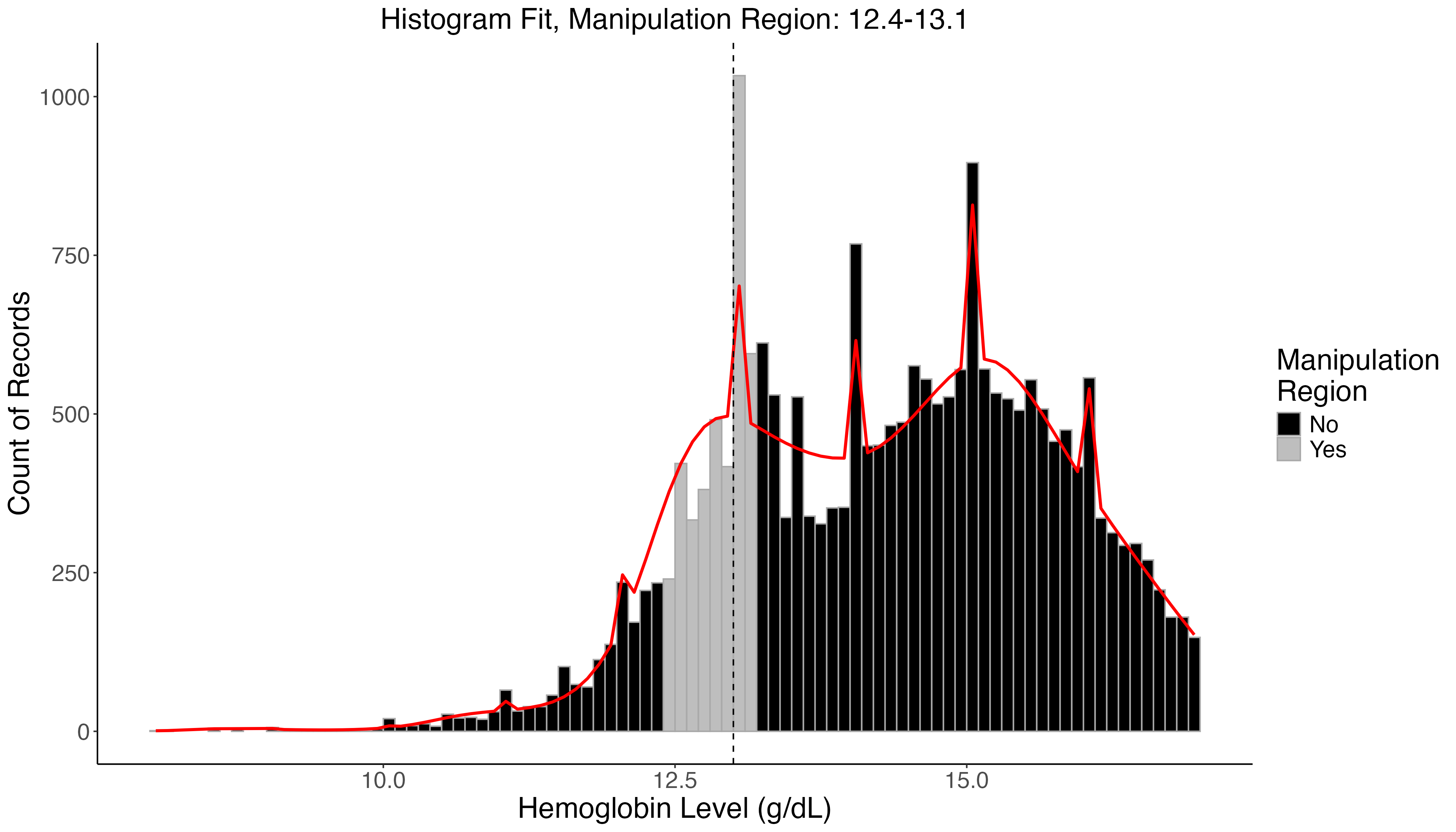}
\caption{Histogram of the distribution of h-level with the estimated un-manipulated histogram (in red) excluding whole blood donations. \label{fig:Men_Plasma_hist}}
\end{figure}

\begin{figure}[ht]
	\centering
        \includegraphics[width=0.49\textwidth]{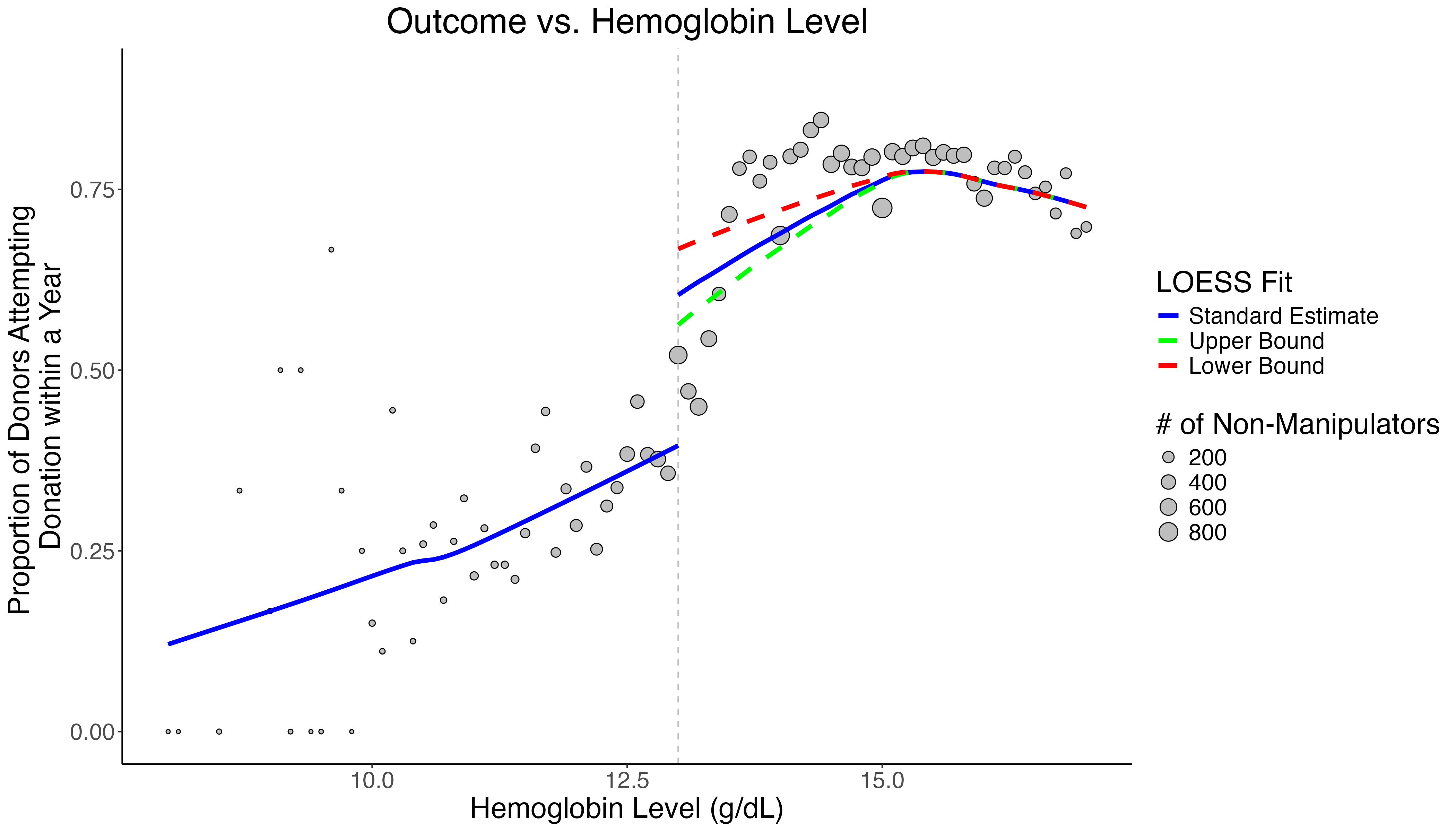}
        \includegraphics[width=0.49\textwidth]{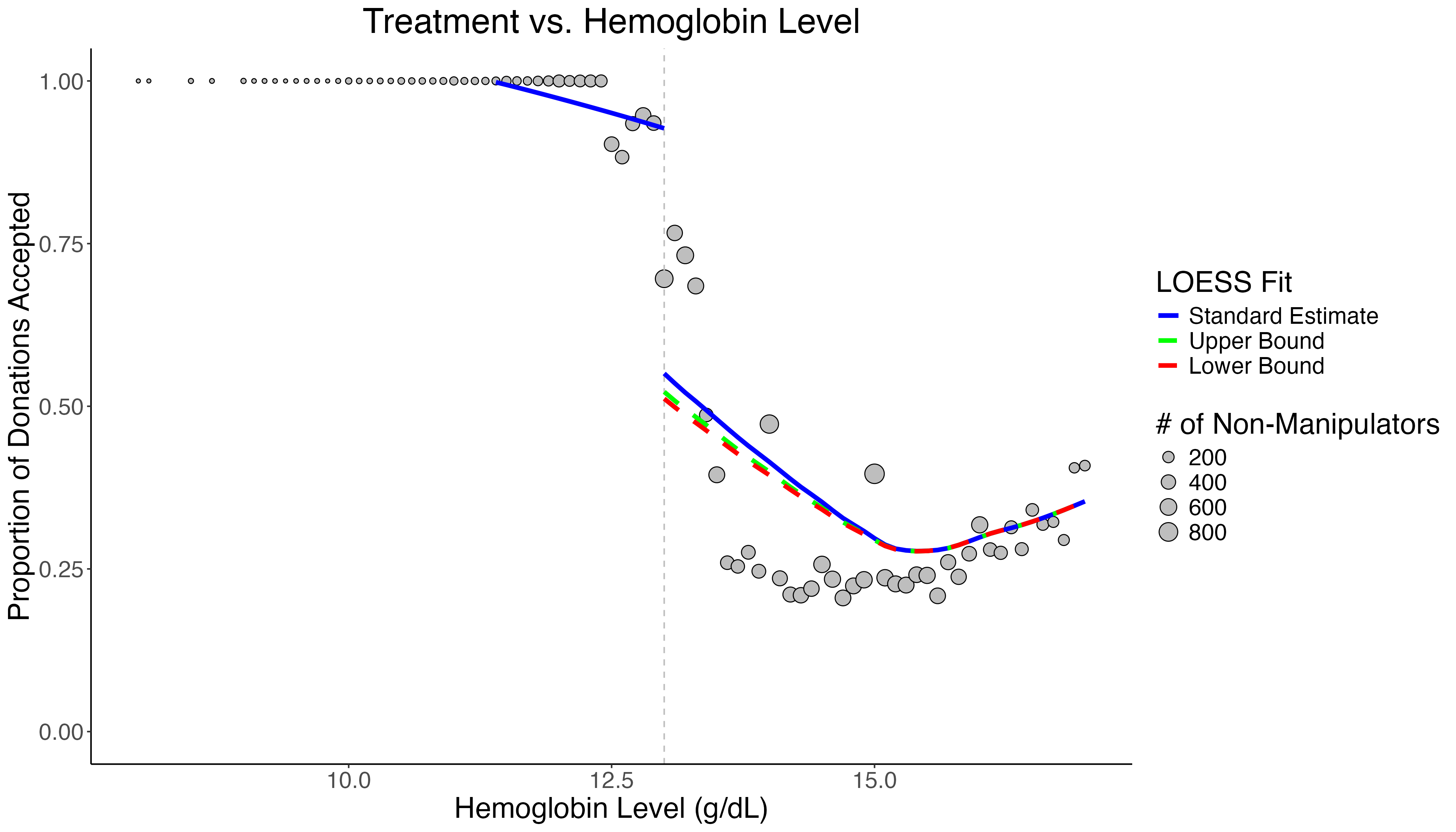}
\caption{Effect of being deferred for plasma donors. Estimation made at the 13g/DL threshold. On the left, estimation of the denominator (probability to make a successful donation), and on the right the numerator (probability to return in the next 12 months) of the causal effect estimator. The reweighted curves are shown in the green and red dashed lines. \label{fig:Men_Plasma}}
\end{figure}

In this section, we investigate the effect of being deferred at the threshold of 13g/dL, that is, the effect of being allowed to donate plasma vs being deferred. Similarly to the estimation at the whole blood cut-off in Section \ref{sec:resultmen}, we disregard whole blood donations from our sample before running the estimation with a manipulation window of 12.4 to 13.1. We are left with around 23,500 donations.  Figure \ref{fig:Men_Plasma_hist} and \ref{fig:Men_Plasma} show the different estimation steps. The histogram shows modest evidence of manipulation compared to the histograms in Figure \ref{fig:Men_WB_hist}. We estimate the effect to be between $-0.65$ and $-0.41$. 

We find a negative effect for plasma, and the effect is much larger than for whole blood donations. A possible reason is that if a donor is not manipulated up to 13 so that they can donate plasma, then they are unlikely to be able to donate whole blood or plasma in the future. However, if someone misses the 13.5d/DL threshold to donate whole blood, they might still have a chance to donate plasma in the future. Therefore, the temporary deferral for whole blood is less severe than that for plasma. 

As described in Section \ref{sec:resultmen}, in our dataset, we know the outcome of successful donations, but do not know the intended donations for unsuccessful donations. Therefore, when analyzing whole blood donations, we removed the successful plasma donations but did not removes the attempted plasma donations that were unsuccessful. The same issue exists when disregarding successful whole blood donations to estimate the effect at the plasma's threshold. Since the vast majority of donors in our sample are whole blood donors, any bias introduced by including unsuccessful plasma donors in the whole blood donation analyses is likely to be very small whereas any bias introduced by including unsuccessful whole blood donors in the plasma donation analyses might be larger.

\section{Figure Bruhin et al. (2020)}

\begin{figure}[ht]
	\centering
        \includegraphics[width=0.99\textwidth]{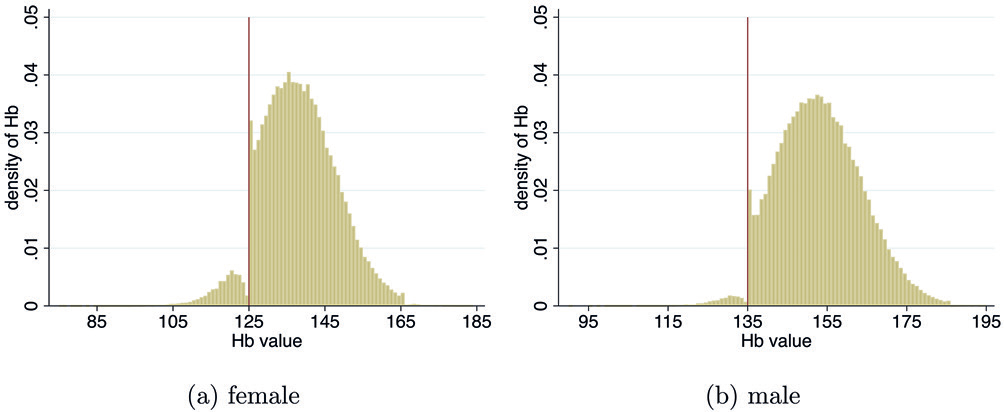}
\caption{Histogram of h-level in Bruhin et al. (2020).\label{fig:Histogram_Bruhin}}
\end{figure}


\end{document}